\RequirePackage{fix-cm}
\documentclass[smallextended]{svjour3}       % onecolumn (second format)
\smartqed  % flush right qed marks, e.g. at end of proof
\usepackage{graphicx}
\usepackage{subfigure}
\usepackage{amsmath}
\usepackage{amssymb}
\usepackage{url}
\usepackage{natbib}

\begin{document}

\title{X-ray reverberation around accreting black holes%\thanks{Grants or other notes
}


\author{P.~Uttley \and E.~M.~Cackett \and A.~C.~Fabian \and E.~Kara \and D.~R.~Wilkins}


\institute{P. Uttley \at Anton Pannekoek Institute, University of Amsterdam, Postbus 94249, 1090 GE Amsterdam, The Netherlands \\
 \email{p.uttley@uva.nl}\\
 \and
E. M. Cackett \at Department of Physics \& Astronomy, Wayne State University, 666 W. Hancock, Detroit, MI 48201, USA \\
%              Tel.: +123-45-678910\\
%              Fax: +123-45-678910\\
              \email{ecackett@wayne.edu}\\           %  \\
\and
 A. C. Fabian, E. Kara \at Institute of Astronomy, Madingley Road, Cambridge, CB3 0HA, UK\\
 \email{acf@ast.cam.ac.uk}\\
 \email{ekara@ast.cam.ac.uk}\\
 \and 
 D. R. Wilkins \at Department of Astronomy \& Physics, Saint Mary's University, Halifax, NS, B3H 3C3, Canada\\
\email{drw@ap.smu.ca}
}
\date{Received: date / Accepted: date}
\maketitle
\begin{abstract}
Luminous accreting stellar mass and supermassive black holes produce power-law
  continuum X-ray emission from a compact central corona. Reverberation time lags occur due to light travel time-delays between changes in the direct coronal
  emission and corresponding variations in its reflection from the
  accretion flow. Reverberation is detectable using light curves made
  in different X-ray energy bands, since the direct and reflected
  components have different spectral shapes. Larger, lower frequency,
  lags are also seen and are identified with propagation of fluctuations through the accretion flow and associated corona. We review the evidence for X-ray
  reverberation in active galactic nuclei and black hole X-ray binaries, showing how it can be best measured and how it may be  modelled. 
The timescales and energy-dependence of the high frequency
  reverberation lags show that much of the signal is originating from
  very close to the black hole in some objects, within a few
  gravitational radii of the event horizon.  We consider how these signals can be studied in the future to carry out X-ray reverberation mapping of the regions closest to black holes.
\end{abstract}

\section{Introduction}

Accreting black holes illuminate their surroundings, thereby making
both near and distant gas detectable. If, as is usually the case, the
luminosity varies with time, then the response from the surrounding
gas will also vary, but after a time delay due to light crossing time.
This delay or {\em reverberation} lag ranges from milliseconds to many
hours for irradiation of the innermost accretion flow at a few
gravitational radii ($r_{\rm g}=GM/c^2$) around black holes of mass $M$
ranging from 10 to $10^9 M_{\odot}$, respectively.

A prominent feature of most unobscured Active Galactic Nuclei (AGN) is
the Broad Line Region (BLR) consisting of clouds orbiting at
thousands~km/s at distances of light-days to light-months as the
black hole mass ranges from $10^6 - 10^9 M_{\odot}$. \citet{blandmckee82} showed how the resultant reverberation of the emission
lines following changes in the central ultraviolet continuum can be
combined with models for the gas velocity and ionization state of the
broad-line clouds to map their geometry via the {\it impulse response}\footnote{In Blandford \& McKee, and some subsequent optical and X-ray reverberation mapping work (including by the authors of this review), the impulse response is also called the {\it transfer function}.  However, impulse response is the formally correct signal processing term to describe the time domain response of the system to a delta-function `impulse', which is what we intend here (in signal processing terminology, the transfer function is in fact the Fourier transform of the impulse response).}, which encodes the geometry to relate the input light curve to the output reprocessed light curve.  Measurement of the resulting lags combined with the line velocity widths could yield the mass of the central black hole. Such work has culminated in the measurement of black hole masses for a wide range of AGN (e.g. \citealt{grier12} and see \citealt{petersonbentz06} for a review) and is now leading to the measurement of the detailed structure of the BLR, identifying its inclination to the observer as well as whether the gas is simply orbiting, outflowing, inflowing or some combination of all three \citep{bentz10,pancoast11,pancoast13,grier13}.

\begin{figure}
\begin{center}
\includegraphics[width=12cm,angle=0]{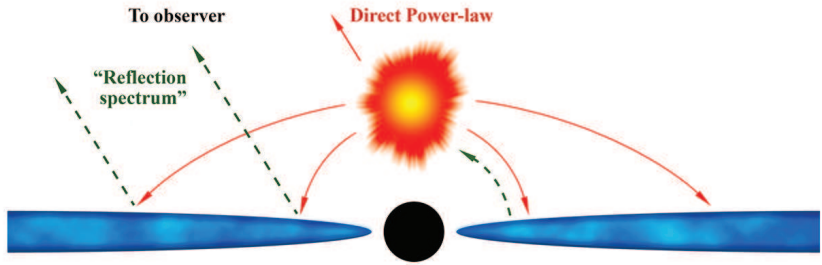}
\caption{Schematic diagram showing the power-law emitting corona
  (orange) above the accretion disc (blue), orbiting about a central
  black hole. The observer sees both the direct power-law and its
  ``reflection'', or back-scattered spectrum.  The black hole causes
  strong gravitational light bending of the innermost rays. The
  reverberation signal is the time lag introduced by light-travel time differences between observed variations in the direct power-law and the corresponding changes in the reflection spectrum.}
\label{fig:disk_corona}
\end{center}
\end{figure}

Emission lines occur from the innermost accretion flows in the X-ray
band, produced by the process of X-ray reflection (see \citealt{fabian10} for a review). The term reflection here means backscattered and
fluorescent emission as well as secondary radiation generated by
radiative heating of the gas. The primary emission is usually a
power-law continuum produced by Compton-upscattering of soft disc
photons by a corona lying above the accretion disc (see Fig.~\ref{fig:disk_corona}). A
prominent emission line in the reflection spectrum is usually that of
Fe K$\alpha$ at 6.4-6.97~keV, depending on ionization state. At low
ionization this is because iron is the most abundant cosmic element
with a low Auger yield. X-ray reflection around black holes was
discussed by \citet{guilbert88}. The resulting reflection
continuum was computed by \citet{lightman88}, followed by the line
emission by \citet{georgefabian91} and \citet{matt92,matt93}. \citet{rossfabian93} showed how disc photoionisation also leads to significant reflection features at soft X-ray energies.

\begin{figure}
\begin{center}
\includegraphics[width=10cm]{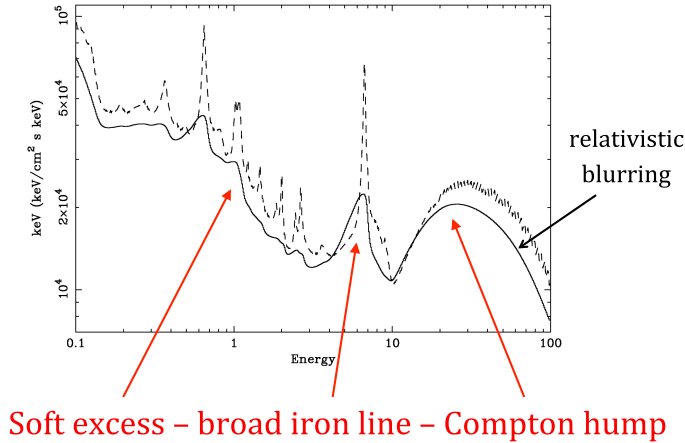}
\caption{Relativistically-blurred reflection spectrum from an ionized disc compared with its local (unblurred) counterpart, shown as a dashed line. The reflection spectrum typically has 3 characteristic parts: a soft excess, broad iron line and a Compton hump.}
\label{fig:reverb_schematic}
\end{center}
\end{figure}

\citet{fabian89} demonstrated how reflection from the inner accretion
disc around a black hole leads to the emission lines being broadened
by the Doppler effect caused by the high velocities and skewed by the
strong gravitational redshifts expected close to the black
hole. Fig.~\ref{fig:reverb_schematic} shows a more recent example using a model reflection
spectrum \citep{rossfabian05}.  \citet{fabian89} also mentioned
that reverberation may be detectable within the
line wings, the broadest of which are produced at the smallest radii.
The wings should respond first followed by the line core which
originates further out. This concept was explored by \citet{stella90} and
modelled for a disc around a Schwarzschild black hole by \citet{campana95}.
1995 was also the year that the first
relativistically-broadened iron line was detected, in the AGN
MCG--6-30-15 with the {\it ASCA} satellite \citep{tanaka95}.

X-ray reverberation was modelled further for Kerr black holes by \citet{reynolds99} and by \citet{young00}, who also made predictions for the appearance of reverberation signatures with future, large-area X-ray observatories.  At the time, the detection of reverberation from the inner disc was
considered to require the next generation of X-ray telescopes with
square metres of collecting area.  This idea was based on the assumption that we would detect the response of the disc reflection to {\it individual} continuum flares, to directly reconstruct the impulse response.  A subsequent search for reverberation using time-domain methods did not find any signals \citep{reynolds00,vaughan01}, implying that reverberation was out of reach to current instrumentation.  A crucial measure for the detection of the effect is the ratio of the number of detected photons to the
light-crossing time of the gravitational radius of the source. When
considering this ratio, the typically higher brightness of
stellar-mass black holes in Galactic X-ray binaries (BHXRB) compared with AGN,
does not compensate for the $10^5$ or more fold increase in light
crossing time for the detection of reverberation lags.  However, the detection of X-ray lags on significantly longer time-scales than the light-crossing time was facilitated in X-ray binaries (XRB) by the enormous number of cycles of variability (scaling inversely with black hole mass) that could be combined using time-series techniques to yield a significant detection.

\begin{figure}
\begin{center}
\includegraphics[width=10cm]{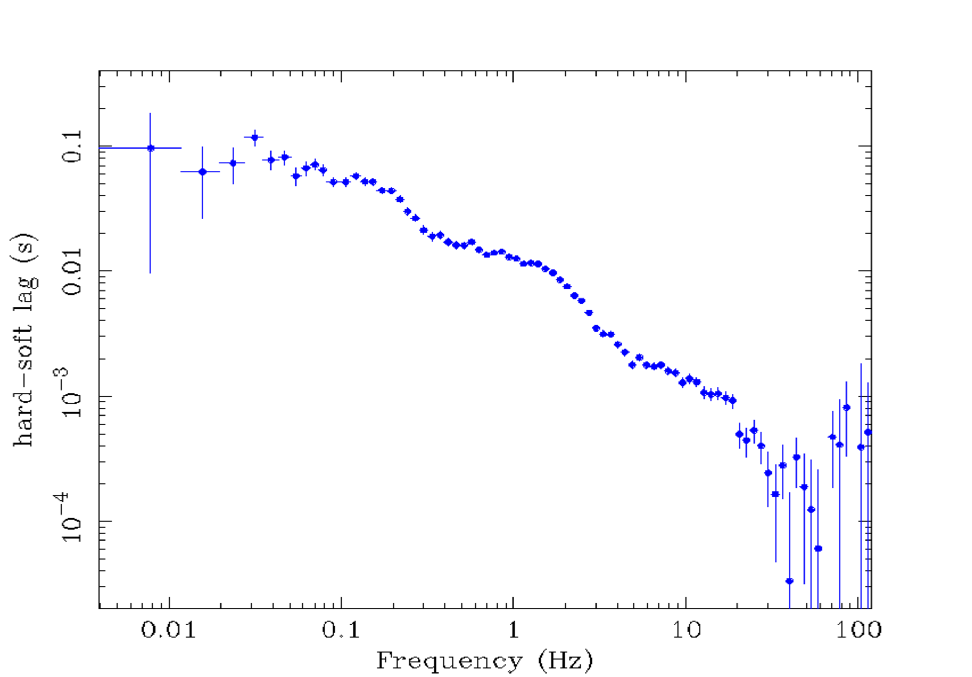}
\caption{Time lag (8--13~keV relative to 2--4~keV) versus frequency for a hard state observation of Cyg~X-1 obtained by {\it RXTE} in December 1996.  The trend can be very roughly approximated with a power-law of slope $-0.7$, but note the clear step-like features, which correspond roughly to different Lorentzian features in the power spectrum \citep{nowak00}.}
\label{fig:cygx1lagfreq}
\end{center}
\end{figure}

Lags from accreting stellar-mass black holes were first studied using Fourier techniques in the X-ray binary system Cyg~X-1 by \citet{miyamoto88}, although they had been observed earlier using less-powerful time-domain techniques \citep{page85}. The observed lags were {\it hard}, in that variations in hard photons lagged those in soft photons, and {\it time-scale dependent}, in that the time delay increases towards lower Fourier-frequencies (longer variability time-scales).  An example of the lag-frequency dependence in Cyg~X-1 is shown in Fig.~\ref{fig:cygx1lagfreq}.  Crucially, the time lags can reach 0.1~s which is much larger than expected from reverberation unless the scattering region is thousands of $r_{\rm g}$ in size. Nevertheless some interpretations of those lags did invoke enormous scattering regions and explained the spectral development in terms of Compton upscattering: harder photons scatter around in a cloud for longer \citep{kazanas97}.  However, given the large low-frequency lags seen in BHXRB data obtained by the {\it Rossi X-ray Timing Explorer}, these mechanisms were considered to be unfeasible when taking into account the energetics of heating such a large corona \citep{nowak99}.  To get around this difficulty \citet{reig03} and later \citet{giannios04} developed a model where the hard lags are produced by scattering at large scales in a focussed jet, which solves the heating problem, but this model suffers from other significant difficulties, not least in explaining the observed relativistically broadened reflection (see \citealt{uttley11} for a discussion).  

Coronal upscattering models predict a log-linear dependence of time-lag versus energy, and such a dependence is {\it approximately} observed \citep{nowak99} but the detailed lag-energy dependence shows significant `wiggles' notably around the iron line \citep{kotov01}.  As noted however, reverberation cannot explain the large hard lags observed at low frequencies \citep{kotov01,cassatella12}.  Thus, \citet{kotov01} proposed a {\it propagation} model for the lags (later explored in detail by \citealt{arevalouttley06}), where they are interpreted in terms of the inward propagation of variations in the accretion flow through a corona which becomes hotter at smaller radii (thus harder emission is produced more centrally, leading to hard lags).  Similar models where the spectrum of the emission evolves on slower time-scales than light-crossing were proposed by \citet{poutanenfabian99}, invoking the evolution of magnetic reconnection flares and \citet{misra00}, discussing waves through an extended hot accretion flow, but these models still have difficulties explaining the largest lags.   In the propagation model, the delays scale with radial inflow (i.e. viscous) time-scales and hence the largest delays can be produced from relatively small radii (tens of $r_{\rm g}$ or less) where we expect coronal power-law emission to be significant.  Significant support for the propagation model came from the discovery of the linear rms-flux relation in XRB (and AGN) X-ray variability \citep{uttley01}, which is easily explained by intrinsic accretion flow variability and its propagation through the flow \citep{lyubarskii97}.  The observed non-linear, lognormal flux distribution in BHXRBs (and apparent non-linear variability in AGN) can also be simply explained by the propagation of variations through the flow, which multiply together as they move inwards \citep{uttleymchv05}.

While significant progress was being made in the understanding of BHXRB X-ray variability, work was also underway to study lags in AGN X-ray light curves using Fourier techniques. The first time-scale dependent hard lags at low frequencies were discovered by \citet{papadakis01}, with similar lags found in several other AGN \citep{vaughanfn03,mchardy04,arevalo06,markowitz07,arevalo08}.  These AGN hard lags showed a time-scale and energy dependence consistent with those seen in BHXRBs, suggesting a similar physical origin.  Prompted by the similarity with BHXRBs, \citet{mchardy07} compared the lag-frequency dependence of the AGN Ark~564 over a broad range of Fourier frequencies with that seen in BHXRBs, finding evidence for a step-change in the lags rather than a continuous power-law dependence of lag and Fourier-frequency, similar to what is seen in hard and intermediate state BHXRBs, and suggestive of different processes producing the lags on different time-scales.  The high-frequency lag was short and negative, i.e. a soft lag, and was suggested to be caused by reprocessing by the disc.  However, the small negative lag was detected below $3\sigma$ significance ($-11\pm4$~s) and this initial hint at the existence of soft lags was only briefly remarked upon and not subsequently followed-up.

The first robust ($>5\sigma$) detection of a short (30~s) soft, high-frequency lag was seen in a 500~ks long observation of 1H0707-495 and was identified as reverberation from the photoionised disc \citep{fabian09}. This source, a narrow-line Seyfert 1 galaxy, has prominent, bright, reflection components in soft X-rays where the count rate is high, providing good signal-to-noise to detect the reverberation signal. The lag timescale is appropriate for the light crossing time of the inner disc of a few million $M_{\odot}$ black hole.  This discovery led to a flurry of detections of soft, high-frequency lags in AGN and crucially also the first Fe~K$\alpha$ lags \citep{zoghbi12}, providing a clear signature of reverberation from reflected emission.  Following a detailed discussion of methods and an introduction to the impulse response in Sect.~2, we discuss these discoveries, including the evidence for reverberation in BHXRBs, in Sect.~3.   Results are presented for lags both as a function of Fourier frequency and energy.

Sect. 4 explores models for the lag behaviour encoded in
the impulse response. We consider the impulse response for a source
situated above an accretion disc. Initially the source is assumed to
be point-like, then we outline the expected behaviour of extended
sources. The 2D behaviour of the impulse response in energy and time
is explained.  After a brief summary in Sect.~5, we explore future directions for research on time lags due to reverberation around black holes in Sect.~6. Reverberation is a
powerful tool with which the geometry of the inner accretion flow and
its relation to the primary X-ray power source can be studied. In some
sources it is already revealing the behaviour within the innermost
regions at a few $r_{\rm g}$ of rapidly spinning black
holes. Reverberation techniques will allow us to understand the inner
workings of quasars, the most powerful persistent sources in the Universe, and in BHXRBs uncover the changes in emitting region and accretion flow structure associated with jet formation and destruction.

%\documentclass[11pt,letterpaper]{article}
%\usepackage{graphicx,sidecap,multicol,natbib}
%\usepackage[letterpaper,margin=2cm,centering]{geometry}

%\input{defn.tex}

%\parindent 0pt
%\parskip 0.25cm

%\begin{document}

\section{Analysis Methods}\label{sec:analysis}
In this section we will review the methods used to study variability and search for time-lags in X-ray light curves.  These methods are crucial for the discovery and exploitation of the lags, so we aim to provide important background as well as recipes for the measurement of different spectral-timing quantities.  Those readers who are familiar with these methods or who wish to focus on the observations and modelling can skip over this section, which is mainly pedagogical.  We will first consider the basic Fourier techniques at our disposal, and then describe how these techniques are put into practice, as well as some practical issues regarding the determination of errors and the sensitivity of lag measurements.  Finally we will introduce the concept of the impulse response, which shows how we can model the observed energy-dependent variability properties in terms of reverberation and other mechanisms, which will be considered in more detail in Sect.~\ref{sec:modelling}.
\subsection{The toolbox: Fourier analysis techniques}
Time-series analysis can be done in the time- or frequency-domain.  Optical reverberation studies of AGN use time-domain techniques to measure time-lags, specifically the cross-correlation function \citep[e.g.][]{peterson93,petersonetal04,bentz09,denney10}.  More recently, stochastic modelling and fitting of light curves has been applied to more accurately model the optical reverberation data \citep{zu11,pancoast11}.  The situation for X-ray time-series analysis is different however: here the focus to date has been on Fourier analysis techniques (with just a few exceptions, e.g. \citealt{maccarone00,dasgupta06,legg12}).  There are several reasons for this difference.  Firstly, the Fourier power spectrum is an easy way to describe the underlying structure of a stochastic variable process, e.g. dependence of variability amplitude on time-scale, with statistical errors that are close to independent between frequency bins and therefore is easily modellable (unlike time-domain techniques such as the autocorrelation function or structure function where errors are correlated between bins, e.g. see \citealt{emma10}). 

Secondly, although aliasing effects make it difficult to cleanly apply Fourier techniques to irregularly-sampled and `gappy' data from long optical monitoring campaigns, (Fast) Fourier techniques lend themselves particularly well to analysis of the very large, high-time-resolution light curves used to study the very rapid variability in X-ray binaries.  The same approaches are easily applicable to the contiguous light curves obtained from `long-look' observations of AGN by missions such as {\it EXOSAT} (e.g. \citealt{lawrence87}) and currently, {\it XMM-Newton} (e.g. \citealt{mchardy05}).  Another important factor is that there has been a significant focus on comparisons of XRB and AGN X-ray variability, e.g. to uncover the mass-scaling of variability time-scales, as well as comparing spectral-timing properties \citep{mchardy06,arevalo06}.  Thus it is useful to use a common approach to data from both kinds of object in the X-ray band, and these efforts have also stimulated the development of techniques to study irregularly-sampled AGN data in the Fourier domain \citep{uttley02,markowitz03}, or more recently, through a combination of Fourier and time-domain approaches \citep{lancemiller10,kelly11,zoghbi13_gaps}.  Also important is the fact that Fourier-techniques allow for the easier interpretation of complex data by decomposing the data in terms of the variations on different time-scales.
\subsubsection{The discrete Fourier transform and power spectral density}
The X-ray light curves of AGN and XRBs are best described as stochastic, {\it noise processes}\footnote{Even quasi-periodic oscillations seen in XRBs follow the same statistics as noise \citep{vanderklis97}.}. A well-known type of noise process is observational, i.e. Poisson noise, but here the observed intrinsic variations themselves are also a type of noise and thus inherently unpredictable at some level.  The extent to which we can predict the flux at one time based on the flux at another depends on the shape of the underlying power spectral density function or PSD, $\mathit{P}(f)$, which describes the average variance per unit frequency of a signal at a given temporal frequency $f$.  Strongly autocorrelated signals, where variations between adjacent time bins are small and increase strongly with larger time-separation (e.g. `random walks' or Brownian noise) correspond to steep {\it red noise} PSDs ($P(f)\propto f^{\alpha}$ with $\alpha \leq -2$).  Above some limiting time-scale the size of variations must reach some physical limit and the PSD below the corresponding frequency flattens, typically to {\it flicker noise} with $\alpha\simeq -1$.  This characteristic bend in the PSD is detected in AGNs and BHXRBs, and occurs at a frequency corresponding to a characteristic time-scale which scales linearly with black hole mass and possibly also inversely with accretion rate (\citealt{uttley05,mchardy06} but see also \citealt{gm12} which confirms the linear mass-dependence but suggests a weaker accretion rate dependence).  At even lower frequencies, in BHXRBs and also detected in the AGN Ark~564, the PSD can flatten again to become {\it white noise} ($\alpha=0$) so that the total variance becomes finite and there is no long-term memory in the light curve on those time-scales.  Poisson noise is also a form of white noise.

The PSD can be estimated from the periodogram, which is the modulus-squared of the discrete Fourier transform of the light curve.  The discrete Fourier transform (DFT) $X$ of a light curve $x$ consisting of fluxes measured in $N$ contiguous time bins of width $\Delta t$ is given by:
\begin{equation}
\label{eqn:dft}
X_n = \sum_{k = 0}^{N - 1} x_k \exp\left( 2\pi i n k / N \right)
\end{equation}
where $x_k$ is the $k$th value of the light curve and $X_n$ is the discrete Fourier transform at each Fourier frequency, $f_n = n/(N\Delta t)$, where $n=1,2,3,...N/2$.  Thus the minimum frequency is the inverse of the duration of the observation, $T_{\rm obs}=N\Delta t$ and the maximum is the Nyquist frequency, $f_{\rm max} = 1/(2\Delta t)$.   

The periodogram is simply given by:
\begin{equation}
\label{eqn:periodogram}
|X_n|^2 = X_n^{*} X_n
\end{equation}
Where the asterisk denotes complex conjugation.  In practice the periodogram is further normalised to give the same units as the PSD:
\begin{equation}
\label{eqn:psd}
P_n=\frac{2\Delta t}{\langle x \rangle^{2} N} |X_n|^{2} 
\end{equation}
where the value in angle brackets is the mean flux of the light curve and the normalised periodogram $P_{n}$ is thus expressed in units of {\it fractional} variance per Hz \citep{belloni90,miyamoto92}, so that this normalisation is often called the `rms-squared' normalisation.  Thus, integrating the PSD with this normalisation over a given frequency range and taking the square root gives the fractional rms variability (often called $F_{\rm var}$) contributed by variations over that frequency range. 

For a noise process, the observed periodogram is a random realisation of the underlying PSD, with values drawn from a highly-skewed $\chi^{2}_{2}$ distribution with mean scaled to the mean of the underlying PSD.  It is thus important to note that the `noisiness' of the observed periodogram is in some sense intrinsic to the `signal', i.e. the underlying variability process.  Since the underlying PSD of the process is the physically interesting quantity, we usually bin up the periodogram to obtain an estimate of the PSD (in frequency and also over periodograms measured from multiple independent light curve segments), so that the PSD in a frequency bin $\nu_{j}$ averaged over $M$ segments and $K$ frequencies per segment is given by:
\begin{equation}
\label{eqn:binnedpsd}
\bar{P}(\nu_{j}) =\frac{1}{KM} \sum_{n=i,i+K-1} \sum_{m=1,M} P_{n,m}
\end{equation}
where $\bar{P}(\nu_{j})$ is the estimate of the PSD obtained from the average of the periodogram in the bin $\nu_{j}$ (henceforth, we will refer to the measured quantity as the PSD) and $P_{n,m}$ is the value of a single sample of the periodogram measured from the $m$th segment with a frequency $f_n$ that is contained within the frequency bin $\nu_{j}$ (which contains frequencies in the range $f_i$ to $f_{i+K-1}$).  Note that here we use $\nu$ to denote frequency {\it bins}, rather than the underlying frequency $f$.  The standard error on the PSD $\Delta \bar{P}(\nu_{j})$ can be determined either from the standard deviation in the $KM$ samples (divided by $\sqrt{KM}$ to give the standard error on the mean) or simply from the statistical properties of the $\chi^{2}_{2}$ distribution, which imply that, for a large number of samples\footnote{It is important to bear in mind that due to the highly skewed nature of the $\chi^{2}_{2}$ distribution, errors on the PSD only approach Gaussian after binning a large number of samples ($KM>50$).  An alternative approach, which converges more quickly to Gaussian-distributed errors, is to bin $\log(P_{n,m})$, which also necessitates adding a constant bias to the binned log-power, see \citet{papadakis93,vaughan05} for details.}:
\begin{equation}
\label{eqn:psderror}
\Delta \bar{P} (\nu_{j}) = \frac{\bar{P}(\nu_{j})}{\sqrt{KM}}
\end{equation}
Poisson noise leads to flattening of the PSD at high frequencies with a normalisation depending on the observed count rate, which is easily accounted for by subtracting a constant $P_{\rm noise}$ from the observed PSD.  In the case where Poisson statistics applies and the fluxes are expressed in terms of count rates:
\begin{equation}
\label{eqn:noiselevel}
P_{\rm noise}=\frac{2\left(\langle x \rangle + \langle b \rangle \right)}{\langle x \rangle^{2}}
\end{equation}
where $\langle b \rangle$ is the average background count rate in the light curve (we assume that the background is already subtracted from $x$).  For non-contiguous sampling, the noise level must be increased in line with the reduced Nyquist frequency \citep{markowitz03,vaughan03}.  If fluxes are not given in count rates (or the data are expressed as a count rate but the statistics are not Poissonian), then the equivalent noise level can be determined using $P_{\rm noise}=\langle \Delta x^{2} \rangle / \left( \langle x \rangle^{2} f_{\rm Nyq} \right) $, where $\langle \Delta x^{2} \rangle$ is the average of the squared error-bars of the light curve, and $f_{\rm Nyq}$ is the Nyquist frequency ($f_{\rm Nyq}=f_{\rm max}=1/(2\Delta t)$ for contiguous sampling).
\begin{figure}
\centering
\includegraphics[width=0.33\textwidth]{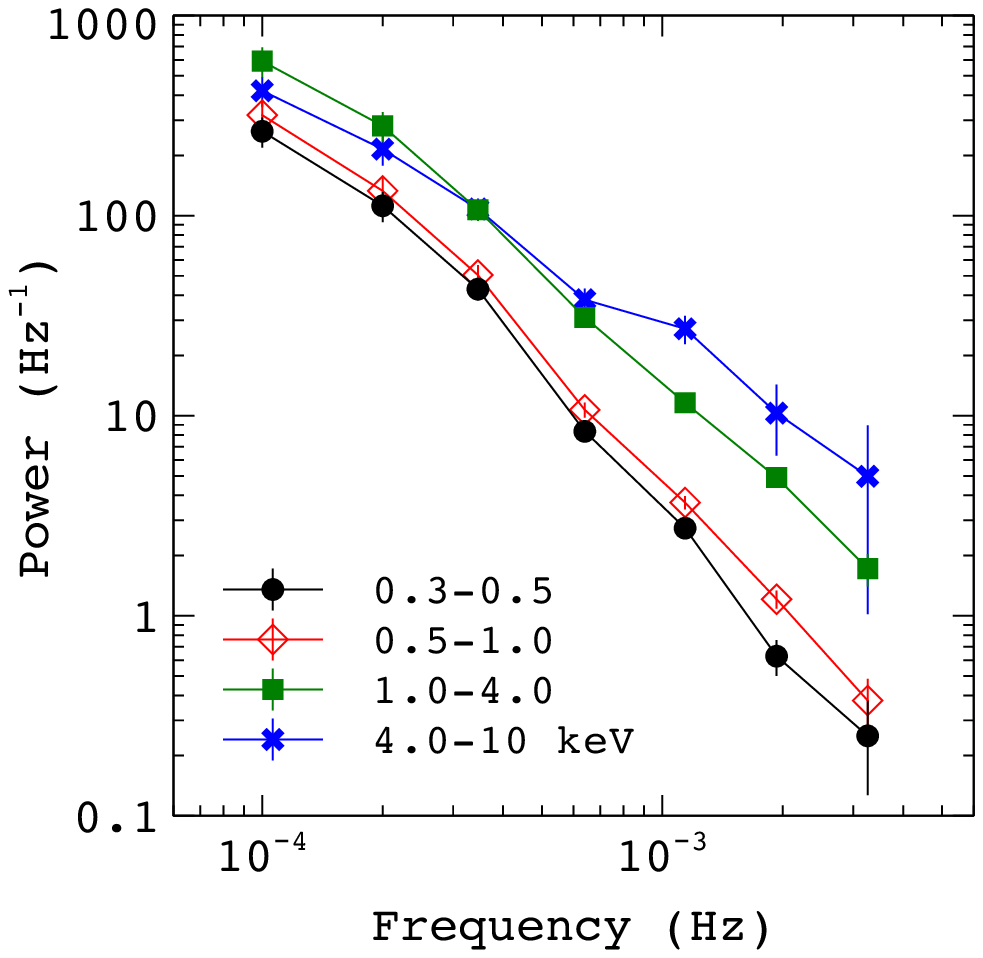}
\includegraphics[width=0.57\textwidth]{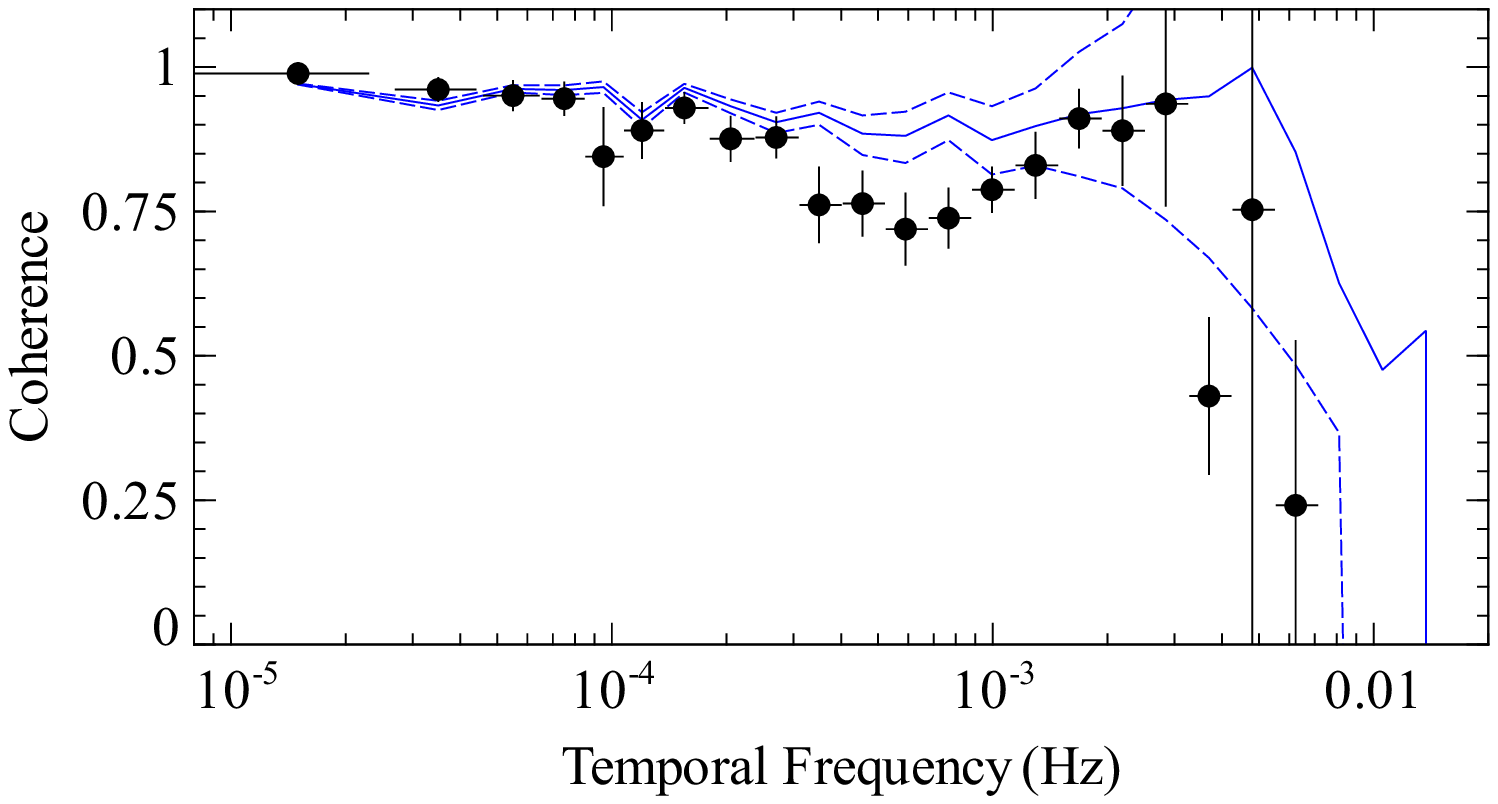}
\caption{{\it Left:} The Poisson-noise subtracted PSDs of the NLS1 AGN 1H0707-495 (taken from \citealt{zog11_1h0707}).  {\it Right:} 1H0707-495 frequency-dependent coherence between the 0.3--1~keV and 1--4~keV bands.  The solid and dashed blue lines give the median and upper and lower 95~per~cent confidence levels for the coherence obtained from simulations of correlated (unity intrinsic coherence) light curves with the same flux levels, variance and PSD shape as the data.  Note the dip, suggesting a changeover between two processes.  At high frequencies, the coherence is consistent with unity, as expected from simple reverberation.  Figure taken from \citet{zoghbi10}.}
\label{fig:1h0707psdcoh}
\end{figure}

Some example noise-subtracted PSDs for the light curve of 1H0707-495 measured in different energy bands are shown in Fig.~\ref{fig:1h0707psdcoh} (left panel).  All bands show the characteristic red-noise shape at high frequencies (although not clear from the figure, there is evidence for a break to a flatter slope at $\sim 1.4\times 10^{-4}$~Hz, see \citealt{zoghbi10}).  However, the higher energies show a flatter PSD, showing that there is relatively more rapid variability at these energies.  Similar energy-dependent behaviour is also seen in BHXRBs (e.g. \citealt{nowak99}).

\subsubsection{The cross spectrum and lags}
\label{sec:crossspeclags}
The Fourier cross spectrum between two light curves $x(t)$ and $y(t)$ with DFTs $X_n$ and $Y_n$ is defined to be:
\begin{equation}
\label{eqn:crossspec}
C_{XY,n}=X_n^{*} Y_n
\end{equation}
We can see how the cross-spectrum is used to derive the frequency-dependent (phase) lag between two bands by considering the complex polar representation of the Fourier transform $X_n=A_{X,n}\exp\left[i\psi_n\right]$, where $A_{X,n}$ is the absolute magnitude or amplitude of the Fourier transform and $\psi_n$ is the phase of the signal (which for a noise process is randomly distributed between $-\pi$ and $\pi$) at the frequency $f_n$.  Thus, a linearly correlated light curve $y(t)$ with an additional phase-shift $\phi_n$ at frequency $f_n$ has a Fourier transform $Y_n=A_{Y,n}\exp\left[i(\psi_n+\phi_n)\right]$.  Multiplying by the complex conjugate of $X_{n}$, the phase $\psi_n$ cancels and the cross-spectrum is given by:
\begin{equation}
C_{XY,n} = A_{X,n}A_{Y,n}\exp\left(i\phi_{n}\right)
\end{equation}
with the phase of the cross-spectrum giving the phase lag between the light curves, as expected.

In principle the cross-spectrum may also be normalised in the same way as the periodogram, except that instead of dividing by $\langle x \rangle^{2}$ to obtain fractional rms-squared units, we must divide by the product of light curve means, $\langle x \rangle \langle y \rangle$.  Note that due to the well-known linear rms-flux correlation in AGN and XRB light curves, different results can be obtained if the lags depend on the flux level (see Sect.~\ref{sec:frontiers}) and either the cross-spectrum is normalised by the means of each light curve segment {\it before} averaging segments, or a single combined mean value for all segments is used {\it after} averaging \citep{alston13}.

In the presence of any uncorrelated signal between the two light curves (e.g. due to Poisson noise, but there may also be an {\it intrinsically incoherent} signal, perhaps due to the presence of an additional independently-varying component in one energy band but not the other), the cross-spectrum should be averaged over Fourier frequencies in a given frequency bin $\nu_{j}$, as well as over multiple light curve segments, to reduce the effects of noise:
\begin{equation}
\label{eqn:binnedcrossspec}
\bar{C}_{XY} (\nu_{j}) = \frac{1}{KM} \sum_{n=i,i+K-1} \sum_{m=1,M}  C_{XY,n,m}
\end{equation}
Here we assume the same notation as in Equation~\ref{eqn:binnedpsd}.  The phase of the resulting binned cross-spectrum (i.e. the {\it argument} of the complex cross-spectrum vector) $\phi(\nu_{j})$, gives the average phase lag between the two light curves in the $\nu_{j}$ frequency bin.  The time lag is thus given by:
\begin{equation}
\label{eqn:timelag}
\tau(\nu_{j}) = \phi(\nu_{j})/\left(2\pi \nu_{j}\right)
\end{equation}
Conversion of phase to time-lags is often carried out for ease of interpretation of the data, but since the phase is limited to the range $-\pi$ to $\pi$, caution should be taken when interpreting any time-lags corresponding to phase lags close to these limits (e.g. so-called phase-wrapping across the phase boundaries can lead to sudden `flips' in the lag).  Furthermore, for broad frequency bins, it is not always clear what value should be used for the bin frequency, $\nu_{j}$ to obtain the time-lag, e.g. should the frequency be weighted according to the power contributing from each sample frequency in the bin, or should the bin centre or unweighted average frequency be used?  The more common procedure is to use the bin centre, but the question is in some sense a matter of taste and convention, because conversion of phase lags to time-lags is done mainly for the convenience of expressing the phase lag in terms of a physically useful time-scale.  The choice of frequency will lead to some small biases in the observed time-lag, but this effect can be easily accounted for when modelling the lags, or by fitting models to the cross-spectrum or phase-lags directly.

Examples of lag-frequency spectra are shown for a BHXRB and AGN respectively in Fig.~\ref{fig:cygx1lagfreq} and Fig.~\ref{1h07_lag}.

\subsubsection{The coherence and errors}
\label{sec:coherr}
The {\it coherence} $\gamma^{2}$ in the frequency bin $\nu_{j}$ is defined as:
\begin{equation}
\label{eqn:coherence}
\gamma^{2}(\nu_{j}) = \frac{|\bar{C}_{XY}(\nu_{j})|^{2}-n^{2}}{\bar{P}_{X} (\nu_{j}) \bar{P}_{Y}(\nu_{j})}
\end{equation}
where the binned PSD and cross-spectrum have been normalised in the same way (see above).  We stress here that it is only meaningful to measure the coherence of the binned cross-spectrum (i.e. averaged over segments and/or frequency), so that uncorrelated noise cancels during the binning, otherwise observed coherence will always be unity, by definition.  The $n^{2}$ term is a bias term, which arises because the Poisson noise-level contributes to the modulus-squared of the cross-spectrum\footnote{$n^{2}=\left[(\bar{P}_{X} (\nu_{j})-P_{X,{\rm noise}})P_{Y,{\rm noise}}+(\bar{P}_{Y}(\nu_{j})-P_{Y,{\rm noise}})P_{X,{\rm noise}}+P_{X,{\rm noise}}P_{Y,{\rm noise}}\right]/KM$, where we assume that the binned PSDs are not already noise-subtracted.  See \citet{vaughan_nowak97} for further details.}.  Formally, the coherence indicates the fraction of variance in both bands which can be predicted via a linear transformation between the two light curves \citep{vaughan_nowak97}.  A reduction in coherence can also occur due to a non-linear transformation (see the discussion in \citealt{vaughan_nowak97,nowak99}).  An example plot of coherence versus frequency is shown for 1H0707-495 in Fig.~\ref{fig:1h0707psdcoh} (right panel).

Geometrically, the coherence gives an indication of the scatter on the cross-spectrum vector that is caused by the incoherent components \citep{nowak99}.  It can thus be used to derive the error on the phase lag \citep{bendat10}:
\begin{equation}
\label{eqn:lagerror}
\Delta \phi(\nu_{j}) = \sqrt{\frac{1-\gamma^{2}(\nu_{j})}{2\gamma^{2}(\nu_{j})KM}}
\end{equation}
and the time-lag error is simply $\Delta \tau = \Delta \phi/(2\pi \nu_{j})$. It is easy also to work out the equivalent expression in terms of the contributions to the cross-spectrum of the signal and noise components of the Fourier transform, which we will consider later, when we discuss the sensitivity of lag measurements. 

One important point to note here is that the coherence used to estimate the lag error is the {\it raw} coherence, where the PSD values in the denominator of Equation~\ref{eqn:coherence} have not had the Poisson noise level subtracted.  The advantage of this approach is that the lag errors can be easily determined from the data, without any other assumptions, since the quantities used are simply the observed cross-spectrum and PSD.  Problems can arise once the raw coherence becomes comparable to the $n^{2}$ bias term, since oversubtraction of the bias would lead to negative values of coherence, preventing the standard estimation of lag errors using the coherence.  In that case, if bias is not subtracted then in the limiting case of negligible intrinsic variability, the raw coherence becomes $1/KM$ and the estimated error on the phase lag will saturate at $\Delta \phi= 1/\sqrt{2}$ (the real error, accounting for bias, is substantially larger than this).

If the {\it intrinsic} coherence is required for a study of the variability properties of the source, i.e. the coherence attributable to the source itself, correcting for observational noise, the noise-levels should be subtracted from the PSDs in the denominator of Equation~\ref{eqn:coherence}.  Errors on the intrinsic coherence can be estimated using the approach outlined in \citet{vaughan_nowak97}, but it is important to bear in mind that the errors on intrinsic coherence are difficult to correctly determine for low variability $S/N$.  This problem does not affect the lag-error estimation however, since this is based on the raw coherence.

\subsection{Practical Application}
We now consider the practical application of the Fourier methods outlined above to real data.
\subsubsection{Frequency-dependent spectral-timing products}
The cross-spectral approach outlined above can be used to measure the {\it lag-frequency spectrum}: the phase or time-lag between two broad energy bands plotted as a function of Fourier frequency, as well as the coherence, if required.  This - now standard - approach was first applied to data from X-ray binaries and also used to study lower frequencies in AGN before eventually leading to the discovery of soft lags and reverberation at high frequencies \citep{fabian09}.  As we have already noted, the key advantage of studying lags in the Fourier domain, rather than the time-domain (e.g. with the cross-correlation function), is that complex time-scale-dependent lag behaviour associated with multiple physical processes can easily be disentangled. Hence, this approach has become the workhorse for the discovery of soft lags in AGN, as we will see in Sect.~\ref{sec:obs}.

The practical steps to calculate the lag-frequency spectrum and other frequency-dependent products are:
\begin{enumerate}
\item Create light curves with identical time-sampling in two, separate energy-bands.
\item Split the light curves into $M$ continuous segments of equal duration, choosing the segment duration based on the lowest frequency to be sampled.  Small gaps in the light curves (up to a few per cent of the light curve duration) can be interpolated over (with random errors added to match the observational errors) without significant distortion of the resulting cross-spectrum, providing that the variability on time-scales comparable or shorter than the gap size is small compared to the amplitude on longer time-scales (e.g. the gap time-scale corresponds to the steep red-noise part of the PSD).  Larger gaps should be avoided using a suitable choice of segment size or a method that accounts for their effects on the data and errors \citep{zoghbi13_gaps}.
\item For each segment, obtain the PSDs in both energy bands (Equations~\ref{eqn:dft}--\ref{eqn:binnedpsd}) and the cross-spectrum (Equations~\ref{eqn:crossspec}--\ref{eqn:binnedcrossspec}), averaging them to form PSDs and the cross-spectrum binned over segments.
\item Further bin the PSD and cross-spectrum over frequency, as required.  It is often convenient to bin geometrically in frequency, i.e. from frequency $f$ to frequency $Bf$ where $B$ is the selected binning factor ($B>1.0$), so that the frequency bins have the same width in log-frequency.  If there are $K$ frequencies sampled in a bin, the number of samples is then equal to $K\times M$. 
\item Use the phase of the binned cross-spectrum to calculate the phase-lag and/or time-lag (Equation~\ref{eqn:timelag}) for each frequency bin. 
\item Calculate the {\it raw} coherence from the binned PSD and cross-spectrum (Equation~\ref{eqn:coherence}).
\item Use the coherence in each frequency bin to calculate the error bar on the lag in each frequency bin (Equation~\ref{eqn:lagerror}).
\item If desired, obtain the intrinsic coherence, following the prescription in \citet{vaughan_nowak97}.
\end{enumerate}
Note that the choice of segment size may be important if `leakage' effects - due to the finite sampling window - are a problem in the data set being considered.  For example, \citet{alston13} have shown that an effect of a finite segment size, which is more pronounced for shorter segments, is that lags can leak across frequencies (the effect is related to the problem of `red-noise leak' seen in PSDs, e.g. see \citealt{uttley02} for discussion).  This effect is strongest when the gradient of the phase lag-frequency spectrum is largest.  The leakage effect does not qualitatively affect any of the reverberation results discovered so far, but may become important as more detailed models are applied.  It can be reduced by choosing longer segments (so that the flatter part of the PSD is sampled, leading to less leakage from low frequencies), or by `end-matching' the data to take out any long-term trend \citep{alston13}, although this can lead to other biases, which should be modelled.  One approach to maximising the segment length (since choosing a fixed segment length can lead to some data being ignored, when observations with different durations are combined) is to measure the Fourier transforms for the entire contiguous light curves for each observation, and combine them by binning in frequency\footnote{Since the measured Fourier frequencies depend on segment length, binning is best done by making a frequency-ordered list of frequencies and power or cross-spectral value from all segments and then binning the power/cross-spectra according to frequency.}. This approach is best used when leakage is not expected to be significant and/or segment lengths are relatively similar, since combining data with different leakage effects (due to different segment lengths) can lead to confusing results which are difficult to model.

Following the recipe presented here will produce the range of commonly-used frequency-dependent spectral-timing products: the PSD in both bands, the coherence vs. frequency and the lag vs. frequency.  Multiple bands can be used to compare the frequency-dependent spectral-timing properties as a function of energy.  However, new insights can be gained by using a similar approach to make energy-dependent spectral-timing products, which we outline below.

\subsubsection{The lag-energy spectrum}
Although the standard lag-frequency measurement can give valuable insights, it is especially useful to look at the dependence of the lag on energy at higher spectral resolution, to reveal which spectral components contribute to the lags, and thus the {\it causal relationship} between the different spectral components.

To plot the {\it lag-energy spectrum}, it helps to optimise the signal-to-noise in two ways.  Firstly, we must select a broader frequency range to average the cross-spectrum over than the narrower bins used for lag-frequency spectra.  The frequency range may be chosen to select some particularly interesting behaviour for the measurement of the lag-energy spectrum, e.g. the measurement of the spectrum corresponding to negative or 'soft' lags.  However, as noted by \citet{lancemiller10}, it is important to bear in mind that impulse responses corresponding to a single physical component may have a complex effect over a wide range of frequencies, so that it is not always obvious that a given frequency selection also corresponds to the selection of distinct physical components or effects.  Lag-energy spectra may therefore be treated as being suggestive of certain physical effects, but ultimately, fully self-consistent modelling of the data over a wide-range of frequencies will be the best way to test our physical understanding.

Secondly, to further increase signal-to-noise, we can choose a broad {\it reference} energy band with which to measure a cross-spectrum for each of the individual energy bins or {\it channels-of-interest} (CI).  By selecting a common reference band, we can measure the lag of each energy bin relative to the same reference band, to obtain a lag-energy spectrum (see also \citealt{vaughan94} for an alternative route to the lag-energy spectrum).  Provided that each energy bin shares the same intrinsic coherence with the reference band, it is easy to then interpret the relative lag {\it between} each energy bin, as the lag between the variations at those energies which are also correlated with the reference band.  In principle, any choice of reference band can be made but this can potentially convey different information, if the intrinsic coherence changes between energy bins.  In the simplest case where variations are intrinsically fully-coherent between energies, it is simplest to use the broadest possible reference band, i.e. across all energies with good $S/N$.  Using a broad reference band has the advantage that the error associated with Poisson noise in the reference band is minimised.  This effect is especially important for the study of XRBs, as we shall see.

The steps for calculation of the lag-energy spectrum are as follows:
\begin{enumerate}
\item Choose a reference energy band and make its light curve.
\item Make a light curve for each channel-of-interest using the same time-sampling as the reference band.  Before making the spectral-timing products, {\it if the channel-of-interest is contained in the same energy range as the reference band and samples the same data, i.e. is not from a separate detector}, then subtract off the channel-of-interest to leave a CI-corrected reference light curve.
\item Using the CI-corrected reference light curve with the CI light curve, obtain the PSDs and cross-spectrum following steps 2-3 of the approach for making frequency-dependent spectral-timing products.
\item Average the PSDs and cross-spectra over the selected broad frequency range, and obtain lags, raw coherence and lag errors associated with the channel-of-interest.  The same data may also be used to make rms and covariance spectra, outlined in the following subsection.
\item Repeat for all the channels-of-interest to make a lag-energy spectrum.
\end{enumerate} 
Note that subtracting the channel-of-interest light curve from the reference band is necessary in order to subtract off the part of the Poisson noise in the CI which is correlated with itself in the reference band, thereby contaminating the cross-spectrum with a spurious zero-lag component.  An equivalent procedure is to subtract the CI PSD from the cross-spectrum.  Technically, doing this correction means that each CI is correlated with a slightly different reference light curve (with a slightly different average energy) and hence the relative lags are not strictly equivalent to the true relative lag between each CI.  However, provided the CI bins are relatively narrow and thus contain only a small fraction of reference band photons, the effect on the lag-energy spectrum is small \citep{zog11_1h0707}.

Examples of lag-energy spectra are shown throughout Sect.~\ref{sec:obs}.

\subsubsection{The rms and covariance spectrum}
\begin{figure}
\begin{center}
\includegraphics[width=8cm]{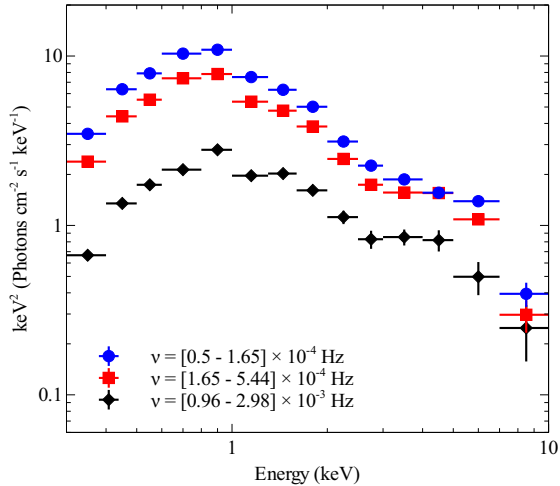}
\caption{Covariance spectra of 1H0707-495 in different frequency ranges.  The covariance spectrum of the frequency range over which the high-frequency soft lags are detected is shown in black and shows a harder spectrum than is seen at lower frequencies, consistent with the energy-dependent PSD behaviour.  Variable reflection features appear to be present on all time-scales, as expected if the continuum drives variable reflection (figure taken from \citealt{kara13a}).}
\label{fig:1h0707covspec}
\end{center}
\end{figure}

The first application of a broad reference band to obtain detailed energy-dependent spectral-timing data was for the {\it covariance spectrum}, used by \citet{wilkinson09} to uncover the variability of accretion disc emission in the hard state BHXRB GX~339-4 (see Sect.~\ref{sec:bhb}), and subsequently used in a number of other analyses of data from AGN and XRBs \citep{uttley11,middleton11,cassatella12swift,kara13a,cackett13}.  The covariance spectrum\footnote{Strictly speaking, the `covariance' spectrum measures the {\it square-root} of the covariance of each channel with the reference band.} is the cross-spectral counterpart of the rms-spectrum, which measures the rms amplitude of variability as a detailed function of energy.  By selecting a frequency range (with width $\Delta \nu$) over which to integrate the PSD in each energy band, a Fourier-resolved rms-spectrum can be obtained \citep{revnivtsev99,gilfanov00}.  In the same way, we can use the cross-spectrum to obtain a Fourier-frequency resolved covariance spectrum, which shows the spectral shape of the components which are correlated with the reference band.  Thus a careful choice of reference band can be used to identify those spectral components which vary together in a given frequency range, and those which do not.  The rms and covariance spectra can be determined as follows\footnote{Note that the description of the calculation of the covariance spectrum and its errors given here should be used instead of that given in \citet{cassatella12swift}, which contains several typos.  We would like to thank Simon Vaughan for bringing these errors to our attention.}:
\begin{enumerate}
\item Follow steps 1--4 for the calculation of the lag-energy spectrum described above.
\item Subtract the noise level (Equation~\ref{eqn:noiselevel}) from the CI PSD and the CI-corrected reference PSDs which have been averaged over the frequency-range of interest.  
\item Multiply the CI PSD by the frequency range width $\Delta \nu$, to obtain the variance in that frequency range.  If the PSD uses the fractional rms-squared normalisation and an absolute counts spectrum is desired, multiply the CI PSD by the squared-mean of the CI light curve, to obtain the variance in absolute units.  Take the square root to obtain the (fractional or absolute) rms of the CI, which can then be used to make the rms spectrum. 
\item Repeat the above step for the CI-corrected reference band PSD.
\item Take the amplitude of the binned bias-subtracted cross-spectrum \newline ($\sqrt{|\bar{C}_{XY} (\nu_{j})|^{2}-n^{2}}$) and multiply it by $\Delta \nu$.  Then multiply by the product of reference and CI light curve means if the normalisation is to be corrected into absolute units.
\item Divide the resulting value by the (Poisson-noise subtracted) rms of the CI-corrected reference band.  The final resulting value is the value of the covariance spectrum in the channel-of-interest.  It has the same units as the equivalent rms spectrum.
\end{enumerate}
Mathematically then, the covariance spectrum in absolute flux units (which may then be fitted and interpreted in a similar way to a time-averaged X-ray spectrum) is given by:
\begin{equation}
\label{eqn:covspec}
Cv(\nu_{j})=\langle x \rangle \sqrt{\frac{\Delta \nu_{j} \left(|\bar{C}_{XY}(\nu_{j})|^{2}-n^{2}\right)}{\bar{P}_{Y}(\nu_{j}) - P_{Y,\rm noise}}}
\end{equation}
where $Y$ denotes the CI-corrected reference band, $\Delta \nu_{j}$ is the frequency width of the $\nu_{j}$ bin and we assume that the cross and power-spectra use the fractional rms-squared normalisation (otherwise multiplication by the $\langle x \rangle$ is not required).  Note that the covariance spectrum is closely related to the coherence\footnote{The covariance spectrum can also be calculated directly from the coherence, using $Cv(\nu_{j})=\langle x \rangle \sqrt{\gamma^{2}(\nu_{j})(\bar{P}_{X}(\nu_{j}) - P_{X,\rm noise})\Delta \nu_{j}}$.}, and in the limit of unity coherence, the covariance spectrum should have the same shape as the rms spectrum.  However, even in this case, the signal-to-noise of the covariance spectrum is substantially better than that of the rms spectrum, since the reference band light curve is effectively used as a `matched filter' to pick out the correlated variations in each channel-of-interest.  Examples of covariance spectra measured for different frequency ranges for 1H0707-495 are shown in Fig.~\ref{fig:1h0707covspec}, and for the hard state BHXRB GX~339-4 in the insets of Fig.~\ref{fig:gx339lagvsen}.

The errors on the rms spectrum must be calculated based on the errors expected due to Poisson noise (e.g. see \citealt{vaughan03}), since they are independent between different energy bins in the rms spectrum, whereas the errors in the rms due to intrinsic stochastic variability (i.e. the intrinsic $\chi^{2}_{2}$ scatter in the PSD) are correlated between energy bins if the coherence between bins is non-zero.  In the limit where the intrinsic coherence is unity, the errors on the (absolute) rms-spectrum are given by:
\begin{equation}
\label{rmsspecerr}
\Delta \sigma_{X}(\nu_{j}) = \sqrt{\frac{2 \sigma^{2}_{X}(\nu_{j}) \sigma^{2}_{X,\rm noise} + (\sigma^{2}_{X,\rm noise})^{2}}{2KM\sigma^{2}_{X}(\nu_{j})}}
\end{equation}
where $\sigma_{X}(\nu_{j})$ is the noise-subtracted rms in the channel-of-interest $x$ and $\sigma^{2}_{X,\rm noise}$ is the absolute rms-squared value obtained from integrating under the Poisson noise level of the bands, i.e. $\sigma^{2}_{X}(\nu_{j})=(P_{X} (\nu_{j}) - P_{X,\rm noise})\langle x \rangle ^{2} \Delta \nu_{j}$, $\sigma^{2}_{X,\rm noise}=P_{X,\rm noise} \langle x \rangle ^{2} \Delta \nu_{j}$, where the PSDs are in the fractional rms-squared normalisation.  See \citet{vaughan03} for discussion of the rms-spectrum and its error (determined from numerical simulations), and \citet{wilkinson11} for a formal derivation of the error.

For unity intrinsic coherence, the error on the covariance spectrum is given by:
\begin{equation}
\label{covspecerr}
\Delta Cv(\nu_{j}) = \sqrt{\frac{\left[Cv(\nu_{j})\right]^{2} \sigma^{2}_{Y,\rm noise} + \sigma^{2}_{Y}(\nu_{j})\sigma^{2}_{X,\rm noise} + \sigma^{2}_{X,\rm noise} \sigma^{2}_{Y,\rm noise}}{2KM\sigma^{2}_{Y}(\nu_{j})}}
\end{equation}
where $\sigma^{2}_{Y}(\nu_{j})$ is the noise-subtracted absolute rms-squared of the reference band and $\sigma^{2}_{Y,\rm noise}=P_{Y,\rm noise} \langle y \rangle ^{2} \Delta \nu_{j}$, where the PSDs are again in the fractional rms-squared normalisation.  Note that $\left[Cv(\nu_{j})\right]^{2}$ is used instead of the absolute rms-squared of the CI ($\sigma_{X}(\nu_{j})$) in this formula, because we assume unity intrinsic coherence, in which case $\left[Cv(\nu_{j})\right]^{2}$ gives a significantly more accurate measure than the equivalent value obtained from the rms-spectrum.

Note that the errors on the covariance spectrum can be substantially smaller than those of the rms spectrum provided that $\sigma^{2}_{Y}(\nu_{j}) > \sigma^{2}_{Y,\rm noise}$ {\it and} $\sigma^{2}_{X}(\nu_{j}) < \sigma^{2}_{X,\rm noise}$.  This is likely to be the case in many situations where the reference band contains substantially more photons than the channels of interest, e.g. through covering a broader energy range and/or lower X-ray energies.
\subsection{Sensitivity and signal-to-noise considerations}
\label{sec:sensitivity}
We now consider the sensitivity of lag measurements as well as the accuracy and limitations of the equation for describing the error on the lag (Equation~\ref{eqn:lagerror}).  To test the efficacy of the error equation and examine the statistical uncertainty of the lag as a function of flux and number of samples measured, we carry out Monte Carlo simulations of light curves with a fairly typical PSD for an NLS1 AGN, with slope -1 breaking to -2 above a frequency of $10^{-4}$~Hz.  The PSD normalisation in fractional rms-squared units is 200~Hz$^{-1}$ at the break frequency, yielding a typical fractional rms in 100~ks of a few tens of per~cent.  For each flux level selected, we used the method of \citet{timmerkoenig95} to simulate light curves which were then simply shifted by 100 seconds to yield a lagging light curve (the simulated light curves were also exponentiated, to produce the appropriate log-normal flux distribution, see \citealt{uttleymchv05}).  Poisson noise was then added to both light curves according to the chosen count rates: the `driving' light curve was chosen to have a count rate 30 times higher than the lagging light curve (analogous to the differences in count rate expected when measuring lag-energy spectra using a broad reference band).  The lag was measured by averaging the cross-spectra over the frequency range $1$--$3\times10^{-3}$~Hz, and the raw coherence and hence lag error were estimated in the standard way.  Using the same underlying light curves, we regenerated the Poisson noise on both light curves $10^{4}$ times, to examine the distribution of the observed lag and the estimated lag error.

\begin{figure}
\begin{center}
\includegraphics[width=7cm,angle=-90]{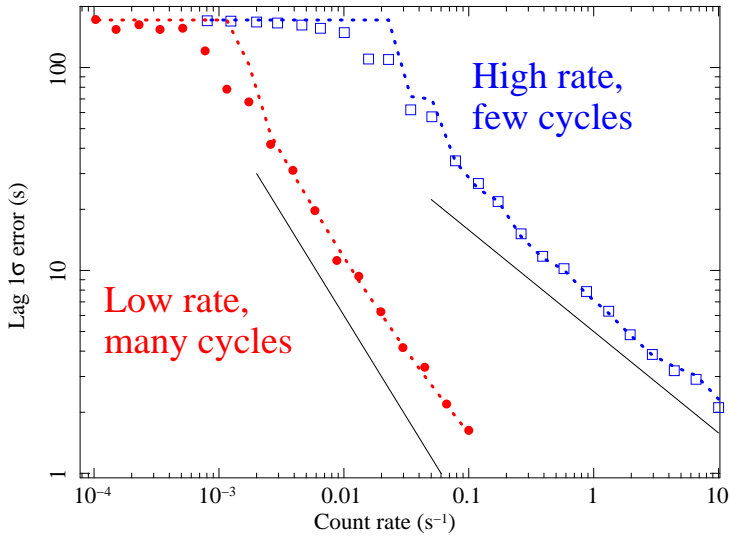}
\caption{Errors on the measurement of an $\sim80$~s lag in simulated light curves, comparing the error bar obtained using Equation~\ref{eqn:lagerror} (dotted lines, note that the median error obtained from the 300 or $10^{4}$ simulated light curves was used in each case) with the equivalent $1\sigma$ error obtained from the distribution of simulated lag measurements (open and filled data points).  Count rates are given for the lagging band (the 'channel-of-interest') while the driving band (the `reference band') is given a count rate which is 30 times larger.  Since the same PSD is assumed for all cases, the count rate can be related to counts per cycle of variability, simply by multiplying the rate by the time-scale being probed ($\sim500$~s in this case).  The blue dotted line and open squares show the results for high counts per cycle, low number of measured cycles (corresponding to AGN), while the red dotted line and filled circles show the low counts per cycle, large number of cycles case (which is more similar to BHXRBs).  Note the excellent match of the simulated distribution of lags with the analytical error estimate, over a wide range of lag errors.  The solid black lines show the slopes expected for lag signal-to-noise scaling linearly with count rate (low rate, many cycles case) and with the square-root of count rate (high rate, few cycles case), expected for the two signal-to-noise regimes.  See text for details of the simulations.}
\label{lagerror}
\end{center}
\end{figure}

The results of our simulations are shown in Fig.~\ref{lagerror}.  Note that due to the choice of the frequency bin centre as the frequency for conversion of phase to time lag (see discussion in Sect.~\ref{sec:crossspeclags}), the observed lag is only $\sim80$~s.  The figure shows the equivalent $1\sigma$ range of the distribution of observed lags\footnote{I.e. corresponding to half the separation in lag between the 15.87 and 84.13 percentile values of the distribution, which is equivalent to the standard deviation for a Gaussian distribution.} and the median analytical estimate of the error for a variety of different count rates (plotted versus the lagging light curve count rate) and also for two different regimes.  Since the same PSD is used in all cases the count rates can also be multiplied by the time-scale being probed (e.g. in this case $\sim500$~s for a lag in the range $1$--$3\times10^{-3}$~Hz) to give the {\it counts per cycle} of variability, meaning that the results can be extrapolated to objects with different black hole masses but the same mass-scaled PSD.  The high count rate regime (blue) uses lags calculated by averaging the cross-spectrum from 3 light curve segments, each 40960~s long (bin size is chosen to be 10~s), i.e. similar to what might be typical for AGN observations.  The low count rate regime (red) uses 1000 segments, also each 40960~s long\footnote{The distributions in the low count rate regime are generated using only 300 realisations instead of $10^{4}$, due to computational speed limitations.}, thus compensating for the lower count rates.  This latter situation could never be realised for AGN with realistic X-ray observatories, but is closer to the situation in X-ray binaries, where we observe many more cycles of variability, but at much lower count rate per cycle.  

It is interesting to note that the two signal-to-noise regimes considered here show different dependences of the lag error on the flux, also highlighted by the solid black lines which show the corresponding functional forms: the high-rate (per cycle), few cycles regime shows the error scaling with $1/\sqrt(flux)$, as is familiar from conventional spectroscopy.  However, the low-rate (per cycle), many cycles regime shows the error scaling with $1/flux$.  To understand this behaviour, we recall that the phase-lag error depends simply on the observed `raw' coherence.  If we consider the case where the intrinsic coherence is unity, it is easy to show that the raw coherence is given by:
\begin{equation}
\label{eqn:simplecoherence}
\gamma^{2}(\nu_{j}) = \frac{(P_{X}(\nu_{j}) - P_{X,\rm noise})(P_{Y}(\nu_{j}) - P_{Y,\rm noise})}{P_{X}(\nu_{j}) P_{Y}(\nu_{j})}
\end{equation}
which simplifies to:
\begin{equation}
\gamma^{2}(\nu_{j}) = \left[\left(1+\frac{P_{X,\rm noise}}{P_{X, \rm signal}}\right)\left(1+\frac{P_{Y,\rm noise}}{P_{Y, \rm signal}}\right)\right]^{-1}
\end{equation}
where we define the intrinsic signal power in the channel-of-interest as $P_{X, \rm signal}= P_{X}(\nu_{j}) - P_{X,\rm noise}$ and similarly for the reference band.  Thus Equation~\ref{eqn:lagerror} can be expressed as:
\begin{equation}
\label{eqn:simplelagerror}
\Delta \phi(\nu_{j}) = \sqrt{\left(\frac{P_{X,\rm noise}}{P_{X, \rm signal}}+\frac{P_{Y,\rm noise}}{P_{Y, \rm signal}}+\frac{P_{X,\rm noise}P_{Y,\rm noise}}{P_{X, \rm signal}P_{Y, \rm signal}}\right)/2M}
\end{equation}
Therefore the error on the lag depends on the ratio of the Poisson noise level to the intrinsic variability power in the frequency range $\nu_{j}$.  Moreover, the error contains terms within the square-root which are linear in this ratio, and a non-linear, squared term.  When the linear terms dominate the lag error, the signal-to-noise of the lag measurements scales with the square-root of the count rate, since (in fractional rms-squared units), the Poisson noise level scales inversely with count rate.  

If we assume that the reference band signal-to-noise is always larger than that of the channels-of-interest, it is easy to see that the non-linear term dominates the error when $\frac{P_{Y,\rm noise}}{P_{Y, \rm signal}}\gg 1$.  This situation will occur when there are few photons per variability time-scale in the reference band, and/or the rms-amplitude of variability is small.  In particular, the former condition is typically satisfied when we observe X-ray binaries at time-scales where we expect reverberation signatures to dominate the lags.  In this regime, it is easy to see that the signal-to-noise of any lag measurements scales linearly with count rate, as shown from our simulations.  Thus, large improvements can be gained by studying brighter sources or using X-ray detectors with larger effective area.  This point will be important when we consider the future of reverberation studies, in Sect.~\ref{sec:future}.   It is also interesting to note that Equation~\ref{eqn:simplelagerror} implies that in the AGN case where the linear terms inside the square-root dominate the error, the error on the lag measured over an equivalent {\it mass-scaled} frequency range (for the same exposure) is independent of black hole mass.  This is because, for a source with the same mass-scaled PSD, the PSD amplitude scales increases linearly with decreasing frequency, while the number of cycles measured in a fixed exposure decreases, so that the two changes cancel out.  AGN with very massive black holes will remain difficult to study however, since the frequency range where reverberation lags are expected will be shifted to time-scales longer than the exposure time (e.g. see \citealt{demarco13}).
\begin{figure}
\begin{center}
\includegraphics[width=7cm,angle=-90]{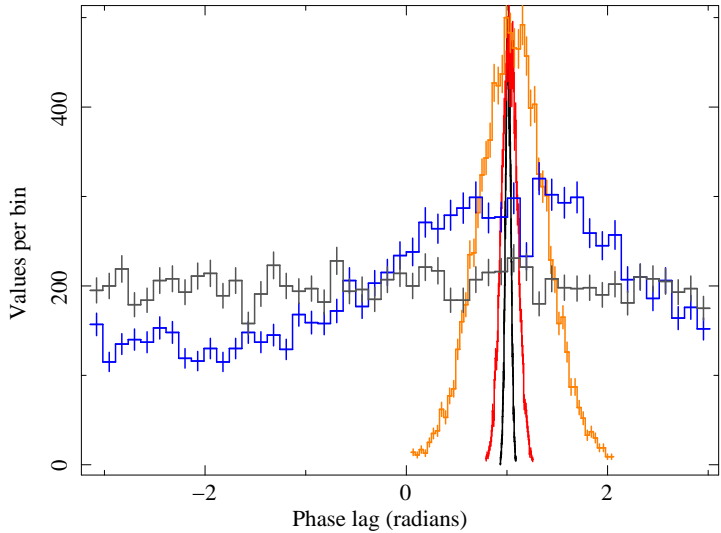}
\caption{Phase lag distributions for the $10^{4}$ simulations carried out for five different count rates in the high rate, few cycles regime (10, 1.3, 0.12, 0.01, $8\times10^{-4}$ count~s$^{-1}$, corresponding to successively broader distributions).}
\label{lagerrordist}
\end{center}
\end{figure}

We can use our simulations to consider the limitations of the standard lag error formula.  Fig.~\ref{lagerror} shows that the error formula performs extremely well in both signal-to-noise regimes over a wide range of count rates.  However, it starts to break down for large errors.  This effect is linked to the fact that the phase lag is bound to lie between $-\pi$ and $\pi$ and thus the Gaussian shape of the distribution of measured lags breaks down for large errors.  We demonstrate this effect in Fig.~\ref{lagerrordist}, which shows how for small lag errors, the distributions are close to Gaussian, but for large errors the lags start to wrap around in phase until at the lowest count rates the phase lag distribution is completely uniform between $-\pi$ and $\pi$.  This limit corresponds to the limiting value of the lag errors, seen in Fig.~\ref{lagerror}, of 171~s (which corresponds to a phase lag of $\sim 0.34\times 2\pi$, or $1\sigma$).  Of course, in this situation, the lags are completely undetermined.  Note that the analytical lag error reaches this limit more quickly because of the bias term that must be subtracted from the numerator of the coherence (see Sect.~\ref{sec:coherr}).  The subtracted term is the {\it expectation value} of the bias on the modulus-squared of the cross-spectrum.  This bias is distributed as a $\chi^{2}_{2}$ distribution and hence is positively skewed with a median value less than the expectation value, leading to coherence which is frequently negative and so cannot be used for lag estimation (hence the error is set to the limiting value).  

The simulations show that the phase lag distribution starts to become significantly non-Gaussian for phase lag errors exceeding $\Delta \phi\simeq 0.75$.  This limit should apply regardless of the signal-to-noise regime or observed count rates.  Thus, care should be taken when fitting lag-frequency or lag-energy spectra with errors this large, since the standard assumption of Gaussian errors will no longer apply.  Ideally, the cross-spectrum should be fitted directly in this case (since its errors on the real and imaginary components will remain Gaussian), or simulations should be used to assess the uncertainty on fitted model parameters.

Finally we point out that there is some small scatter on the observed phase or time lag for the different realisations of the {\it underlying} light curves which are used as the basis of the simulations for each different count rate.  This effect causes the point-to-point scatter around the expected smoother trends in Fig.~\ref{lagerror}, and the small differences in centroid of the lag distributions (even though the intrinsic lag between the light curves is the same), including for the cases where the distribution is close to Gaussian, as seen in Fig.~\ref{lagerrordist}.  The effect arises because the red-noise PSD of the underlying variability is intrinsically noisy.  Thus, for different realisations of the underlying light curves, power is redistributed differently between frequencies, causing a change in the average phase or time-lag measured over the chosen frequency range.  This effect is small compared to the expected errors and if necessary can be mitigated by using narrower frequency ranges to measure the lags (e.g. as in a lag-frequency spectrum), or averaging over many cycles of variability.  It is also important to bear in mind that, provided that the intrinsic variations in different bands are well-correlated (i.e. intrinsic coherence is close to unity), errors such as this are {\it systematic} and apply a similar fractional shift to the lags measured at all energies.  Thus the shape of the lag-energy spectrum should not be changed by this effect.

\subsection{Introducing the impulse response}
\label{sec:imprespintro}
Finally, as a prelude to the discussion of more detailed models in Sect.~\ref{sec:modelling}, and also a taster for the consideration of the actual physical `meaning' of the observations presented in the next section, we briefly consider how timing behaviour can be interpreted in terms of the {\it impulse response}, which enables us to link models for variability with models for the emission, and connect these models to the observed spectral-timing properties.  
\subsubsection{Basic concepts}
Consider continuous light curves measured in two bands, $x(t)$ and $y(t)$.  Let us imagine that the variations in both bands are driven by the same underlying variable signal, which is described by a time-series $s(t)$.  The variable signal does not necessarily emit radiation itself, e.g. it may correspond to fluctuations in mass accretion rate which themselves drive variable emission.  On the other hand, the variable signal may correspond to variations of some driving continuum source (e.g. in the case of reverberation). Depending on the physics of the variability process and the emission mechanism, the emission which we see in each band may be related to the underlying driving signal by a linear impulse response, which represents the response of the emission in that band to an instantaneous flash ({\it i.e.} a delta-function impulse) in the underlying driving signal.  Thus, if the impulse responses of bands $x$ and $y$ are $g$ and $h$ respectively, the signals in these bands are obtained by integrating over all time delays $\tau$:

\begin{equation}
x(t) = \int_{-\infty}^{\infty} g(\tau) s(t - \tau) \; d\tau
\label{eq:tf1}
\end{equation}

\begin{equation}
y(t) = \int_{-\infty}^{\infty} h(\tau) s(t - \tau) \; d\tau
\label{eq:tf2}
\end{equation}
I.e. the observed light curve is the convolution of the underlying driving signal time-series with the impulse response for that energy band. It is important to note that for all but the sharpest impulse responses, the effect of the convolution is not only to delay the underlying time-series, but also to smear it out on time-scales comparable to or shorter than the width of the impulse response.  Thus the information about the impulse response is encoded not only in time-lags, but also in the time-scale-dependent amplitude of variability in each band.  The impulse response itself depends on the physical process for variability and emission.  Many emission mechanisms will incorporate delays in the observed emission. For example a delta-function `flash' of seed photons which are upscattered by thermal Comptonisation will be delayed and smeared out by the time-taken to be scattered and escape the scattering region, with escaping higher energy photons subject to longer delays since they undergo more scatterings.  The propagation of signals through the accretion flow will lead to much longer delays associated with the radial (viscous) propagation time through the flow. 

\subsubsection{A simple example: reverberation from a spherical shell}
\label{sec:revsphershell}
\begin{figure}
\centering
\includegraphics[angle=270,width=0.8\textwidth]{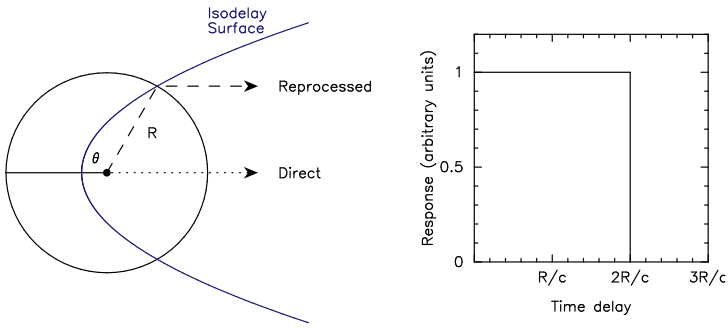}
\caption{{\it Left:} Schematic diagram showing reprocessing by a thin spherical shell of radius $R$.  The path length difference between the direct emission (dotted line) and the reprocessed emission (dashed line) is $R(1 + \cos\theta)$, and hence the time delay for a given position on the sphere is  $\tau = (1 + \cos\theta ) R / c$.  An isodelay surface is shown in blue.  {\it Right:} The corresponding impulse response is a simple top-hat function extending from the minimum delay ($\tau = 0$) to the maximum delay at $\theta = 180^\circ$, which is $\tau = 2R/c$. }
\label{fig:tophat}
\end{figure}

In reverberation scenarios, we have a driving continuum light curve that irradiates the accretion disc leading to reflected emission.   The delay is primarily set by the light travel time between the source of the irradiating emission and the location of the reflected emission.  One can build up a simple impulse response for a known geometry based on the path-length difference between the direct emission from the driving light curve and the reflected emission from each reprocessed region.  At a given time $\tau$ after a delta-function flare, reprocessing from a spherical shell will be seen from any region intersecting with an isodelay surface given by
\begin{equation}
\tau = (1 + \cos\theta ) r / c
\end{equation} 
where $r$ is the radius from the source of the flare and $\theta$ the angle measured from the observer's line of sight.  We demonstrate this for a simple spherical shell in Fig.~\ref{fig:tophat} (left panel).  From this figure, it can be seen that the path length difference between the direct and reprocessed emission will be $r(1 + \cos\theta)$, and the time delay will simply be the path length difference divided by the speed of light. The region of the shell which we see emission from at a time $\tau$ after observing a flare, is the region intersected by the isodelay surface at $\tau$.   Of course, in reality we do not have delta-function flares, but continuous variability, and thus our reflected light curve will be like the driving light curve but smoothed and delayed by the range of time delays possible from different regions of the sphere.

The impulse response for a thin spherical shell is also shown in Fig.~\ref{fig:tophat} (right panel).  Assuming that the shell reprocesses the emission equally at all places, the impulse response will simply be given by a top-hat function that extends from the minimum delay ($\tau = 0$ in this case for reprocessed emission directly along the line of sight on the near side of the sphere) to the maximum delay ($\tau = 2R/c$ which corresponds to emission along the line of sight from the far side of the sphere).  The reprocessed light curve, then, will simply be the driving light curve convolved with the top-hat function following Equations~\ref{eq:tf1} and \ref{eq:tf2}.
\subsubsection{The effects on spectral-timing measurements}
\label{sec:spectimeffects}
From the convolution theorem of Fourier transforms, it is easy to see that the Fourier transform of an observed light curve is equal to the Fourier transform of the driving signal multiplied by the Fourier transform of the impulse response:
\begin{equation}
X(\nu)=S(\nu)G(\nu)
\end{equation}
\begin{equation}
Y(\nu)=S(\nu)H(\nu)
\end{equation}
It is then simple to determine the effect of the impulse response on the PSD:
\begin{equation}
|X(\nu)|^{2} = S^{*}(\nu)G^{*}(\nu)S(\nu)G(\nu) = |S(\nu)|^{2}|G(\nu)|^{2}
\end{equation}
Thus, after applying the appropriate normalisation, we find that the observed PSD is equal to the PSD of the driving signal, multiplied by the modulus-squared of the Fourier transform of the impulse response.  The impulse response acts as a filter on the variabilty, modifying the shape of the PSD.  Energy-dependent impulse responses will result in PSDs with energy-dependent shapes.  In particular, a more extended impulse response will suppress power at high frequencies.  Thus, the flattening of the high-frequency PSD which is frequently observed at higher energies may indicate a sharper impulse response at these energies \citep{kotov01}.

It is also simple to understand the meaning of the cross-spectrum and phase/time-lags in terms of the impulse response:
\begin{equation}
C(\nu)=S^{*}(\nu)G^{*}(\nu)S(\nu)H(\nu)=|S(\nu)|^{2}G^{*}(\nu)H(\nu)
\end{equation}
So that after applying the appropriate normalisation, the observed, binned cross-spectrum is equal to the driving signal PSD {\it multiplied by the cross-spectrum of the impulse response}.  Thus the cross-spectrum encodes detailed information about the shape of the impulse response as a function of energy and thus the phase/time lags in each energy band.

It is important to note that the impulse responses considered here and in more detail in Sect.~~\ref{sec:modelling} are all linear, in that they represent systems which respond linearly to the driving signal.  Reverberation should be well-described by a linear impulse response in cases where variations in the illuminating continuum do not significantly alter the {\it shape} of the reflected spectrum (e.g. due to significant changes in the reflector ionisation state).  We will briefly discuss systems which may be modelled using non-linear impulse responses in Sect.~\ref{sec:future:models}.
%\bibliographystyle{apj}
%\bibliography{agn}

\section{Observations of Reverberation Signatures}\label{sec:obs}
We now consider the observational evidence for X-ray reverberation, focussing first on the evidence for reflection reverberation in Active Galactic Nuclei, before discussing the evidence for disc thermal reverberation in black hole X-ray binaries.

\subsection{Active Galactic Nuclei}

\subsubsection{The discovery of the soft lag}

\begin{figure}
\begin{center}
\includegraphics[width=6.5cm,angle=-90]{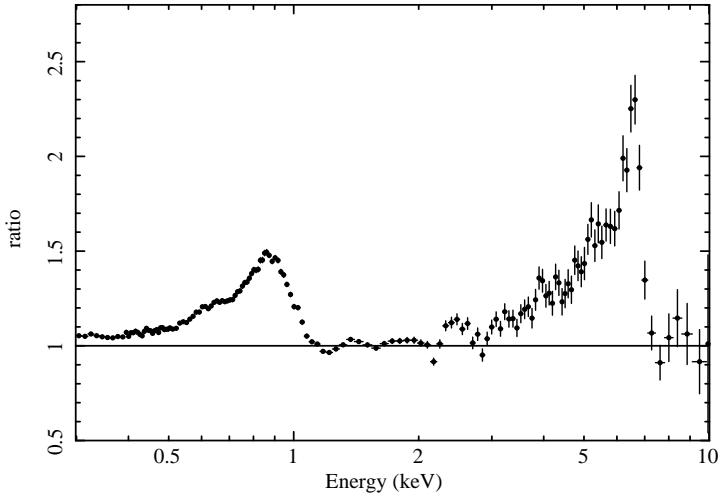}
\caption{The ratio spectrum of 1H0707-495 to a continuum model \citep{fabian09}.  The broad iron K and iron L band are clearly evident in the data.  The origin of the soft excess below 1~keV in this source had been debatable, but in this work was found to be dominated by relativistically broadened emission lines.}
\label{1h07_spec}
\end{center}
\end{figure}

Reverberation lags in the X-rays were first robustly observed by \citet{fabian09} through a
500~ks {XMM-Newton} observation of Narrow-line Seyfert I galaxy,
1H0707-495.  In this work, Fabian et al. found that the soft excess (a
traditionally contentious part of the energy spectrum) was composed of
relativistically broadened reflection features, namely the iron L
emission line, and possibly the oxygen line, as well.
Fig.~\ref{1h07_spec} shows the ratio of the spectrum of 1H0707-495 to
a phenomenological model, consisting of an absorbed power law plus
black body emission from the accretion disc. The broad asymmetric peak at 6.4~keV is the Fe~K
emission line, which had been observed many times before
\citep[e.g.][]{tanaka95} because it is in a relatively `simple' part
of the spectrum with little absorption, and because iron is the most
cosmically-abundant element with a high fluorescent yield. The broad line at
$\sim 0.7$~keV is identified as the iron L line (with possibly some
contribution from oxygen at $\sim 0.5$~keV). The lines in the soft
excess are observationally more difficult to detect, as there are many
emission lines at soft energies, contributions from the tail of the
disc black body emission and possible absorption effects.

\begin{figure}
\begin{center}
\includegraphics[width=6.5cm,angle=-90]{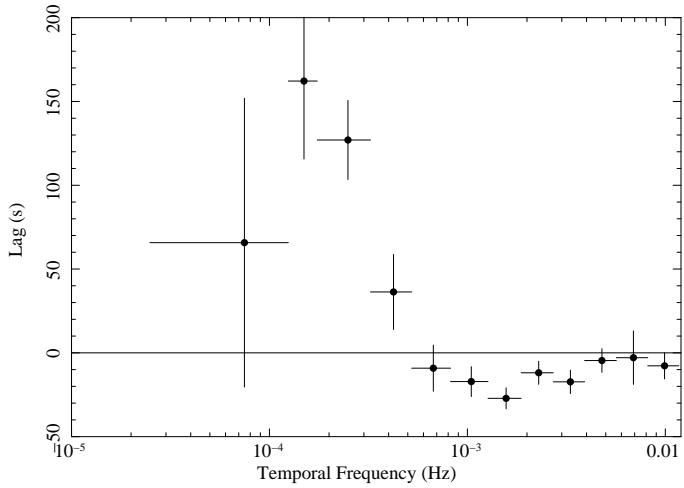}
\caption{The frequency-dependent lags in 1H0707-495 between the continuum dominated hard band at 1--4~keV and the reflection dominated soft band at 0.3--1 keV.}
\label{1h07_lag}
\end{center}
\end{figure}

\begin{figure}
\begin{center}
\includegraphics[width=\textwidth]{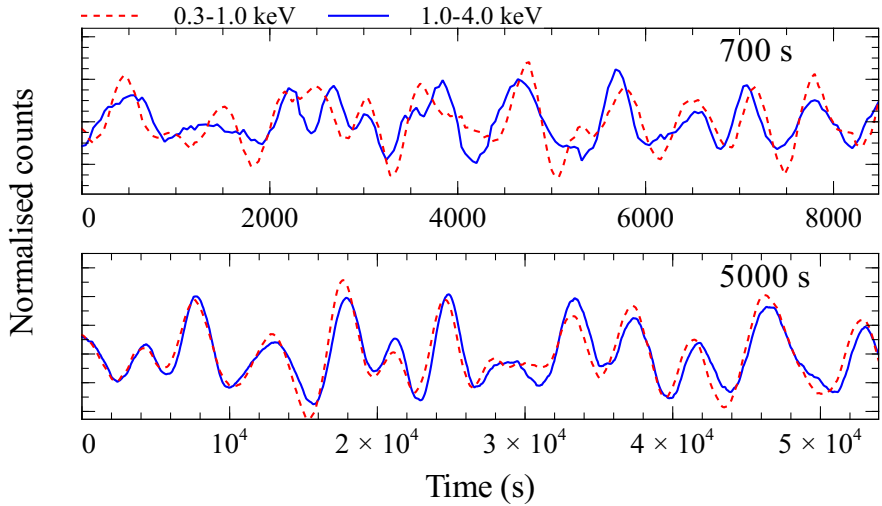}
\caption{Illustration of the soft lag (top) and hard lag (bottom) in the time domain, by filtering the light curve on short time-scales (to probe the high frequency soft lag) and long time-scales (to probe the low-frequency hard lag).  This is completely analogous to the Fourier analysis shown in Fig.~\ref{1h07_lag}. Figure from \citet{zoghbi10}.}
\label{1h07_lc}
\end{center}
\end{figure}

The clear determination of disc reflection in the soft band motivated the search for time lags between the soft band
and the continuum-dominated band at 1--4~keV.  Using the Fourier
timing techniques described in the previous section, the
frequency-dependent lag was measured between these two bands.  A clear `soft
lag' of $\sim 30$~s was discovered at $\sim 10^{-3}$~Hz
(Fig.~\ref{1h07_lag}). In other words, soft band variations on the
order of 1000~s followed the corresponding hard band variations by an
average of 30~s.  A follow-up study of the reverberation lags in
1H0707-495, \citep{zoghbi10} illustrated this frequency-dependent
result, by using light curves filtered to highlight variations on different time-scales, thus showing the lags in the time domain (Fig.~\ref{1h07_lc}).
Similar to Fig.~\ref{1h07_lag}, we are looking at the time delays
between the soft band (0.3--1~keV) and the continuum-dominated band
(1--4~keV).  The top panel shows the light curve variations on the
time scale of 700~s (corresponding to a frequency of $\sim 1 \times
10^{-3}$~Hz).  The soft band lags the hard band at this time scale
(corresponding to the negative lag in Fig.~\ref{1h07_lag}).  However,
looking at variations on the order of 5000~s (frequency of $2 \times
10^{-4}$~Hz), the hard band lags the soft.  These light curves show
the same effect we see in the frequency domain, but also elucidate why
it is much easier to do this type of analysis in the frequency domain, where distinct lags on different time-scales can be cleanly separated without any pre-selected filtering of the light curves.

Given the spectral results that the soft band is dominated by reflection, and the harder, 1--4~keV band by the continuum emission from the corona, the soft lag is naturally interpreted in terms of the average light-travel time delay between the compact corona and the inner accretion disc.  For 1H0707-495, the average lag was measured to be $\sim 30$~s, which, knowing the mass of the black hole, corresponds to a source height distance of $< 5 r_{\mathrm{g}}$ above the accretion disc. Note that this is the average lag, and does not account for the fact that both energy bands contain contributions from both the directly observed continuum and the reflection. The hard and soft bands are taken to be direct proxies for the continuum and reflected emission, respectively, whereas in reality, the contributions of both components to each band will reduce the measured value or `dilute' the measured lag. Neither does this account for the fact that the corona is likely extended, or inclination effects (since we are actually seeing the lag caused by a {\it path-length difference} between the paths from corona-observer and corona-disc-observer). These effects will be discussed further in Sect.~\ref{sec:freqdep}.    

At frequencies below $10^{-3}$~Hz, the lag was observed to switch sign to become a `hard lag'.  This hard lag had been observed previously in several NLS1 galaxies (e.g. \citealt{vaughanfn03,mchardy04,arevalo06}) and first in galactic black hole X-ray binaries \citep{miyamoto89,nowak99}. The time-scales of this lag negated the possibility that it was a Comptonisation delay, but rather suggest that the lag is caused by fluctuations in the mass accretion rate in the disc that get propagated inwards on the viscous time-scale, causing the outermost soft X-rays in the corona to respond before hard X-rays at smaller radii.  See Sect.~\ref{sec:bhb} for more on the hard lag and its interpretation in black hole X-ray binaries.

\begin{figure}
\begin{center}
\includegraphics[width=7.5cm]{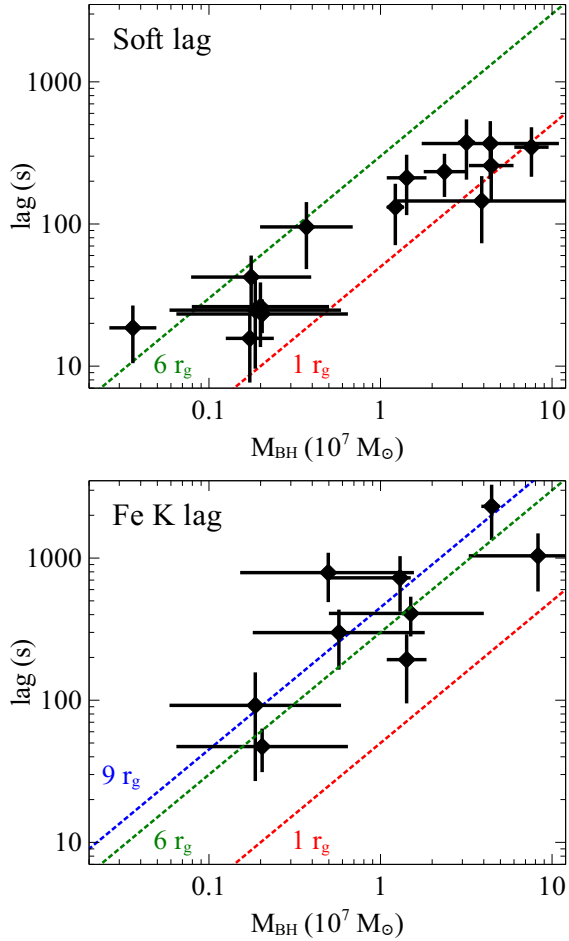}
\caption{{\em Top: }A sample of 15 Seyfert galaxies with significant soft lag measurements. The amplitude of the lag was found to scale with black hole mass, and suggested that the X-ray emitting region was very close (within $10~r_{\mathrm{g}}$) to the central black hole. Figure adapted from \citet{demarco13}. {\em Bottom: } The current sample of iron K lags (defined as being the lag of 6--7~keV relative to 3--4~keV) for nine AGN (see Table~\ref{lagtable}), plotted with the corresponding black hole mass.  The short amplitude lag found for both the iron K and soft lags suggest that they originate from the same small emitting region.}
\label{lag_mbh}
\end{center}
\end{figure}

After the initial discovery of reverberation lags in 1H0707-495, soft lags were discovered in a number of other NLS1 sources \citep{emma11, zog11_rej1034, demarco11, cackett13, fabian13}.  \citet{demarco13} conducted a systematic search through the XMM-Newton archive for variable Seyfert galaxies with sufficiently long observations, and found significant high-frequency soft lags in 15 sources.  Plotting the amplitude of the lags with their best-estimated black hole masses\footnote{Black hole masses used by \citet{demarco13,kara13c} and in Fig.~\ref{lag_mbh} were obtained from the literature, and estimated primarily using optical broad line reverberation. In a few cases masses were estimated using the scaling relation between optical continuum luminosity and broad line region radius, which can be used to estimate black hole mass when combined with optical line width (e.g. \citealt{kaspi00,grier12}), or the correlation between black hole mass and host galaxy bulge stellar velocity dispersion (e.g. \citealt{gebhardt00}).}, revealed that the amplitude of the lag scales approximately linearly with mass\footnote{The observed scaling is flatter than expected from a linear relationship, but this can be explained as a bias due to the fact that we only sample the higher frequency end of the soft lag range in the highest mass objects, which leads to systematically shorter lags than would be seen if we could sample the maximum amplitude of soft lags seen at lower frequencies \citep{demarco13}.}, and suggested that the X-ray source was compact and within a few gravitational radii of the accretion disc (top panel of Fig.~\ref{lag_mbh}).  The plot has been updated with the additional soft lag measurements that have been found since publication.   

\subsubsection{The iron K lag}

\begin{figure}
\begin{center}
\includegraphics[width=7.5cm]{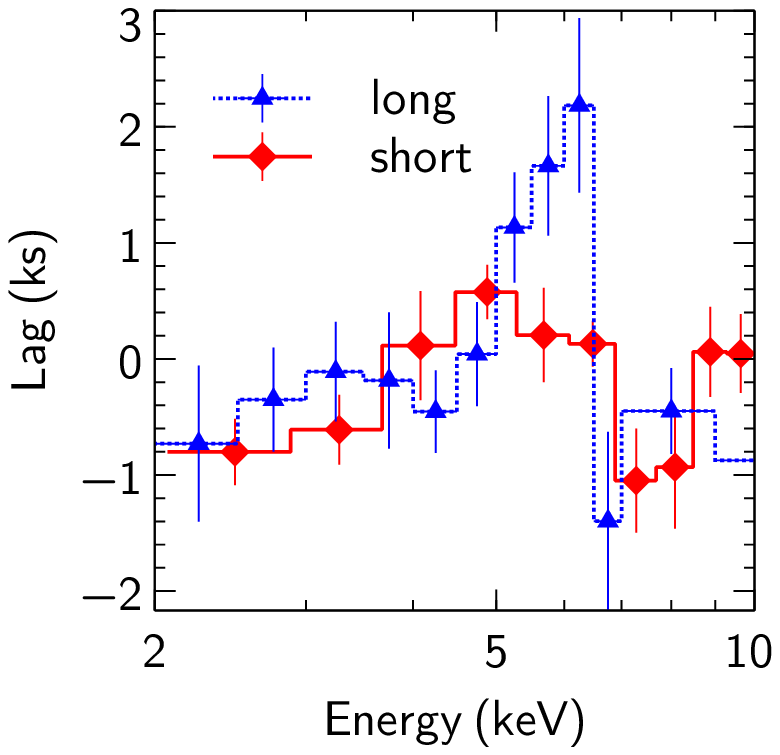}
\caption{The iron K lag in NGC~4151.  The lag-energy  spectrum is read from bottom to top, such that the continuum emission at 2~keV responds first, followed by the red wing of the iron line 1000~s later, and lastly the blue horn of the line 1000~s after that.  The higher frequency lags ($5$--$50\times10^{-5}$~Hz) in red show emission from smaller radii, and therefore are dominated by the red wing of the line, while the blue shows the lower frequency lags ($1$--$2\times10^{-5}$~Hz), which show the line from larger radii, closer to the rest-frame energy. Figure from \citet{zoghbi12}.}
\label{ngc4151}
\end{center}
\end{figure}
Up to this point, reverberation lag studies focussed on the reflection-dominated soft band, and the continuum band at 1--4~keV because the signal-to-noise is higher at these low energies.  The ultimate test of reverberation, though, is in the iron K band. The broad iron K line peaks at 6.4~keV, but is asymmetrically broadened due to special relativistic beaming and Doppler shifts from a rotating accretion disc and the gravitational redshift close to the black hole.  Because the iron line has the main advantage of being in a `clean' part of the spectrum, where it is not affected by the reprocessed black body emission or by other broadened lines, it is our best indicator of the relativistic effects caused by the black hole, and therefore was a natural place to search for reverberation lags.  This was first accomplished by \citet{zoghbi12} for the bright Seyfert, NGC~4151 (Fig.~\ref{ngc4151}).  In this work, the continuum emission was found to respond first, followed by the red wing of the line (from the very innermost radii) and finally by the rest frame of the line (produced at farther radii).  This showed not just the average time lag from all light paths from corona to disc, but actually identified the reverberation from light echoing from different parts of the accretion disc.

In addition to finding the first Fe~K lag, \citet{zoghbi12} discovered the frequency-dependence of the Fe~K lag.  At low frequencies ($1$--$2\times10^{-5}$~Hz), the lag-energy spectrum shows the prominent core of the Fe~K line, peaking at close to the rest energy.  However, at higher frequencies the line core is suppressed in the lag-energy spectrum and we see systematically shorter lags peaking more in the red wing of the line.  A likely explanation for this behaviour is that we are seeing the first evidence of the systematically shorter length-scales which produce the red wing of the line, from closer to the black hole where gravitational and transverse Doppler shifts are strongest.  At high Fourier frequencies, the larger-scale rest-frame emission should be washed out by the larger light-crossing times, and effectively removed from the lag-energy spectrum, revealing the red-wing of the line which is preserved in the lag-energy spectrum due to the short light-travel times from the compact central corona to the inner disc (see Sect.~\ref{sec:model:endeptf} for a more detailed explanation of this effect).  Hints of frequency dependence of the Fe~K lag have been seen in other sources \citep{zoghbi13,kara13d}, but only NCG~4151 with its high count rate, shows a significant frequency dependence so far.  

\begin{figure}
\begin{center}
\includegraphics[width=9cm]{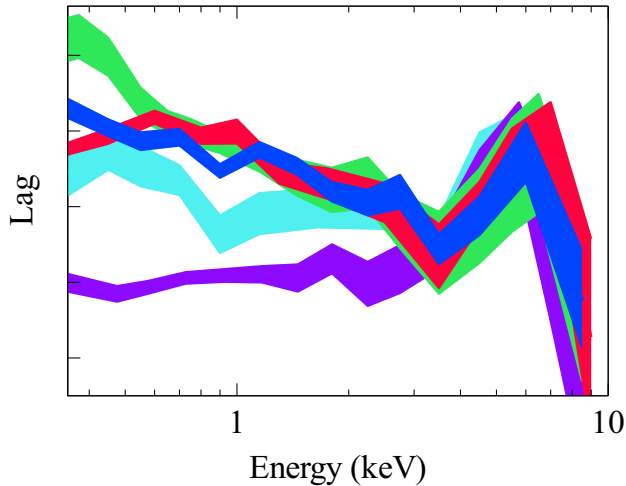}
\caption{The lag-energy spectra overplotted for five of the published sources with Fe~K lags.  The amplitude of the lag has been scaled such that the lags between 3--4~keV and 6--7~keV match for all sources.  The sources shown are: 1H0707-495 (blue), IRAS 13224-3809 (red), Ark~564 (green), Mrk~335 (cyan) and PG~1244+026 (purple).  While the shape of the Fe~K lag is similar in all these sources, the lags associated with the soft excess vary greatly.}
\label{lagen_stack}
\end{center}
\end{figure}

The iron K lag is a powerful tool for understanding the geometry and kinematics of the inner accretion flow, as it encodes spectral and timing information about the reflected emission (See Sect.~\ref{sec:modelling} for more on modelling lags in the iron K band).  Since the initial discovery in NGC~4151, iron K lags have been found in nine sources, including the original reverberating source, 1H0707-495 \citep{kara13a}.  Fig.~\ref{lagen_stack} shows five of those sources, overplotted on the same axis. As they all have different black hole masses (and therefore different Fe K lag amplitudes), the lags have been scaled such that the relative lag matches between 3--4~keV and 6--7~keV.  It is clear from this figure, that the shape of the Fe~K lag is similar in all these maximally-spinning black hole systems, but the lags associated with the complex soft excess vary greatly.  This indicates the importance of the Fe~K line in understanding the effect of strong gravity on the reverberation lag.  Future work in understanding the soft lag is also important, and may help break degeneracies in spectral modelling of the soft excess.
 
The bottom panel of Fig.~\ref{lag_mbh} shows the amplitude of the iron K lag for all the published measurements thus far plotted against black hole mass \citep{kara13c}.  Similar to the soft lags in the top panel of Fig.~\ref{lag_mbh}, the iron K lags show a linear dependence on mass, and confirm that both the soft excess in these sources and the X-rays illuminating the accretion disc producing the iron K line, originate from a small emitting region close to the central black hole.  The fact that the lags in the iron K band are generally larger than the soft lags is likely due to greater dilution by the continuum in the soft band, which can be accounted for in modelling of the lag (see Sect.~\ref{sec:modelling} for a discussion of dilution).  As we gain a clearer understanding of the geometry of the systems we are probing, we can use reverberation lags as an indicator of black hole mass.  The lag measures the size of the region in physical units (i.e. in metres rather than in gravitational units), and so if we understand how far the source is from the accretion disc in gravitational units, we can use the lag to make a measurement of the black hole mass.  

\begin{figure}
\begin{center}
\includegraphics[width=8cm]{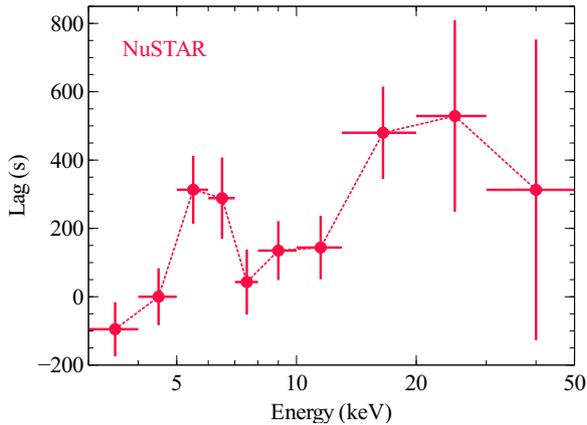}
\caption{{\it NuSTAR} lag-energy spectrum of SWIFT~J2127.4+5654.  Figure taken from Kara et al. (in prep.).  Note the increase at energies above the Fe~K line, consistent with reverberation of the disc reflection continuum.}
\label{fig:nustarlags}
\end{center}
\end{figure}

Finally, it is worth emphasising that the high-frequency iron~K lags represent a model-independent confirmation of the interpretation of broad iron K lines as signatures of relativistic reflection from a compact reflector close to the black hole: these results are independent of any models fitted to the time-averaged spectra and demonstrate the reality of broad iron K features as distinct emission components.  This picture is supported by the first detections of reverberation lags with the {\it NuSTAR} hard X-ray observatory (\citealt{zoghbi14}, Kara et al. in prep.).  Fig.~\ref{fig:nustarlags} shows the {\it NuSTAR} lag-energy spectrum for SWIFT~J2127.4+5654 (Kara et al. in prep.).  The discovery of Fe~K reverberation in {\it NuSTAR} data confirms that the detections by {\it XMM-Newton} are not in any sense `pathological' to that mission.  Furthermore, there is a clear indication of lags expected from the disc reflection continuum at even higher energies.  

\subsubsection{Confirming small scale reverberation}
\label{sec:smallscale}
\begin{figure}
\begin{center}
\includegraphics[width=10.5cm]{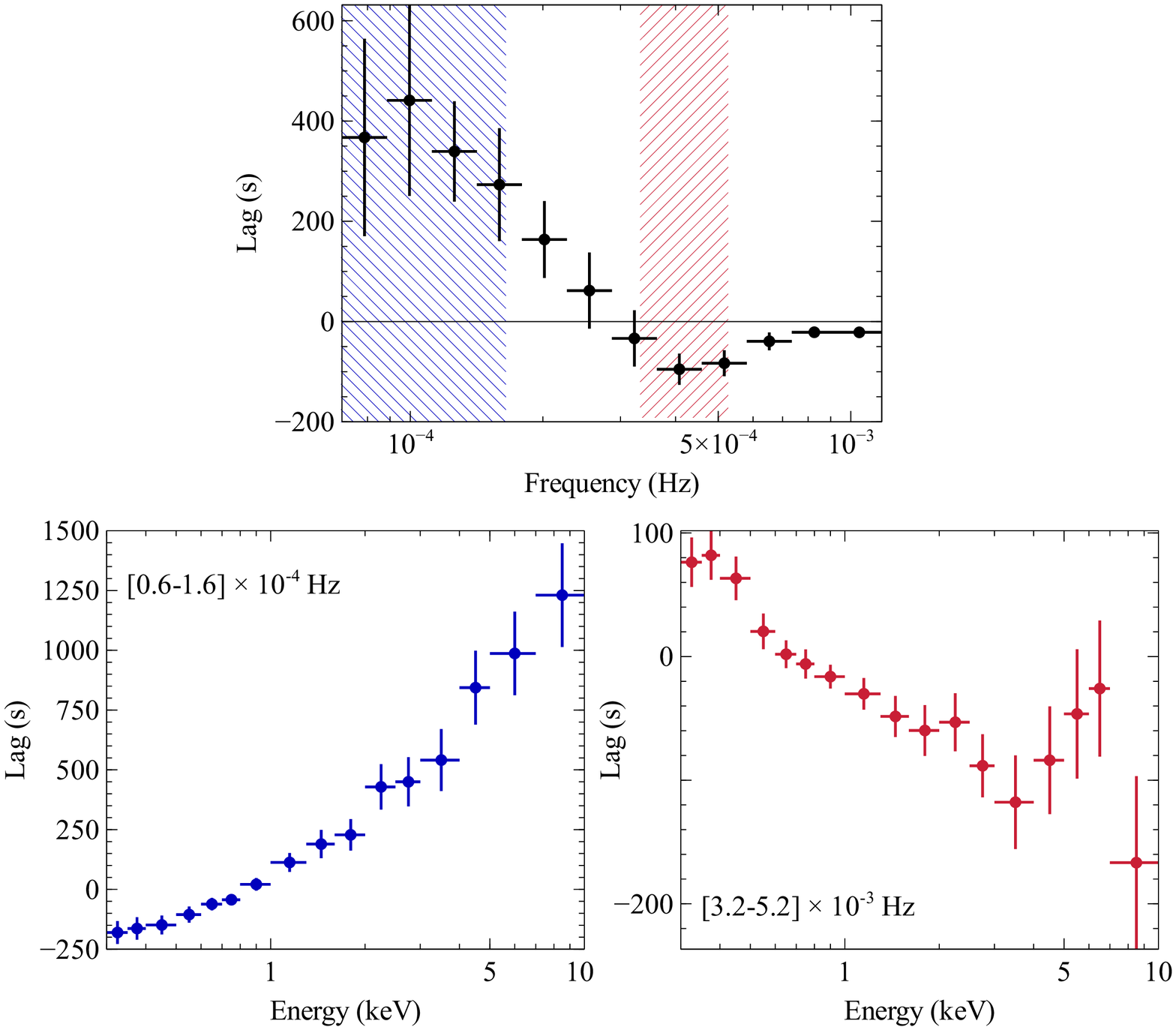}
\caption{The lag-energy spectrum of Ark~564 for low frequencies (left) and high frequencies (right), using frequency ranges highlighted in the top panel, which shows the lag-frequency dependence for 1.2--4~keV relative to 0.3--1~keV.  The iron K reflection feature is found at high frequencies, while the low frequencies show a featureless lag, increasing with energy. This shows that the low-frequency lags are not due to reflection. See Sect.~\ref{sec:bhb} for discussion of the origin of the low-frequency hard lag.}
\label{ark564}
\end{center}
\end{figure}
An alternative to the small-scale disc reverberation model for the soft lags was discussed by \citet{lancemiller4051} (and see also \citealt{legg12}), who suggested that the low-frequency hard lags are the real reverberation signatures, produced by reflection from more distant circumnuclear material (possibly associated with a disc wind or other large-scale absorbing/scattering gas), arguing that the negative, soft lags seen at higher frequencies were an artefact of the high-frequency oscillations expected from a top-hat like impulse response, which are caused by a `phase-wrapping'-like effect (see also Sect.~\ref{sec:freqdep}).   The physical interpretation of such an impulse response is that the lags can be understood as scattering/reflection from material covering a large solid-angle at a range of size scales, from a hundred to a couple of thousand light seconds from the central source, or by invoking a more distant reflector which must be aligned close to the line of sight, to account for the short time delays observed.  However, work by \citet{zog11_1h0707} and \citet{emma11} showed that the broad frequency range of the soft lag in many sources cannot be explained by the oscillatory effects expected for simple impulse reponses (see also Sect.~\ref{sec:freqdep}).  The unique line-of-sight argument is also inconsistent with the ubiquity of the soft lag in NLS1 sources \citep{demarco13}.   

\citet{lancemiller10} developed the large-scale scatterer/reflector model for the lags further in the case of 1H0707-495, proposing instead that the broad frequency range of soft lags (which cannot be explained by oscillations in the impulse response Fourier transform) could be explained if the soft band scattering impulse response is narrower with a smaller centroid than that seen in hard X-rays.  However, the interpretation of the low-frequency hard lags as a reverberation signature cannot explain the differences in the lag-energy spectra at low and high frequencies \citep{zog11_1h0707,kara13a}. As discussed above, the lag-energy spectrum at high frequencies has a clear signature of reflection in the iron K lag, however, as we probe lower frequencies, the lag shows a featureless increase with energy. The differences between the low and high frequency lag-energy spectra of Ark~564 can be seen in Fig.~\ref{ark564} \citep{kara13c}.  The low-frequency lag-energy spectrum, with no strong spectral features, is very similar to that of the hard lags found in black hole binaries, which can be understood as intrinsic lags associated with the continuum, possibly linked to the propagation of variations in the accretion flow which modulate the coronal emission \citep{kotov01,arevalo06,uttley11}.  Furthermore, low-frequency hard lags have recently been found in NGC 6814, a source that is well described by just an absorbed power law, with very little neutral reflection \citep{walton13}, suggesting that the hard lag is due to changes in the continuum, and not produced by large scale reflection as in the \citet{lancemiller10} model.  All this evidence strongly supports the picture that the high-frequency lags are caused by small scale reflection, while the hard lags are some separate process, unrelated to reflection.

\subsubsection{The frontiers of reverberation}
\label{sec:frontiers}
\begin{figure}
\begin{center}
\includegraphics[width=7.5cm]{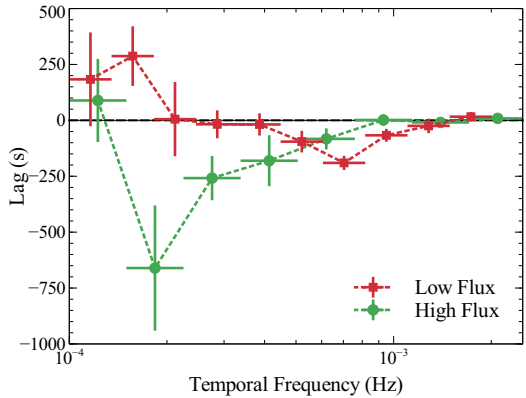}
\caption{The flux-dependent lags of IRAS~13224-3809.  The low-flux intervals (red) reveal shorter amplitude reverberation lags at a higher frequency, suggesting that the X-ray source is closer to the central potential and more compact at low fluxes.  The high-flux intervals (green) show a larger lag on longer time-scales, consistent with a more extended emitting region.  Figure taken from \citet{kara13b}.}
\label{iras}
\end{center}
\end{figure}

We are now delving into a regime where we can begin to disentangle the many contributions to the reverberation lag, i.e. we can isolate different light paths, and thus map the geometry of the source and inner flow.  In IRAS~13224-3809, we find that the reverberation lag is dependent on flux, which changes with the geometry of the corona \citep{kara13b}.  At low-flux intervals, we see a small amplitude lag at high frequencies (Fig.~\ref{iras}) and infer a compact corona producing the primary X-ray emission.  At high fluxes, the frequency of the negative lag range is lower and the amplitude of the lag is greater, suggesting an overall longer centroid lag of the impulse response (see Sect.~\ref{sec:freqdep}) and hence a longer light-travel time to the disc. We infer from this that as the flux increased, the corona expanded to fill a larger volume.  Interestingly, the low-flux interval shows a very clear iron K line, while lag-structure from the high flux intervals cannot be well constrained. Flux-dependent lags have also been found in NGC 4051 \citep{alston13}, which shows that there is no `hard lag' when the X-ray source is at low fluxes.

The comparison of the high-frequency lag-energy spectra between different objects (Fig.~\ref{lagen_stack}) shows that although iron K lags seem to be fairly common, the picture for the soft lags is more complex.  One possibility is that the soft lags are produced by a variety of mechanisms, some linked to photoionised reflection, but others linked to other components, e.g. the direct continuum itself.  It probably isn't surprising that in AGN with complex spectra that can be explained by strong reflection (e.g. IRAS~13224-3809, 1H0707-495), we see very similar soft lag behaviour \citep{kara13b}, suggesting that in these systems, reflection does dominate the soft lags.  It is worth noting that the variability in these AGN shows a high coherence between hard and soft energies at high frequencies (e.g. see Fig.~\ref{fig:1h0707psdcoh}, right panel) as would be expected if a single process such as reverberation dominates the spectral variability.  However, the AGN PG~1244+026 shows a low coherence between the soft and hard bands, suggesting that the soft excess in that source is not linked to strong photoionised reflection but may be associated with a separate, cool Comptonised component which is uncorrelated with the harder power-law and associated reflection \citep{jin13}.  This possibility is supported by the failure of propagation-plus-reflection models to explain the low-frequency soft lag behaviour in this AGN \citep{gardner14}.  Correspondingly, the lag-energy spectra of PG~1244+026 depend on the reference energy band chosen \citep{kara13d,alston14}, as expected when the variability is produced by distinct and uncorrelated spectral components, with their own lag-energy behaviours.

\begin{figure}
\begin{center}
\includegraphics[width=7cm]{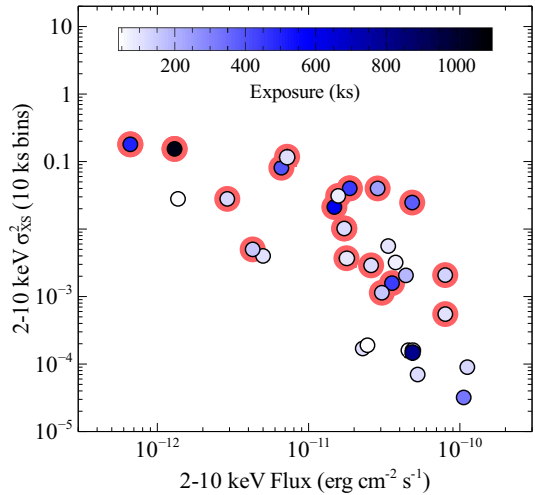}
\caption{The 2--10~keV flux and 2--10~keV excess variance (in 10~ks long segments) for the \citep{ponti12} sample of variable AGN (including also PG~1244+026, IRAS~13224-3809 and SWIFT~J2127.4+5654).  The sources are color coded based on their exposure times with {\em XMM-Newton}.  The sources with known Fe~K or soft lags have been highlighted. The detectability of the lag is based on the sources brightness and variability, and on how long it has been observed. }
\label{agnsample}
\end{center}
\end{figure}

The detectability of reverberation lags is based on the intrinsic phase lag of the reflection (which we assume to be similar from source to source, if lag and variability time-scales both scale linearly with black hole mass) and three other parameters: the flux of the source, the amount of variablility, and the amount of data we have available. In Table~\ref{lagtable}, we highlight the exposure, 2--10~keV flux and 2--10~keV excess variance (measured from 10~ks long light curve segments) for the nine sources with Fe~K lags measured to date.  We compare these sources with other variable AGN to illustrate the detectability of reverberation lags. In addition to PG~1244+026, IRAS~13224-3809 and SWIFT~J2127.4+5654, we compile a sample of variable AGN that are common between the \citet{gm12} sample (which provides the 2--10~keV luminosity and the XMM-Newton exposure as of the date of submission), and the \citet{ponti12} sample (which provides the 2--10~keV excess variance in 10~ks time bins). Fig.~\ref{agnsample} shows the flux, excess variance and exposure for this sample of variable AGN.  The sources with significant reverberation lags (Fe~K or soft lags) are highlighted.   Reverberation lags have been found in many high-flux, strongly variable AGN, and longer observations will uncover lags in a greater range of sources.

\begin{table}
\centering
\begin{tabular}{c|c|c|c}
\hline
{\bf Source} & {\bf Exposure} & {\bf 2--10~keV flux} & {\bf $\sigma^{2}_{\mathrm{XS}}$}\\
& ks & $10^{-12}$ erg cm$^{-2}$ s$^{-1}$ & \\
\hline
NGC~4151$^{1}$ & 99 & 80& 0.00055 \\
1H0707-495$^{2}$ & 1042 & 1.3 & 0.154 \\
IRAS~13224-3809$^{3}$ & 430 & 0.66 & 0.18 \\
MCG-5-23-16$^{4}$ & 130 & 80 & 0.00208 \\
NGC 7314$^{4}$ & 120 &15.7 & 0.031 \\
Ark 564$^{5}$ & 465 & 18.7 & 0.0401 \\
Mrk 335$^{5}$ & 120 & 17.2 & 0.0102 \\
PG 1244+026$^{6}$ & 123 & 2.9 & 0.028 \\
SWIFT J2127.4+5654$^{7}$ & 238 & 29 & 0.04 \\ 
\hline
\end{tabular}
\caption{Parameters for Fe~K Lag detected sources.  References corresponding to subscripts to the source names: 1.  \citet{zoghbi12}; 2. \citet{kara13a}; 3. \citet{kara13b} 4. \citet{zoghbi13}; 5. \citet{kara13c}; 6. \citet{kara13d}; 7. \citet{marinucci14}.}  
\label{lagtable}
\end{table}

Most of the discoveries of X-ray reverberation lags have taken place in supermassive black hole systems, due to the large number of photons per light crossing time (see Sect.~\ref{sec:analysis} for more discussion on this point).  However, Galactic black hole binaries have been influential in the study of the low-frequency hard X-ray time lags, and will be very important for reverberation studies with future telescopes (Sect.~\ref{sec:future}).  We now review the current reverberation observations in black hole X-ray binaries.

\subsection{Black hole X-ray binaries}\label{sec:bhb}

While in recent years significant work measuring reverberation has been undertaken with AGN, X-ray time lags in accreting compact objects were first studied in neutron star and black hole X-ray binaries (e.g. \citealt{vanderKlis1987,miyamoto89,vaughan94,nowak99}).  The lag-studies of the accreting stellar mass black holes revealed fairly ubiquitous hard lags in most cases.  The large amplitudes of the hard lags at low frequencies are difficult to explain with models invoking light-travel time delays, either due to Compton upscattering of photons in an extended corona \citep{nowak99}, or the reverberation delay of reflection from a disc \citep{cassatella12}.  The first detailed study of the lag versus energy dependence in black hole X-ray binaries was carried out on {\it RXTE} data by \citet{kotov01}, who found that although there is some evidence for structure around the energies of Fe~K emission and absorption in the data, this is not consistent with the signature expected from reflection at large radii which would be required to produce the observed lags by reverberation alone.  A likely alternative is that these low-frequency hard lags are associated with the propagation of fluctuations through the accretion flow and/or corona \citep{kotov01,arevalouttley06}, as may also be the case with the hard lags seen at lower frequencies in the AGN data.

\begin{figure}
\includegraphics[width=8cm,angle=-90]{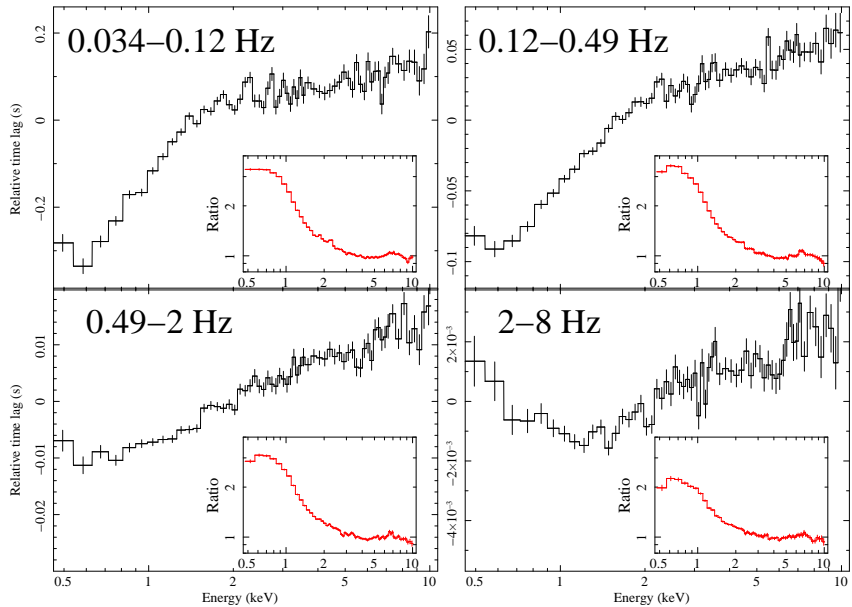}
\caption{{\it XMM-Newton} EPIC-pn lag-energy spectra for four temporal frequency ranges for the hard state of GX~339-4.  The insets show the covariance spectra for the same energy range and frequencies, plotted as a ratio to a Galactic-absorbed power-law with photon index fixed at $\Gamma=1.55$ fitted to the 3--10~keV range, to reveal the disc blackbody emission seen as a clear soft excess.  The disc photons clearly lead the power-law photons at low frequencies, but the behaviour switches at high frequencies, consistent with the switch from propagation to reverberation effects.  See \citet{uttley11} for further details.}
\label{fig:gx339lagvsen}
\end{figure}

Like the AGN, the first clear evidence for X-ray reverberation in black hole X-ray binaries came from the soft X-ray band, made accessible with the high-throughput of the {\it XMM-Newton} EPIC-pn instrument in its fast-readout, timing mode.  The development of the covariance spectrum allowed the study of the detailed energy dependent variability of X-ray binaries in the low-hard state, which revealed that the emission of the accretion disc in that state was significantly variable \citep{wilkinson09}.  This behaviour could be explained if the disc were intrinsically variable, e.g. via fluctuations of the internal viscous heating linked to changes in accretion rate, or alternatively, if the disc variability is driven by heating of the disc by the variable X-ray power-law continuum, i.e. there is {\it thermal reverberation}.  The critical test of these scenarios is to measure the time-lags between the disc and power-law emission, which was done by \citet{uttley11}.  The result, shown in Fig.~\ref{fig:gx339lagvsen} is very clear.  At low frequencies, below $\sim1$~Hz, the disc variations {\it lead} the power-law variations, by tenths of a second, consistent with fluctuations propagating through the disc to the inner corona on a viscous time-scale, requiring that the disc, if thin ($H/R<0.1$), extends to relatively small radii, within $\sim20$~$r_{\rm g}$ of the black hole.  However, at frequencies above 1~Hz, the soft photons start to {\it lag} the power-law photons (at least at intermediate energies), with lags of a couple of ms, consistent with the thermal reverberation picture, with light-travel times out to a few tens of gravitational radii at most.  

So far the best lag-measurements have been limited to a handful of sources in the hard state (since other states are too bright to study with {\it XMM-Newton} EPIC-pn in timing mode), while high-frequency soft lags have been confirmed only in one source (GX~339-4) with the longest observations \citep{uttley11}.  However, the behaviour seen so far is already remarkably similar to that seen in AGN.  At low frequencies we see long hard lags, with substantial evidence that intrinsic disc variability is driving the power-law variations, i.e. supporting the propagating fluctuations model for the low-frequency lags.  Meanwhile, at higher frequencies, the behaviour switches to short, soft lags consistent with reverberation, albeit from the disc blackbody emission and not photoionised reflection, which is to be expected since the high inner disc temperatures in BHXRBs mean that the ion species contributing to strong photoionised reflection signatures in AGN will likely be even more highly-ionised in their stellar-mass counterparts.  It is worth noting here that AGN also show evidence in their X-ray/optical correlations for intrinsic disc variability on long (months-years) time-scales and thermal reverberation on short (days-weeks) time-scales \citep{uttley03,arevalooptxray08}.  In fact, the variations in the optical continuum on day-to-week timescales in AGN, which show short (sub-day) red-lags (e.g. \citealt{cackett07}), can likely only be explained by thermal reverberation of the disc in response to the rapidly-varying central EUV/X-ray source.  It is interesting to note that these time-scales of optical variability are the driving time-scales for optical line reverberation from the broad-line region.  Thus, even optical reverberation signatures in AGN may in some sense be driven by X-ray reverberation effects!

At this point the exact mechanism for the low-frequency hard lags is unclear, but the BHXRB data firmly indicate that they are connected to intrinsic accretion variability, as posited by \citet{kotov01} and \citet{arevalouttley06}.  These accretion variations were originally thought to be variations in some sort of hot, coronal flow, which would naturally lead to hard lags within the power-law continuum, if the coronal temperature increases with decreasing radius.  However, it now seems likely, from both the BHXRB lag-energy spectra and the correlated long-term optical and X-ray variability in AGN, that the cool, optically thick accretion disc itself is responsible for carrying these fluctuations.  A number of possible mechanisms can connect these disc variations to the coronal power-law emission.  The corona may be distributed over the disc, for example, and respond directly to accretion variations in the underlying flow.  However, such a model might have difficulty in explaining the short lags and strong reflection that seem to suggest a centrally concentrated corona in many of the AGN.  An alternative is that the disc variations are communicated to the corona via fluctuations of the `seed' photons from the disc, which are upscattered to produce the power-law emission.  The lags are not introduced by light-travel time effects since light-travel times from disc to corona should not be much larger than the observed short reverberation lags. Instead, the lags in this scenario would be caused by the relative ($\sim$viscous-time-scale) delay between variations of seed photon luminosity (as fluctuations pass through the disc) and the intrinsic coronal heating (as the accretion fluctuations finally enter the corona) which could produce correlated changes in the observed power-law index, leading to hard lags.

Finally, we note that although clear Fe~K reverberation signatures are not yet detected in X-ray binaries, we do not expect them to be seen without better data.  This is because with current instruments, X-ray binaries exist in the regime of low photon signal per characteristic time-scale (see Sect.~\ref{sec:sensitivity}).  However, in this regime improvements with even brighter sources can be significant, provided that those sources are not so bright that they are unobservable with the best current detector for high-resolution lag studies, the EPIC-pn.  It is at least encouraging that variable broad Fe~K emission is seen in the covariance spectra in Fig.~\ref{fig:gx339lagvsen}, so that it is probably a matter of time and good fortune before Fe~K reverberation signatures are seen in black hole X-ray binaries.

\section{Modelling}\label{sec:modelling}

In this section we describe the frequency and energy dependence of both simple and more realistic {\it linear} impulse responses, which is important when comparing with the observed lag properties described in Sect.~\ref{sec:obs}.  We first discuss the frequency dependence expected from simple and more physically motivated impulse responses in Sects.~\ref{sec:freqdep} and \ref{sec:realfreqdep}, before going on to discuss energy dependence in Sect.~\ref{sec:endep}.

\subsection{Frequency dependence of lags: simple impulse responses}\label{sec:freqdep}
 
We described in Sect.~\ref{sec:analysis} the Fourier techniques to
estimate the time lag between two light curves.  The results of such
analyses are time lags as a function of Fourier frequency and energy
(if multiple energy bands are used).  The lag between the variability
in the two energy bands is a function of the frequency of the
underlying variability; i.e. the measured lag may be different for
variability occurring over the longest timescales (the low
frequencies) and for the most rapid variability (the high
frequencies). This may be because variability on these different
timescales and the communication of this variability from one energy
band to another is driven by different underlying physical processes,
or simply because the communication of the variation from one band to
the other occurs over a sufficiently long timescale that it is
blurring out the most rapid variability. In the case of reflection and reverberation, this communication is the passage of the radiation from the corona to the
reflector.  We can understand the frequency dependence of the lags in terms of the shape of the impulse response introduced in Sect.~\ref{sec:imprespintro} and its effects on the Fourier transforms in each band, because the lags are directly determined by the cross-spectrum of the impulse response in each band (Sect.~\ref{sec:spectimeffects}).  In the simple examples discussed below, we assume that the driving continuum band has a simple $\delta$-function impulse response at t=0.  We then examine the effects on the lag-frequency spectrum (and PSD) of different aspects of the impulse response in the `reflection' band, before summarising these effects and considering their implications for explaining the observations.  

\subsubsection{Simple delta-function: time-shift and phase-wrapping}
\begin{figure}
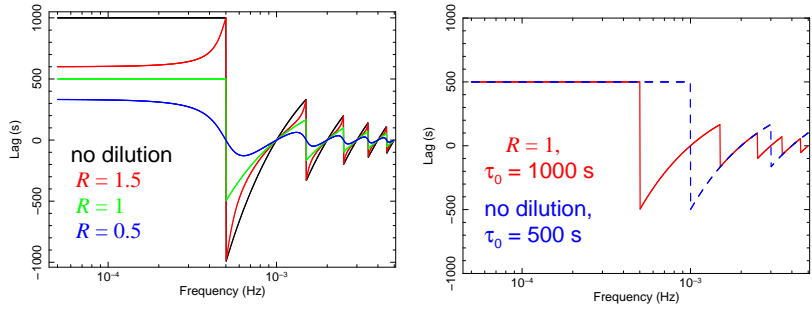

\centering
\includegraphics[angle=270,width=0.45\textwidth]{lagfreqdelfcomp.ps}
\includegraphics[angle=270,width=0.45\textwidth]{lagfreqdfdiltau0comp.ps}
\caption{{\it Left:} The lag-frequency spectra between two light curves related by a simple time-shift of 1000~s (i.e. delta-function impulse responses), for differing amounts of dilution of the lagging `reflection' light curve by a zero-lag direct component.  The relative amplitude of the reflection relative to the diluting direct continuum is $R$.  Phase-wrapping starts to occur at $\nu = \frac{1}{2000} = 5\times10^{-4}$~Hz. {\it Right:} The lag-frequency spectra for a diluted ($R=1$) time-shifted light curve with shift $\tau_0=1000$~s (solid line), compared to an undiluted time-shifted light curve with $\tau_0=500$~s (dashed line).  Although the amplitude of the low-frequency lag is the same, the phase-wrapping frequencies and resulting high-frequency lags are distinct.}
\label{fig:simplelagfreq}
\end{figure}

To understand the basic properties of lags between two light curves we first consider the lag-frequency spectrum between two light curves shifted in time by 1000~s, i.e. besides the driving continuum delta-function at t=0, the impulse response of the responding reflection band is also a delta-function, but at a delay $\tau_0 = 1000$s.  The resulting lag-frequency spectrum is shown as the black curve in Fig.~\ref{fig:simplelagfreq}, left panel. The lag-frequency spectrum shows a constant lag of $\tau$ across all frequencies (variability at all frequencies is delayed by the same time between the two light curves) until the lag time $\tau$ corresponds to a half-wave shift in phase (a phase shift by angle $\pi$) of the Fourier mode of frequency $\nu=\frac{1}{2\tau_0}$. At this point, the waveform could have been shifted either backwards or forwards by half a wave and since the phase of the wave is defined to be in the range $-\pi$ to $\pi$, the phase wraps around from $\pi$ to $-\pi$ and the measured lag becomes negative (so-called `phase-wrapping' which is a type of aliasing; the effect is similar to the
 `wagon-wheel effect' sometimes seen when wheels rotate in movies).  If we consider the behaviour in terms of the Fourier transform, a Dirac delta-function at t=0 has a Fourier transform of unity at all frequencies.  Adding a time-shift $\tau_0$ and applying the shift theorem of Fourier transforms, multiplies this constant value by $\exp(i\omega \tau_0)$, where $\omega=2\pi \nu$ is the angular frequency\footnote{Formally, a time shift multiplies the Fourier transform by $\exp(-i\omega \tau_0)$, but here and throughout we use the convention that a positive phase lag corresponds to a delay, so we correct to this convention by multiplying the imaginary part of the Fourier transform (and hence the resulting phase lags) by -1.}.  In the complex plane, as frequency increases the corresponding vector rotates around the origin at a constant rate with frequency, causing successive crossings of the $\pi/-\pi$ boundary (i.e. positive-negative lag oscillations) at every interval of $\nu = \frac{n}{2\tau_0}$, where $n$ is an odd integer.

\subsubsection{Dilution} 
{\it Dilution} is caused by the presence of emission from the driving continuum in the reflection band (which is expected in most physical scenarios for X-ray reverberation).  One of the main effects is to reduce the amplitude of the lags, but the shape of the lag-frequency spectrum can also be significantly changed.  In the time-shifted light curve case, the direct continuum component simply adds a zero-lag delta-function (representing the direct continuum) to the impulse response of the reflection band.  If we define the relative amplitude of the reflected flux to the direct flux in the reflection band to be $R$, such that $R=1$ represents equal contributions of reflected and direct flux, it is easy to show that the phase lag is given by:
\begin{equation}
\phi(\omega) = \arctan\left(\frac{R\sin(\omega \tau_0)}{1+R\cos(\omega \tau_0)}\right)
\end{equation}
The resulting time lags (i.e. $\phi/\omega$) for different values of $R$ are shown in Fig.~\ref{fig:simplelagfreq}.  Note that the observed lag at low frequencies is reduced, by a factor $R/(1+R)$, but that the zero-crossing point for all $R$ remains the same, so that this zero-point may be used as an indicator of the true lag of the reflected flux component (this effect is shown in Fig.~\ref{fig:simplelagfreq}, right panel).  Note also that the effect of $R>1$ is to make the lag-frequency dependence convex close to the crossing points.  This sharp peak gets narrower as $R$ increases and in the limit $R\rightarrow \infty$ we return to the simple delta-function case.  For $R<1$ the lag-frequency dependence becomes smoother, reducing the frequency where the lags begin to drop (although the zero-crossing frequency is maintained).  For $R<<1$ it is easy to show that the phase lag has a simple sinusoidal dependence on frequency.

\subsubsection{Impulse response shape and width}
\begin{figure}
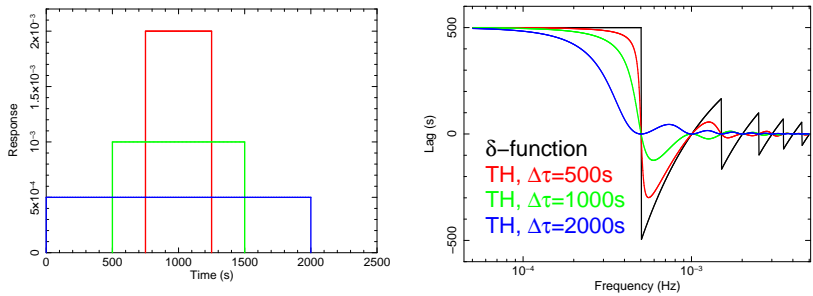

\centering
\includegraphics[angle=270,width=0.45\textwidth]{thfig.ps}
\includegraphics[angle=270,width=0.45\textwidth]{lagfreqthcomp.ps}
\caption{{\it Left:} Top-hat impulse responses.  All have centroid $\tau_0 = 1000$s, yet different widths of $\Delta\tau = 500$~s (red), 1000~s (green) and 2000~s (blue).  All are normalized to have an area of unity.  {\it Right:} The corresponding lag-frequency spectra, together with the lag-frequency spectrum for a delta-function impulse response for comparison.  Dilution is included in all cases, with direct continuum flux set to be equal to the lagging component flux ($R=1$).  Note how the maximum lag and the frequency where the lag first goes to zero are the same for all four impulse responses (both are set by $\tau_0$ and $R$, and only $\tau_0$, respectively).}
\label{fig:tophatwidth}
\end{figure}

In reality, the delayed component in any impulse response is unlikely to be a delta-function, since most physical models for lags produced by light-travel time predict multiple path-lengths.  The lag-frequency behaviour is closely related to the delta-function case however, because an extended and lagging impulse response can be described in the time-domain by a broad function centred at zero time, convolved with a lagging delta-function.  Thus the shift theorem applies and we arrive at a similar functional form for lag-frequency dependence as in the simple delta-function case.  As a simple example, we first consider the case of a top-hat impulse response (see Sect.~\ref{sec:revsphershell}), with centroid lag $\tau_0$ and width $\Delta \tau$.  The Fourier transform of a top-hat is a sinc function.  Assuming again that $R$ is the relative amplitude of reflected to direct emission (which is modelled with a delta-function at $t=0$), it is easy to show that the phase lag is given by:
\begin{equation}
\phi(\omega) = \arctan\left(\frac{R\sin(\omega \tau_0)\mathrm{sinc}(\omega \Delta \tau/2)}{1+R\cos(\omega \tau_0)\mathrm{sinc}(\omega \Delta \tau/2)}\right)
\label{eqn:thlag}
\end{equation}
The lag-frequency dependences for $R=1$ and the same $\tau_0=1000$~s but different values of $\Delta \tau$, together with the corresponding delta-function case, is shown in Fig.~\ref{fig:tophatwidth}.  The sinc function associated with the top-hat imposes an additional oscillatory structure on top of that already present due to the time-shift of the centroid.  Note that in the special case where $\tau_0 = \Delta \tau/2$ the multiplying sinc function is in phase with the oscillations due to the delta-function, causing the sign of the lag to always be positive.

\begin{figure}
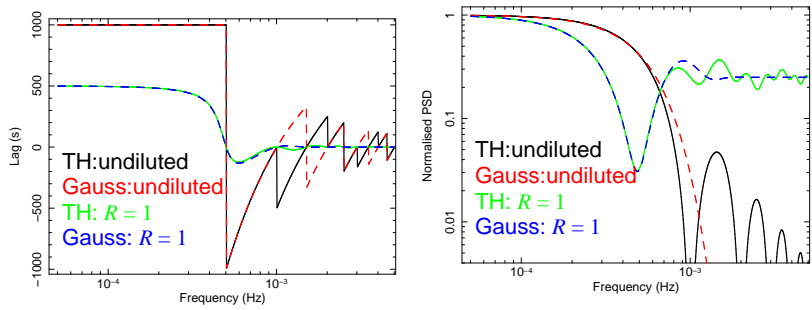

\centering
\includegraphics[angle=270,width=0.45\textwidth]{lagfreqshapecomp.ps}
\includegraphics[angle=270,width=0.45\textwidth]{psdshapecomp.ps}
\caption{{\it Left:} Lag-frequency spectra for different impulse response shapes and dilution.  The Gaussian impulse response case is shown using dashed lines. {\it Right:} The modulus-squared of the Fourier transforms of the top-hat and Gaussian response functions, which shows the `filter' which the driving signal PSD should be multiplied by to yield the output PSD from the reflection band.}
\label{fig:shapedependence}
\end{figure}

We also consider a smoother impulse response for the reflection, in the form of a time-shifted Gaussian with centroid $\tau_0=1000$~s and standard deviation $\sigma=300$~s.  Both types of impulse response are shown in Fig.~\ref{fig:shapedependence}, for both undiluted and diluted ($R=1$) cases. The Fourier transform of a Gaussian is also a Gaussian, which replaces the sinc function in Equation~\ref{eqn:thlag}.  Since the Gaussian is real and always positive, showing no oscillatory behaviour, the lag-frequency spectrum in the undiluted case is the same as for a simple, undiluted delta-function impulse response, in contrast to the top-hat case.  It is important to note that despite its smooth shape, the Gaussian impulse response shows the same sharp oscillations as seen for the other functions, simply as a result of the shifted centroid delay.  Thus a sharp-edged impulse response is not a requirement for sharp oscillatory lag-frequency behaviour.

The 300~s standard deviation of the Gaussian is chosen so that the lag-frequency behaviour in the diluted case closely mimics that of the diluted top-hat impulse response with width $\Delta\tau = 1000$~s.  The only significant difference between the lag-frequency spectra occurs in the high-frequency oscillations, which are heavily suppressed in the Gaussian case.  

\subsubsection{Effects on the PSD}
So far we have only considered the effects of the impulse response on the lag-frequency spectrum.  However, as noted in Sect.~\ref{sec:spectimeffects}, the impulse response also acts as a filter on the driving signal, multiplying its PSD by the modulus-squared of the Fourier transform of the impulse response and thus potentially affecting the shape of the observed PSD.  We show the shape of this filter in Fig.~\ref{fig:shapedependence} (right panel), for the different impulse responses considered in the previous subsection.  It is useful to note that a simple delta-function shows a filter that is unity at all frequencies (i.e. the observed PSD is the same as that of the driving signal, as we would expect).  At low frequencies, all the considered impulse responses also show a unity filter effect, which drops at higher frequencies (the impulse responses act as {\it low pass filters}, suppressing rapid variability).  However, more complex shapes are seen at higher frequencies where the undiluted top-hat and Gaussian cases simply correspond to the modulus squared of a sinc function and zero-centred Gaussian respectively.  The most interesting cases are the diluted top-hat and Gaussian impulse responses, which show a prominent dip in the filter around the frequency where the lag oscillations occur.  This arises as a `destructive interference' effect, since at this frequency the reflected lagging signal is $\pi$ radians out of phase with the input signal which is produced in the same band by the `diluting' delta-function, so the two signals cancel each other out.  At higher frequencies the filter flattens to an approximately constant value equal to $(1+R)^{-2}$, which is due to the direct continuum component (the reflected lagging component is almost completely filtered out).

\subsubsection{Summary and comparison with observations}
We first summarise the key results from our examination of simple impulse responses:
\begin{enumerate}
\item The centroid time ($\tau_0$) of the lagging impulse response component sets the frequency where the lag drops to reach zero ($\nu=1/2\tau_0$).  The centroid (together with dilution) also sets the value of the low-frequency constant time lag.
\item The dilution of the lagging (`reflected') component by the direct continuum (defined by $R$, which is the reflected flux expressed as a fraction of the direct continuum flux in the same band), causes a reduction in the low-frequency constant time-lag, by a factor $R/(1+R)$.  Dilution also affects the shape of the high-frequency drop to zero lag and subsequent oscillatory behaviour, but does not change the frequency of the drop to zero lag from the value set by $\tau_0$.
\item The drop and oscillatory behaviour seen in the lags at high frequencies is a consequence of the time-shift of the lagging component and not an artefact of sharp edges in the impulse response: the same effect is also seen for smooth impulse responses.  However, the shape of the impulse response does affect the more detailed shape of the lags and oscillatory behaviour close to and above the initial frequency set by $\tau_0$.
\item The impulse response shape has strong high-frequency filtering effects on the PSD and can also lead to interference `dips' in the power, e.g. at $\nu=1/2\tau_0$.
\end{enumerate}
Observationally it is important to distinguish variations in lag as a function of frequency which are associated with the time-shift and associated oscillatory effect of a single impulse response, and those which are due to a change in the lag mechanism (in fact, the latter could be modelled in terms of two different impulse responses with different shapes, centroids and widths).  A key point to note is that oscillatory effects make the lag change sign across a narrow, factor~2 range in frequency, which is significantly narrower than the negative lag ranges observed in many AGN.  Nor do we see any clear evidence for interference dips or oscillatory behaviour in the observed PSDs.  Thus the switch from hard to soft lags at high frequencies observed in 1H0707-495 and other AGN is better explained in terms of two different lag mechanisms (i.e. propagation to reverberation), and this inference is supported by the different lag-energy spectra observed at low and high frequencies (see Sects.~\ref{sec:smallscale} and \ref{sec:tophat_en}).

It seems most likely that the complex effects of the impulse response should be looked for at even higher frequencies, where a single process such as reverberation dominates.  E.g. many measurements of lag-frequency spectra show that the high-frequency lags are suppressed at even higher frequencies, potentially giving an indication of the centroid time $\tau_0$ of the reflector impulse response, which is consistent with the observed lag amplitudes and expected dilution.  The associated oscillations and expected features in the PSD might therefore be expected to occur at higher frequencies than can be currently probed.  Since it is the detailed high-frequency behaviour of the impulse response which contains information on its shape, it is unlikely that we can probe it in detail without better data, but we can easily measure the centroid time-shift and compare that with expectations from more physically realistic models.

Another important effect is dilution, which occurs when there is a contribution from both the direct and reflected components in the light curve in a given energy band.  In practice this is always the case, since although the `soft' band may be dominated by X-rays reflected from the disc, it always contains a significant contribution from X-rays seen directly from the corona. For instance, in the X-ray spectrum of 1H\,0707-495 measured by \citet{fabian09}, up to one third of the photons in the soft `reflection' band are direct continuum photons (i.e. $R=2$ in this case). As the measured lag is the average time lag between the correlated variability in the two energy bands, the effect is to `dilute' the measured lag time. This dilution is considered by \citet{kara13a} and \citet{wilkins13} who show that the measured time lag can be reduced by up to 75 per cent once the contributions of the diluting component are considered in both the reflection- and continuum-dominated energy bands.

\subsection{Frequency dependence of lags: realistic impulse responses}
\label{sec:realfreqdep}
Next we consider the effects associated with more realistic impulse responses for reflection, generated using relativistic raytracing, before examining the impact of realistic extended coronal geometries and the first steps towards energy-dependent modelling of the lags.

\subsubsection{The impulse response for an X-ray source above an accretion disc}
\label{sec:enavgtf}
We now introduce the impulse response for an X-ray source above an accretion disc.  As we are discussing an accretion disc around a black hole an appropriate impulse response must take into account the relativistic effects (including Shapiro delay) that will be present around a black hole.  The time-averaged response of the disc is well explored, as we discussed in the Introduction.  The time-resolved response has also been well studied  \citep{campana95,reynolds99,gilfanov00, kotov01,poutanen02,nayakshin02,cassatella12,chainakun12,wilkins13,cackett13_ngc4151,emma14}.

The time-resolved response represents the flux of reflected photons received from the disc as a function of time after an initial, instantaneous flash of radiation from the primary source, and was first calculated by \citet{campana95}.  These authors considered both the cases of the point source lying in the plane of the disc and a uniform spherical source extending up to the inner edge of the accretion disc, accounting for relativistic effects on the energy and flux of the reflection as a function of position in the disc, but adopting a classical approximation for the light travel time. They compute the flux seen in relativistically broadened emission lines from the disc as well as the width and centroid energy of the line as a function of time after the initial flash.

A full treatment of the illumination of an accretion disc around a black hole in general relativity was first completed by \citet{reynolds99} and \citet{young00} who model, in detail, the time and energy evolution of an emitted fluorescence line from the accretion disc due to an instantaneous flash of X-rays from the corona, in order to identify observable signatures from the reverberation of a bright flare of emission from the corona.  Recently, in order to interpret X-ray reverberation measurements in terms of X-rays emitted from the corona reflecting off the accretion disc, \citet{wilkins13} conducted general relativistic ray tracing simulations of X-ray reverberation scenarios, varying the height, vertical and radial extent of the irradiating X-ray source.

\begin{figure}
\centering
\includegraphics[width=0.9\textwidth]{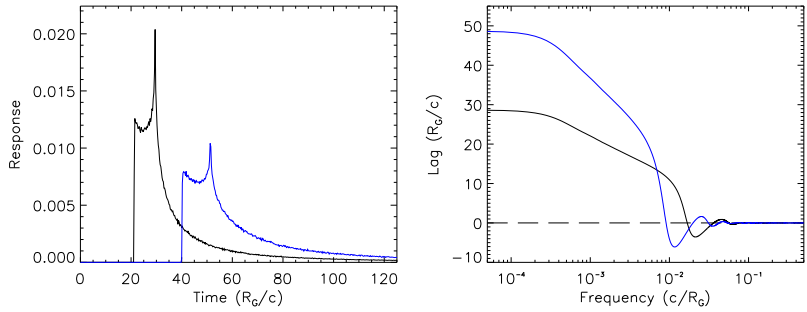}
\caption{{\it Left:} The energy-averaged impulse response for an X-ray source at heights 10~$r_{\rm g}$ (black) and 20~$r_{\rm g}$ (blue) above a maximally spinning black hole, with the accretion disc inclined at $i = 30^\circ$ {\it Right:} The corresponding lag-frequency spectra.  Note that the sign convention means that the reverberation lags here are positive, which would be the case in the measured lag-frequency spectra if the reverberation lags are hard (e.g. Fe K lagging a medium -energy continuum band).  If the reflected component is observed in soft X-rays, the model lag-frequency spectrum would be inverted, according to the convention that positive lags indicate hard variations lagging soft variations.  In other words, the lags at low frequencies in this figure would correspond to the observed (negative) soft lags.  The drop in lag and oscillations at high frequencies thus correspond to the high-frequency suppression of the soft lags, not the switch from hard to soft lags at lower frequencies, which we attribute to a switch between a separate mechanism (probably propagation-driven continuum lags) and reverberation.}
\label{fig:lagfreq_disc}
\end{figure}

We show the response from the accretion disc and corresponding lag-frequency spectrum for a point-source at a height of $10~r_{\rm g}$ above a  maximally-spinning black hole in Fig.~\ref{fig:lagfreq_disc}.  While there are clearly several peaks in the response, it is useful to note that, to first-order, it is approximately a top-hat function.  The corresponding lag-frequency spectrum therefore looks similar to the lag-frequency spectrum of a top-hat function as we described above. 

In more detail, the response of the reflection from the accretion disc to an instantaneous flash of radiation from a point source shows an initial steep rise some time after the initial flash owing to the light travel time to the disc, then gradually decaying to long times as parts of the disc further from the source receive the incoming radiation at later times.

As one might expect, relativistic effects are important when considering reverberation in the strong gravitational field around a black hole \citep{shapiro64}. The first such effect relevant to timing X-rays reverberating from the accretion disc is the delayed passage of photons through the strong gravitational field. The curvature of the spacetime means that while a local observer will measure the light travelling past them at speed $c$ at any location in the field, an outside observer, further from the black hole, will perceive the light travelling through the stronger field as having been slowed down and hence its arrival will be delayed. This is due to a combination of the space being curved, increasing the proper distance the light must travel and the outside observer's time elapsing more quickly than the time measured by an observer in the stronger gravitational field.

\citet{wilkins13} find that the delay of photons travelling through the strong gravitational field close to the black hole is significant in measurements of X-ray reverberation, particularly when the X-ray source is located at low heights. Once the delays to both the reflection and directly observed continuum have been accounted for, this Shapiro delay can cause the X-ray source to appear 0.5$r_{\rm g}$ further from the black hole than it really is if it is incorrectly assumed that the photons are travelling in a flat, Euclidean spacetime, neglecting the effects of strong gravity.

\begin{figure}
\centering
\includegraphics[width=1.0\textwidth]{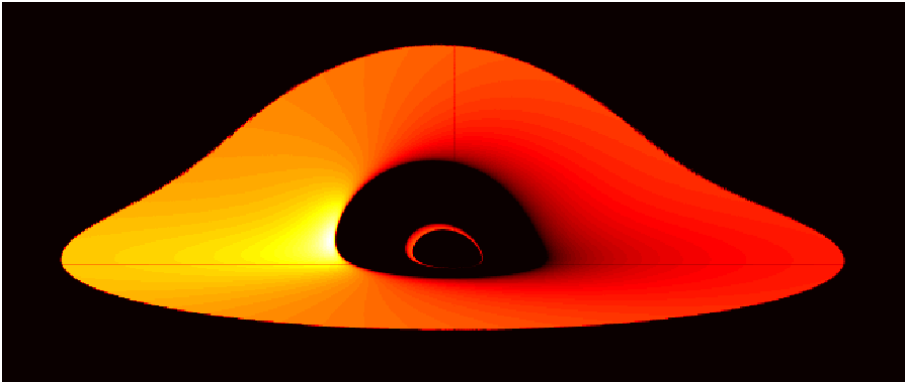}
\caption{Ray-traced image of the accretion disc around the black hole. To an observer on one side of the black hole, the far side of the accretion disc appears to be warped, appearing above the black hole as emitted radiation is bent around the black hole. Shading corresponds to the flux seen from each part of the disc, with a darker colour indicating that both the photon energy and arrival rate is redshifted.}
\label{fig:bhdisc}
\end{figure}

The second significant relativistic effect that influences measurements of X-ray reverberation from the accretion disc is the bending of light in the strong gravitational field around the black hole. There is a secondary sharp peak in the response from the disc. This is the re-emergence of photons reflected from the far side of the accretion disc, behind the black hole that, although classically would be blocked from our view, are lensed by the strong gravitational field around the black hole into our line of sight. Not only are these photons slowed by their passage through the strong gravitational field close to the black hole, but this part of the disc is magnified through gravitational lensing, meaning that the emission in this secondary peak is enhanced, as discussed by \citet{reynolds99}. The source of these re-emerging photons becomes apparent in a ray-traced image of the accretion disc around a black hole. To an observer on one side of the black hole, the far side of the accretion disc appears to be warped appearing above the black hole (Fig.~\ref{fig:bhdisc}). It is these photons that are delayed and redshifted, giving rise to the secondary peak in the impulse response.

Finally, each ray that travels from the disc to the observer will be influenced by gravitational redshifts as well as Doppler shifts due to the orbital motion of the material in the disc. Considering each ray to be a stream of photons emitted at regular intervals, not only are the energies of the photons arriving along a ray shifted, but so too is their arrival rate. Reflection from the approaching side of the disc is enhanced as the emission is beamed towards the observer and reflection from the receding side of the disc is reduced, though since reflection from both sides of the disc is recorded together when computing lag-frequency spectra, this effect will cancel to first order.

There is a reduction in the response seen from the innermost parts of the accretion disc as this emission is gravitationally redshifted to lower energies and photon arrival rates and since many rays emitted from the disc close to the black hole are bent towards the black hole and lost inside the event horizon. This effect, however, is offset by the enhanced illumination of the inner parts of the accretion disc by the coronal X-ray source. Rays are focussed towards the black hole and, hence, on to the innermost parts of the disc. Both of these effects play a role in the observed impulse response of the accretion disc and are computed in detailed ray-tracing simulations, however their apparent effects on the impulse response are much less visible than the Shapiro delay and re-emergence of emission from behind the black hole. The exact shape of the lag-frequency spectrum is described in detail in \citet{cackett13_ngc4151}.

\subsubsection{Beyond point-source models: extended coronae}
While it is instructive to consider the case of a variable X-ray point source above the plane of the accretion disc as a simple model to build intuition for X-ray reverberation around black holes, \citet{wilkins12} find that the illumination pattern of the accretion disc by the coronal X-ray source (the so-called emissivity profile) inferred from the profile of the relativistically broadened iron K$\alpha$ fluorescence line suggests that the corona extends tens of gravitational radii over the inner part of the accretion disc. The corona is inferred to extend to around 35$r_{\rm g}$ over the disc in the narrow line Seyfert 1 galaxy 1H\,0707-495 \citep{wilkins11,wilkins12} and to around 10$r_{\rm g}$ in IRAS\,13224-3809 \citep{fabian13}. To this end, \citet{wilkins13} also model the lag-frequency spectra that would arise from the reverberation of X-rays originating from an extended emitting region using Monte Carlo simulations of rays arising throughout an optically thin extended region, varying both its radial extent over the plane of the accretion disc and its vertical extent perpendicular to the disc plane.

The arrival of the continuum radiation from an extended corona requires careful consideration as there are now multiple ray paths from the continuum source to the observer, so variations in the luminosity of the coronal emission that are correlated with variations in the reflection do not arrive instantaneously at the observer. Furthermore, where there is an extended emission region, it is unphysical for the whole region to vary in luminosity instantaneously, rather the change in luminosity throughout the extent of the corona must propagate causally at or below the speed of light. It is possible to construct an impulse response for the corona describing  the arrival of photons at the telescope after an instantaneous flash either simultaneously across the corona or propagating causally through the corona for an instantaneous injection of energy into the corona. As for the impulse responses corresponding to the reverberation from the accretion disc, this function can be convolved with a light curve representing the underlying variability in the injected energy to produce the light curve that would be observed in the continuum emission. The lag-frequency spectrum is again computed between the light curves for the continuum and reflected X-rays.

\begin{figure}
\centering
\includegraphics[width=0.45\textwidth]{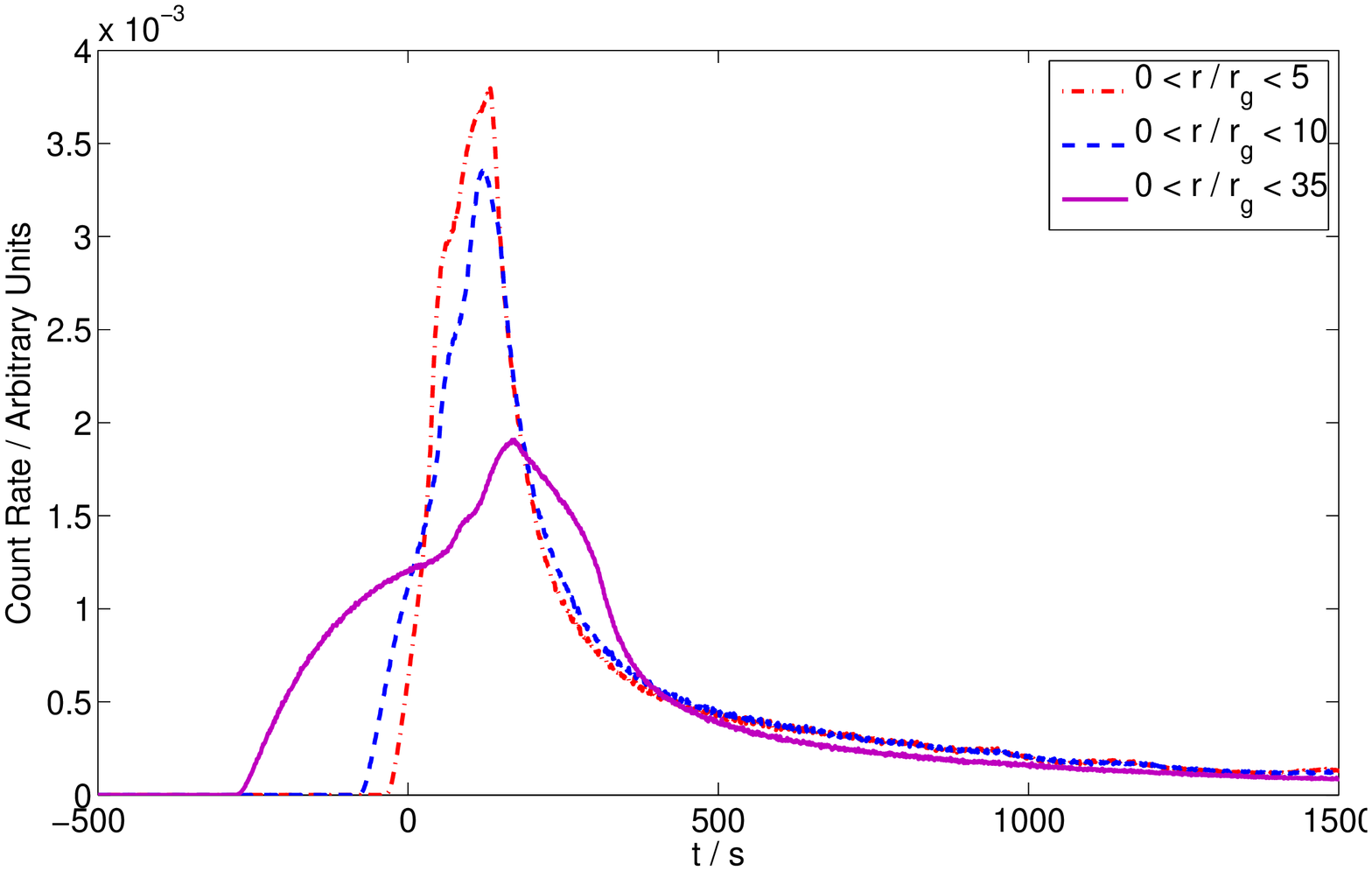}
\includegraphics[width=0.45\textwidth]{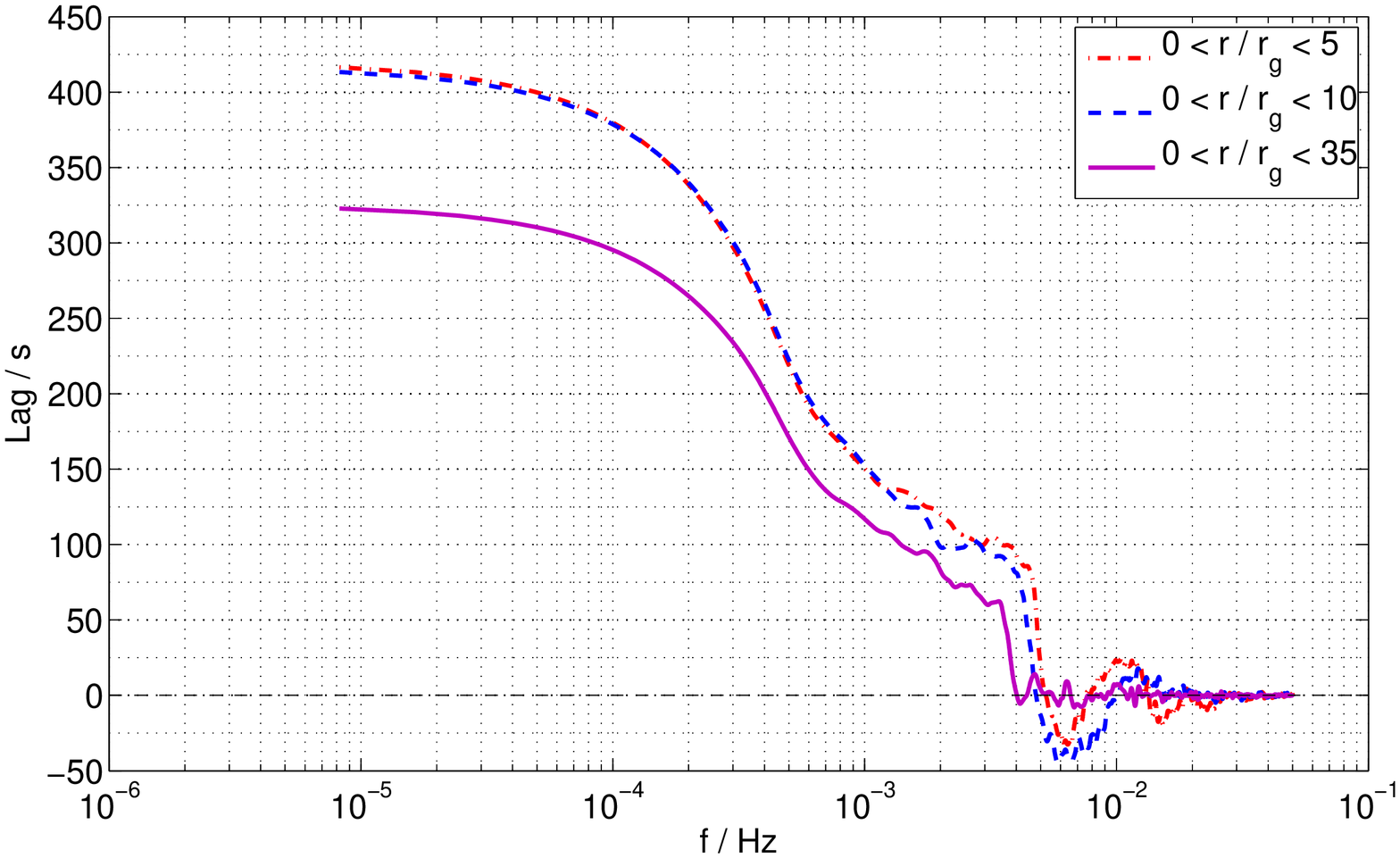}
\caption{{\it Left:} The energy-averaged impulse responses for irradiation of the accretion disc by a radially-extended corona above the plane of the accretion disc.  A $2\times 10^{6}$~M$_{\odot}$ black hole is assumed, so that $r_{\rm g}/c=10$~s.  The impulse responses are given for several different radial extents, and in each case the corona is only weakly vertically extended, between 2 and 2.5$r_{\rm g}$ above the disc.  {\it Right:} The corresponding lag-frequency spectra, showing the relative insensitivity to the radial extent of the X-ray source. However, notice a slight decrease in the reverberation lag time as the radial extent increases as more of the corona is further from the black hole, meaning fewer of the reflected photons have been delayed by the strong gravitational field in the centre.  Note that the same caveat about sign convention applies as in Fig.~\ref{fig:lagfreq_disc}.  Figures taken from \citet{wilkins13}.}
\label{fig:radialextlag}
\end{figure}

Computing the lag-frequency spectra that would be expected for a variety of coronae, \citet{wilkins13} find that the measured lag-frequency spectrum is sensitive to the vertical height and extent of the X-ray emitting region above the plane of the accretion disc. Increasing the extent of the corona above the accretion disc, increases the average light travel time of the rays between the source and the disc, increasing the lag that is measured across all frequencies. On the other hand, the broadband lag-frequency spectrum between a reflection-dominated and continuum-dominated band is not particularly sensitive to the radial extent of the X-ray source, as shown in Fig.~\ref{fig:radialextlag}. Increasing the radial extent of the X-ray source over the disc does not increase the distance from any given part of the source to the nearest position on the disc. There is, however, a slight decrease in the reverberation lag time as the radial extent increases as more of the corona is further from the black hole meaning fewer of the reflected photons have been delayed by the strong gravitational field in the centre.

The result of \citet{kara13b} that the measured reverberation lag increased when the luminosity of IRAS\,13224-3809 increased can now be interpreted in the context of the \textit{vertical} extent of the X-ray emitting corona increasing above the plane of the disc as its luminosity increases. This is not to say that the radial extent of the X-ray source is not also increasing, merely that the dominant effect is due to the increasing vertical extent of the source. It should be noted that like-for-like, the extra lag time due to increasing the vertical extent of the source is, in itself, a greater effect than that of increasing the radial extent of the source since the vertical extent of the source acts to extend the light path to the disc, while increasing the radial extent of the source decreases the effect of the Shapiro delay close to the black hole.

\subsubsection{Towards energy-dependent models}\label{sec:towardsendep}
The lag-frequency spectrum is not just useful as a probe of the corona; measured reverberation lags are sensitive to a number of parameters of the black hole and accretion disc.  \citet{cackett13_ngc4151} model general relativistic impulse responses varying a variety of parameters including the height of the illuminating X-ray point source, the mass and spin of the black hole, the inclination angle at which the accretion disc is seen to the line-of-sight and the relative flux seen in the directly observed continuum compared to that in the reflection. The effects of each of these parameters is discussed in detail in \citet{cackett13_ngc4151} and also applied to fitting the Fe~K lag-frequency spectrum for NGC 4151 (see also Sect.~\ref{sec:ngc4151comparison}).  We discuss these properties in more detail in the next section when considering energy-dependent lags from an accretion disc.

A further study fitting general relativistic impulse responses to frequency-dependent lags was performed by \citet{emma14}, who fit the lags in 12 AGN using the commonly observed soft high-frequency lags signal to constrain the reflector geometry and physical parameters.  They find black hole masses consistent with masses derived independently, as well as obtaining other physical parameters such as height, inclination and spin.  However, the robustness of the spin measurements is unclear given the disagreement with spectral fitting results and the implied unusual coronal geometries (e.g. 1H~0707$-$495, where \citealt{emma14} deduce a coronal height smaller than the disc inner radius implied by their low spin estimate).  Furthermore the effects of several assumptions on spectral shape and reflection fraction need to be further explored for their effects on the measurements.  Nonetheless, it is interesting that the observed lag-frequency spectra can be shown to be consistent with expectations from inner disc reflection and yield black hole masses consistent with independent estimates, even with simplifying model assumptions. 

A key consideration in modelling the frequency-dependent lags is whether the lags at high frequencies are only produced by disc reflection, or whether other components also contribute.  The variety of soft lag-energy shapes (see Fig.~\ref{lagen_stack}) already hints at the latter possibility, suggesting that the Fe~K region is the cleanest part of the spectrum for modelling of frequency-dependent lags.  However, with further consideration of the energy-dependent behaviour and self-consistent modelling of frequency {\it and} energy dependence, it is clear that this can be a powerful technique to uncover physical parameters of the accretion geometry and kinematics.  We now consider the modelling of the energy dependence of the lags in more detail.

\subsection{Energy dependence of lags}\label{sec:endep}

\begin{figure}
\centering
\includegraphics[angle=270,width=0.8\textwidth]{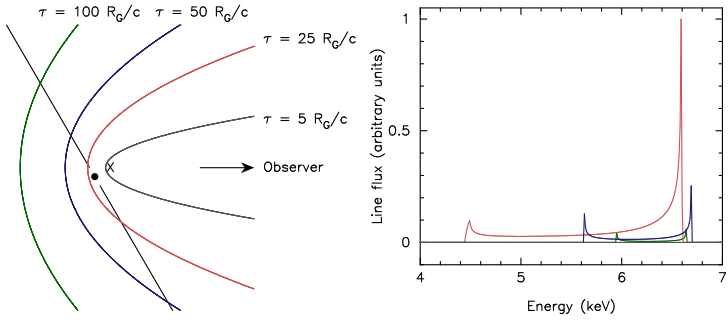}
\caption{{\it Left:} Isodelay surfaces for a delta-function flare from an X-ray source (X) at height 10~$r_{\rm g}$ above a black hole (filled black circle).  The accretion disc is inclined at 30$^\circ$ to the observer.  {\it Right:} Fe~K$\alpha$ line profiles at times corresponding to the isodelay surfaces shown in the left panel.}
\label{fig:isodelay}
\end{figure}

The lag-frequency spectrum helps to determine the average time lag between the direct and reprocessed emission, and hence gives basic information about the geometry and the reprocessing region.  Additional information can be gained if one also considers the energies at which the reflected emission occurs as velocities in the reprocessing gas will be imprinted on the emission lines formed there.  In the classical picture, an emission line from an accretion disc will form a doubled-horned profile with the parts of the disc coming towards the observer being blueshifted and the parts going away from the observer being redshifted.  In the case of an accretion disc around a black hole, relativistic Doppler shifts and gravitational redshifts need to be included which further broaden and skew the line profile leading to a characteristic asymmetric line profile \citep{fabian89}.  Hence, the instantaneous emission line profile at a time $\tau$ after a delta-function flare will be set by the kinematics of the region where the isodelay surface intersects the reprocessing gas.  We demonstrate this in Fig.~\ref{fig:isodelay} where we show the instantaneous Fe K$\alpha$ emission line profile from an accretion disc at several times after a delta-function flare from an X-ray source above the black hole.  As the line profile changes over time the lags will show an energy-dependence.  Before discussing energy-dependent accretion disc impulse responses in more detail we first start by building up some intuition based on simple top-hat impulse responses.

\subsubsection{Simple energy-dependent top-hat impulse responses}\label{sec:tophat_en}

Previously we have discussed how a top-hat impulse response leads to frequency-dependent lags.  Let us now consider a top-hat impulse response that has some dependence on energy.  The energy-dependence can come in several forms.  Firstly one can simply change the relative fractions of direct and reprocessed emission at each energy.  While the lag-frequency spectrum at each energy will have the same overall shape, the dilution effects will mean that the overall lag normalization will change between energies.

\begin{figure}
\centering
\includegraphics[width=0.95\textwidth]{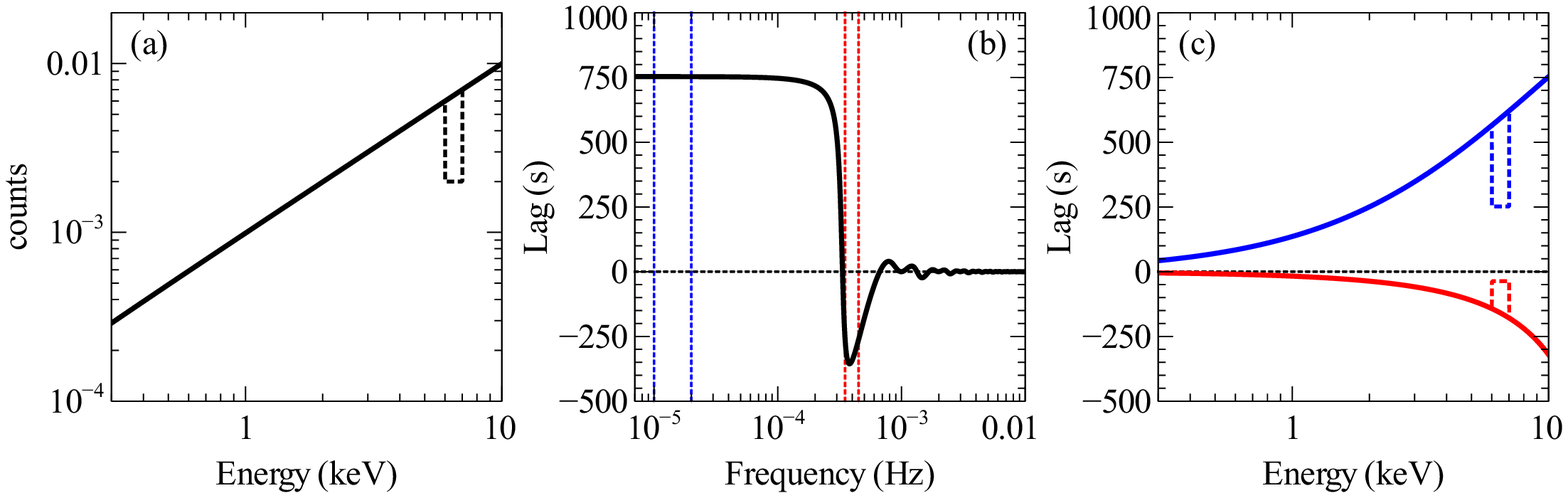}
\caption{The reflection spectrum (a), the lag-frequency spectrum (b) and the low and high frequency lag-energy spectrum (c) shown in blue and red, respectively for a model given by a simple top-hat response function for two test cases: (1) where the reflection fraction increases steadily with energy (solid line), and (2) where the reflection fraction increases steadily with energy except for a demonstrative dip at 6--7~keV (dotted line).  Taken from \citet{kara13c}.}
\label{fig:kara13c_tophat}
\end{figure}

The energy-dependence of such a model was considered by \citet{kara13c}, for comparing the data with models presented by \citet{lancemiller10} and \citet{legg12} which explain the low-frequency hard lags in terms of extended reflection, while the high-frequency soft lags are an artefact of the Fourier analysis.  \citet{kara13c} considered a top-hat impulse response with a width of $\Delta \tau$, and centered at $t_0=1500$~s with $\Delta \tau=1000$~s and 50\% dilution.  They then allowed the reflection component to become steadily more dominant with increasing energy.  The resulting reflection spectrum, lag-frequency spectrum and lag-energy spectrum (at two different frequencies) are shown in Fig.~\ref{fig:kara13c_tophat}.  The low-frequency lag-energy spectrum shows a steady increase in lag with energy because of the decreasing dilution with energy.  A high-frequency lag-energy spectrum taken over the frequency range where the phase wraps and the lags become negative shows the opposite behavior.  It is essentially a mirror image of the low-frequency lag-energy spectrum, though with a lower normalization.  The mirror image behavior when comparing the low-frequency lags with the higher-frequency negative lags is even more obvious when one artificially adds a discrete feature into the reflection spectrum (see the dotted line in Fig.~\ref{fig:kara13c_tophat}).  For instance, by adding a dip in the reflected emission between 6 -- 7 keV.  The lag-energy spectrum at low-frequencies then shows a corresponding dip in the lags between 6 -- 7 keV, whereas the negative high-frequency lags shows the opposite -- an increase in the lag at the same energies.  This is used by \citet{kara13c} as evidence that the low and high frequency lags that have been detected in AGN have different physical origins.  The fact that the low-frequency lags show a monotonic increase with energy while the high-frequency lags show a discrete feature around Fe K cannot readily be explained by a single impulse response, like that considered here.

\begin{figure}
\centering
\includegraphics[width=0.8\textwidth]{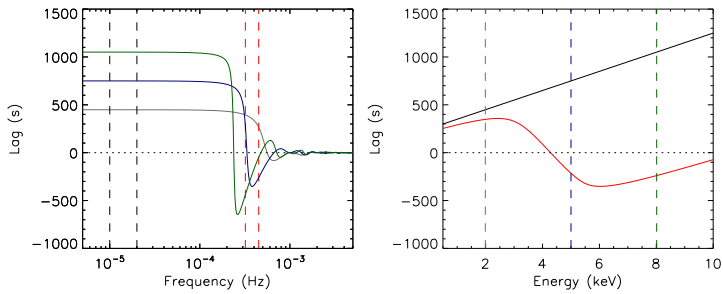}
\caption{Model lag-frequency and lag-energy spectra for a top-hat impulse response with a constant width but with a centroid $\tau_0$ that increases with energy.  {\it Left:} The lag-frequency spectra at 2 keV (gray), 5 keV (blue) and 8 keV (green).  The vertical lines indicate the frequency ranges over width the lag-energy spectra are determined.  {\it Right:} The lag-energy spectra in the two frequency ranges indicated in the left panel.  The vertical dashed lines mark the energies corresponding to the lag-frequency spectra on the left.}
\label{fig:tophat_endep}
\end{figure}

Another simple energy-dependent top-hat impulse response is one where the time corresponding to the center of the top-hat, $\tau_0$, changes with energy, yet the relative contribution from the direct and reprocessed emission is the same at all energies.  We will consider here a top-hat where $\tau_0$ increases linearly with energy, for instance, which might be expected if higher energy photons penetrate further into the reprocessing region.  Now the lag-frequency spectrum at each energy has a different frequency dependence (see Fig.~\ref{fig:tophat_endep}), and thus the frequencies at which phase-wrapping starts and the frequency where the lags become negative will be different for each energy.  The low-frequency lags will show a steady increase in lag with energy.  However, in a given high-frequency band each energy will be at a different stage of phase-wrapping leading to a lag-energy spectrum which switches to negative values and then increases back toward zero.  

Superficially, this behaviour is somewhat reminiscent of what is seen in the observed lag-energy spectra, with lag increasing monotonically with energy at low frequencies and a more complex shape at high frequencies, with both hard and soft X-rays lagging medium-energy X-rays.  However, the energy-dependent frequency switch to negative lags is not seen in the data \citep{kara13c}, and moreover, the enhanced suppression of high-frequencies seen at higher X-ray energies would lead to steeper power spectra at higher energies, not flatter power spectra, as are observed (e.g. Fig.~\ref{fig:1h0707psdcoh}).  Thus, the two-component propagation and reverberation scenario for the broadband frequency-dependence of the lags is a much better match to the data, than a single energy-dependent impulse response component.  However, the intuition developed from the energy-dependent top hat impulse response can be applied to understanding the lag-energy spectra within the frequency range dominated by one of these components.

When we go on to consider the more complicated impulse response for an X-ray point source above an accretion disc we will encounter both types of effect discussed here.  Dilution changes as a function of energy (the relativistic Fe K line has an asymmetric shape), and furthermore both the width and centroid of the impulse response changes as a function of energy, therefore phase-wrapping occurs at different frequencies in different energy ranges.  As we will demonstrate, this leads to clear features in the lag-energy spectrum that change with frequency.

\subsubsection{The energy-dependent impulse response for an X-ray source above an accretion disc} \label{sec:model:endeptf}

\begin{figure}
\centering
\includegraphics[width=0.475\textwidth]{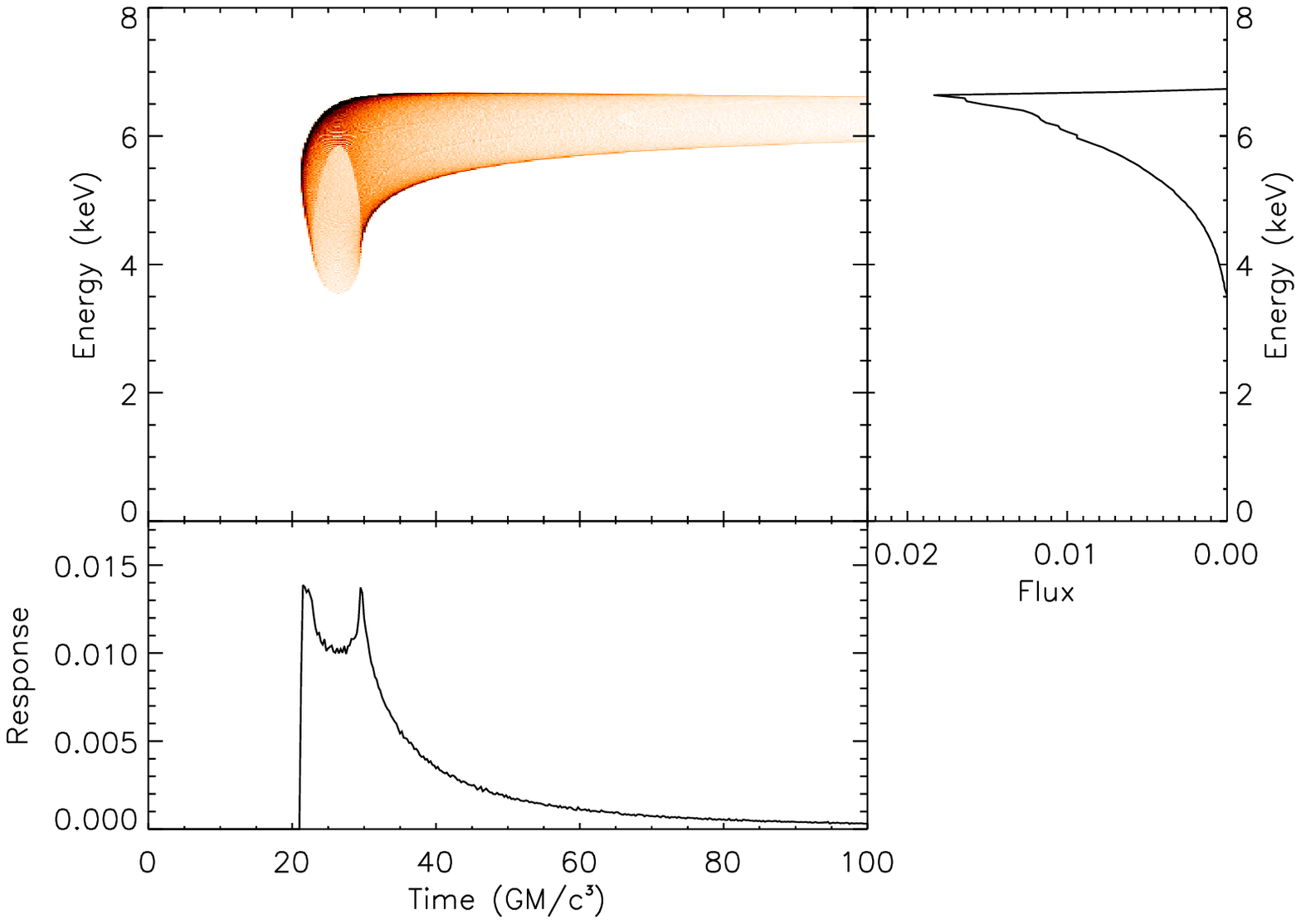}
\includegraphics[width=0.475\textwidth]{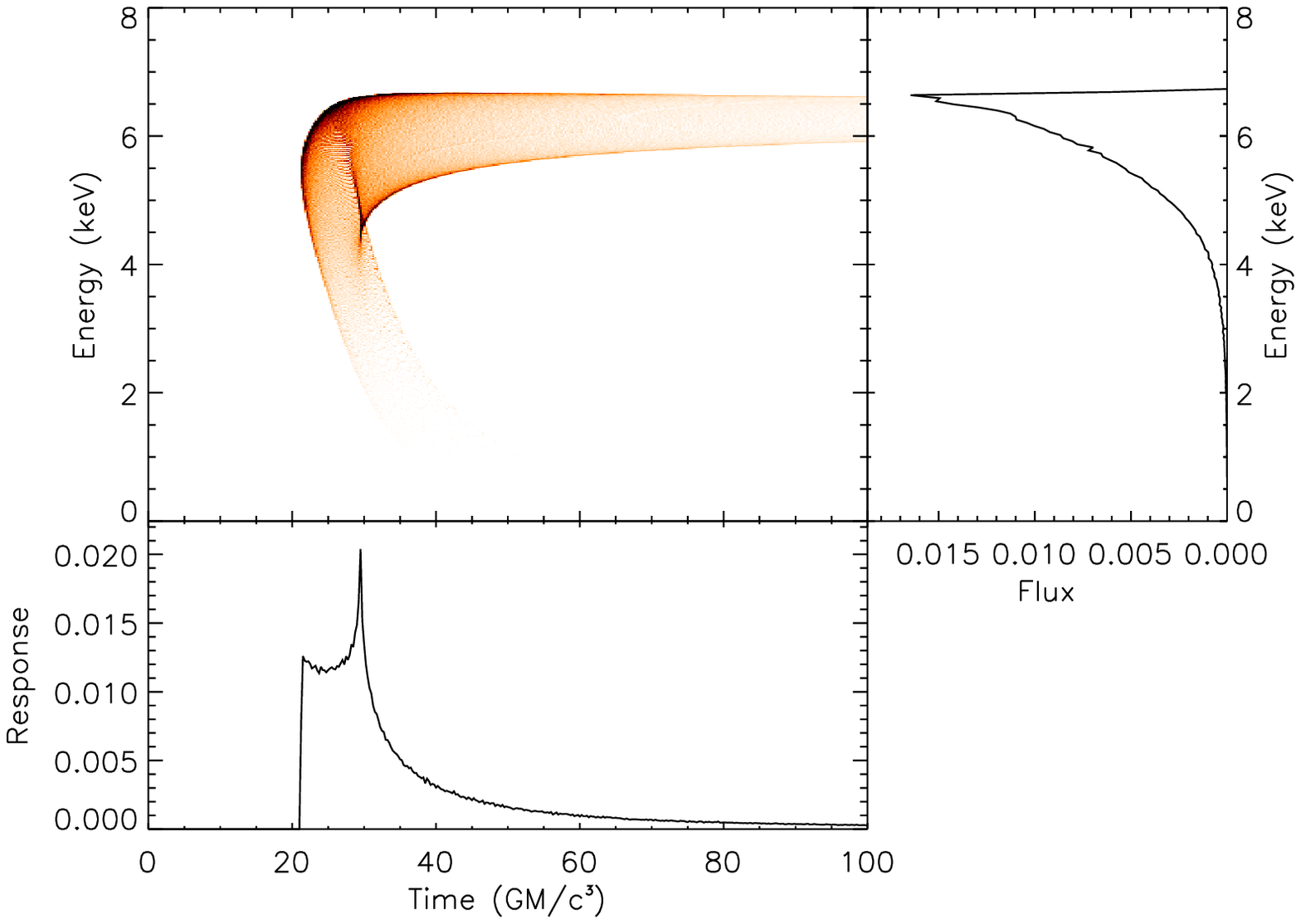}
\includegraphics[width=0.475\textwidth]{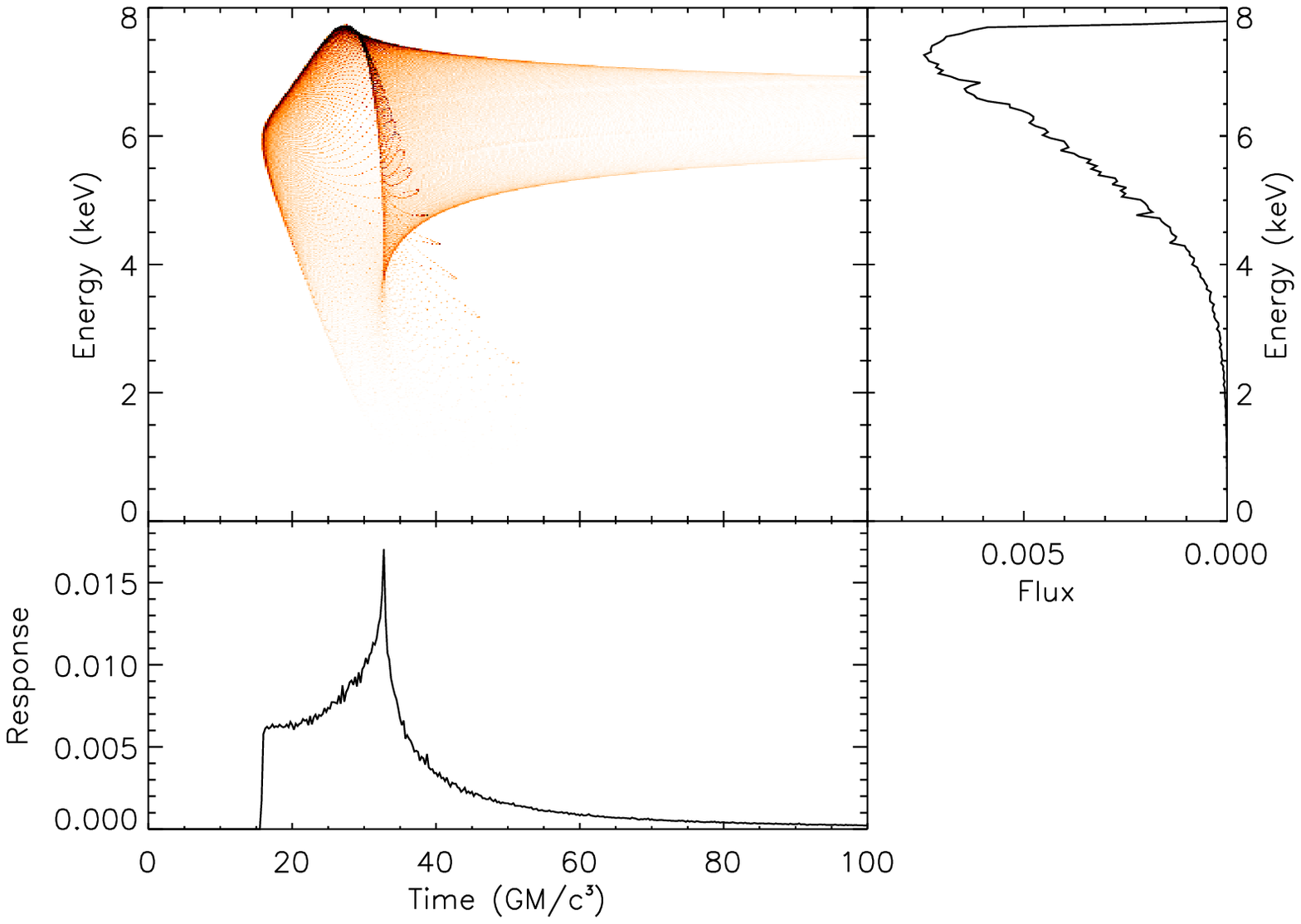}
\includegraphics[width=0.475\textwidth]{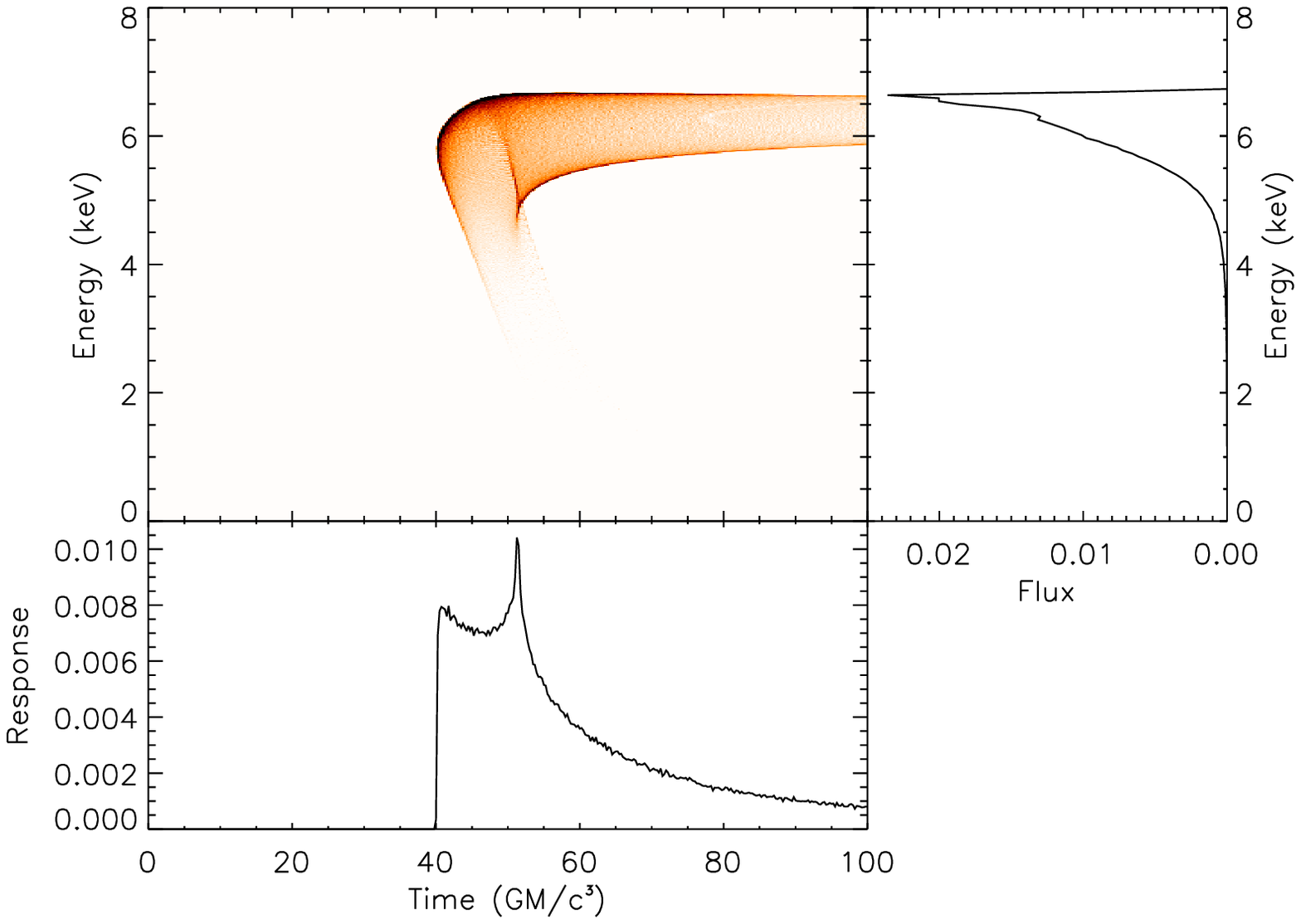}
\caption{Impulse responses for an X-ray source located at a height above a black hole \citep[after][]{reynolds99,cackett13_ngc4151}.  Each diagram indicates a different combination of black hole spin, $a$, height of the X-ray source, $h$, and inclination angle, $i$.  In each diagram the main panel shows the impulse response, the bottom panel shows the energy-averaged response of the Fe K$\alpha$ line, while the right panel shows the time-averaged Fe K$\alpha$ line.  Top left: $a = 0.1$, $h = 10~r_{\rm g}$, $i = 30^\circ$. Top right: $a = 0.998$, $h = 10~r_{\rm g}$, $i = 30^\circ$.  Bottom left: $a = 0.998$, $h = 10~r_{\rm g}$, $i = 60^\circ$. Bottom right: $a = 0.998$, $h = 20~r_{\rm g}$, $i = 30^\circ$.  }
\label{fig:fek_gr_tf}
\end{figure}

We introduced the energy-averaged impulse response for an X-ray source above an accretion disc in Sect.~\ref{sec:enavgtf}.  Now let us consider the energy-dependence of the impulse response.  Here we show the general relativistic impulse responses as calculated by \citet{reynolds99}.  In these models the geometry is assumed to be a simple lamppost model where an X-ray point source is at some height $h$ above a black hole, with the accretion disc inclined at angle $i$ to the observer ($i = 0^\circ$ corresponds to face-on).   In Fig.~\ref{fig:fek_gr_tf} we show four representative impulse responses for different combinations of black hole spin, height of the X-ray source and source inclination\footnote{See also \url{http://stronggravity.eu/public-outreach/animations/reverberation/} for further examples and explanatory animations by M. Dov\v{c}iak.}.

As we noted above, the energy-averaged response of the entire Fe K$\alpha$ line is, to first-order, approximately a top-hat function.  If we then look at the energy-dependence across the Fe K$\alpha$ line we see that both the centroid and the width of the response at a given energy changes across the line.  Thus, as in the simple energy-dependent top-hat examples in section~\ref{sec:tophat_en}, phase wrapping will occur at different frequencies for different energies within the line.  If we look at the time-averaged Fe K$\alpha$ line profile we see the familiar shape \citep{fabian89}.  This means that at each energy within the line we see a different total response, and this will behave like dilution -- the line flux relative to the direct component (which will also be present at these energies) will be strongest where the line peaks, and hence we should expect the longest delays there.  As the strength of the line decreases with decreasing energy we should then also expect the lags to decrease with decreasing energy as the response becomes weaker relative to the direct component.

\begin{figure}
\centering
\includegraphics[width=0.9\textwidth]{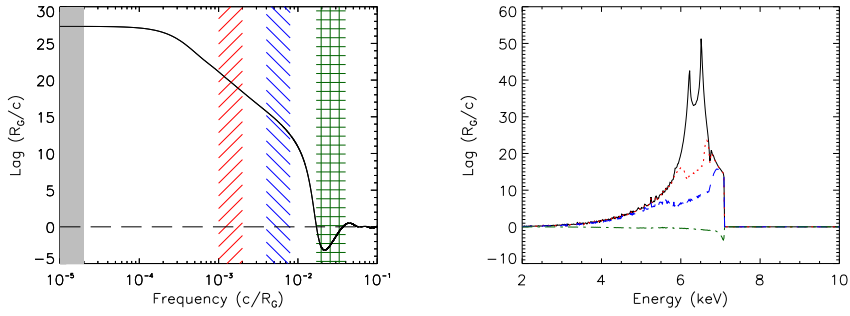}
\caption{{\it Left:} Lag-frequency spectrum for the entire Fe K$\alpha$ line.  The colored regions correspond to the frequency ranges used to calculate the lags as a function of energy. {\it Right:} Lag-energy spectra for four different frequency ranges: $(1 - 2)\times10^{-5}~c/r_{\rm g}$ (black, solid line), $(1 - 2)\times10^{-3}~c/r_{\rm g}$ (red, dotted line), $(4 - 8)\times10^{-3}~c/r_{\rm g}$ (blue, dashed line) and $(1.7 - 4)\times10^{-2}~c/r_{\rm g}$ (green, dot-dashed line).  These are all calculated for $h = 10~r_{\rm g}$, $i = 45^\circ$ and a maximally spinning black hole. Taken from \citet{cackett13_ngc4151}.}
\label{fig:fekft_lagspec}
\end{figure}

Given that to first-order the energy-averaged response of the Fe K$\alpha$ line looks like a top-hat, it should be no surprise that the lag-frequency spectrum for the entire line looks similar to that for a top-hat function (see left panel in Fig.~\ref{fig:fekft_lagspec}).  The exact shape is discussed in detail in \citet{cackett13_ngc4151}.  The energy-dependent lags are less intuitive because of phase wrapping occurring at different frequencies for different energies.  In Fig.~\ref{fig:fekft_lagspec} we show the lag-energy spectrum calculated over a range of different frequencies, and now follow the discussion in \citet{cackett13_ngc4151} describing the behaviour.  We start by looking at the lowest frequencies.  On these longest timescales (lowest frequencies) we will see the response from the entire accretion disc, and thus the lag-energy spectrum shows long lags and a double-horned profile which comes from the outer regions of the disc (see black solid line in Fig.~\ref{fig:fekft_lagspec}).  As we increase the frequency we will start to look at shorter timescales and will begin filtering out the response from the outer part of the disc (the timescale will be short compared to the range of light-travel times to the outer disc).  We therefore start to see a `cut-out' region in the lag-energy spectrum where the lags at energies corresponding to the outer parts of the disc are reduced (see red dotted line).  As the frequency increases further that `cut-out' region gets broader, since we are now looking at increasingly shorter timescales and hence smaller regions of the disc (blue dashed line).  Eventually we get to the frequency where the lags become negative (green dash-dotted line) and we see an inverted lag profile with very small magnitude.  In summary, studying the lags from an Fe~K$\alpha$ line at low frequencies we see the response from the entire disc, while as we go to higher frequencies we progressively filter out the outer parts of the disc, leaving only the response from the inner regions.  Notice that at all frequencies where the lag is positive the red-wing (low energy part) of the lag profile is present -- this is the response coming from the innermost regions of the disc.

As with the lag-frequency spectrum, the lag-energy spectrum is dependent on a number of key parameters.  While for the lag-frequency spectrum the continuum source height and the black hole mass were the most important parameters, when considering the lag-energy spectrum the inclination of the accretion disc is also important in determining the energy where lags are observed.  Black hole spin too has an effect on the lag-energy spectrum.  The exact behaviour as each of these parameters is changed is demonstrated in detail in \citet{cackett13_ngc4151}, and we just briefly comment on each parameter here.  For instance, increasing the height of the X-ray source will increase the overall path length between the direct and reprocessed emission, leading to longer lags. It will also decrease the frequency at which the lags go to zero.  Increasing the black hole mass will increase the overall size-scale of the system, therefore the lags will be larger and the frequency at which the lags go to zero will be lower for a higher black hole mass.  Clearly, both increasing the height and increasing the black hole mass have approximately the same effect on the lag-energy spectrum and thus these parameters are somewhat degenerate -- increasing the black hole mass requires a lower height to achieve approximately the same lag-energy profile.  

The inclination does not have a large effect on the lag-frequency spectrum, however, it does have a big effect on the lag-energy spectrum.  As can be seen from looking at the impulse response in Fig.~\ref{fig:fek_gr_tf}, higher inclinations get a much broader energy response from the disc.  This is simply because the component of the velocity along our line of sight is larger for higher inclinations, and thus Doppler shifts increase with increasing inclination.  This has the most prominent effect on the blue-wing (high energy part) of the lag-energy profile (as will be familiar to anybody who has studied time-averaged iron lines).  As the inclination increases the maximum energy where we see lags increases.  For instance it is at approximately 6.4 keV for nearly face-on systems while for an inclination of 60$^\circ$ it will be at a little greater than 7.5 keV.

Black hole spin also affects the lag-energy spectrum.  Larger black hole spin leads to an innermost-stable circular orbit that is closer to the black hole.  Therefore the surface area of the disc at short lags will increase, and thus we will see more prominent lags from the red-wing.  This region is also where gravitational redshifts are strongest, and therefore extending closer to the disc will decrease the minimum energy (set by the maximum gravitational redshifts) where we see lags.

\subsubsection{Comparison with NGC~4151}\label{sec:ngc4151comparison}
\begin{figure}
\centering
\includegraphics[width=8cm]{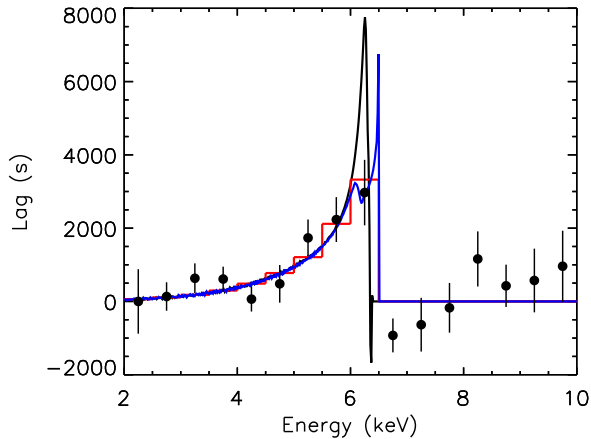}
\caption{Lag-energy spectrum for NGC~4151 in the frequency range $(1 - 2)\times10^{-5}$ Hz \citep[data from][]{zoghbi12}. Best-fitting GR impulse response models for $i = 5^\circ$ (black) and $i= 20^\circ$ (blue) are shown both with $h = 7~r_{\rm g}$, $M = 4.6\times10^7~M_\odot$ and $a=0.998$.  The red line shows the average of the $i = 5^\circ$ model over the same energy bins as the data.  Taken from \citet{cackett13_ngc4151}.}
\label{fig:ngc4151_bestfit}
\end{figure}

Finally, we demonstrate how these models can be applied to real data.  \citet{cackett13_ngc4151} fit the energy-dependent lags in NGC~4151 from \citet{zoghbi12} using the GR impulse responses of \citet{reynolds99} discussed here.  Given the frequency range of the observed lags and the optical reverberation mapping mass for the black hole of $M = 4.6\times10^7~M_\odot$, \citet{cackett13_ngc4151} find the X-ray source to be located at a height of  $h = 7^{+2.9}_{-2.6}~r_{\rm g}$.  We show their best fit in Fig.~\ref{fig:ngc4151_bestfit}.  The fact that the lags are observed to drop sharply above 6.5 keV shows that the inclination $i < 30^\circ$.  Larger values of $i$ would lead to significant lags above 6.5 keV.  Moreover, they found that a maximally spinning black hole was a better fit than $a= 0.1$ at the 1$\sigma$ confidence level.  This first exciting step into fitting the lag-energy spectra already demonstrates the power of such methods.

\newpage

\section{Summary}

The first discovery of X-ray reverberation lags, in 2009, marked the beginning of a new field in X-ray astronomy that is probing the innermost regions around compact objects in a new way.  These short timescale lags measure the physical distance between the compact X-ray source and the inner accretion flow, that will help us understand how these accretion processes work in the strong-field limit.  The main aims of this review are to

\begin{enumerate}
\item Put reverberation lags in the historical context of previous X-ray timing studies, specifically of the low-frequency hard lags found in BHXRBs and AGN.
\item Describe the Fourier timing techniques and the spectral-timing products most often used (i.e., the lag-frequency spectrum, the lag-energy spectrum, and covariance spectrum)
\item Present some of the most significant results, including the soft lag in 1H0707-495,  the Fe~K lag in NGC~4151 and the thermal reverberation in the Galactic black hole X-ray binary, GX~339-4
\item Describe the effects of the impulse response on the lags and outline the general relativistic impulse responses that can model the lag observations and give constraints on the geometry and energetics of the accretion flow.
\end{enumerate}

This review presents the progress made in X-ray reverberation lags since 2009, and we conclude now with a view for the future.

\section{Future Directions}\label{sec:future}

Reverberation time lags are common in X-ray bright, variable,
AGN. They are seen at soft X-ray energies and in the Fe K band. The
presence of Fe K lags is the clearest evidence that reflection
is involved in generating the lags.  
Reverberation lags have now been detected in over two
dozen AGN with timescales correlated with black hole mass. The lag
times correspond to physical lengths of 2 to $20r_{\rm g}$ which means
that at least part of the corona lies within at most $10r_{\rm g}$ of
the black hole; in many cases the implied location is much
closer. Reverberation lags are also now beginning to emerge from
BHXRBs. Modelling gives a consistent picture supporting a compact
corona above the disc of a (possibly rapidly spinning) black hole 
in many of the lag-detected objects.  Now that X-ray reverberation has been confirmed as a physical phenomenon, the next step is to turn the 
measurements of reverberation into a tool to study the close environment of accreting black holes (and in the case of XRBs, neutron stars also).  This step forward will require a combination of further data, new models and model-fitting approaches, and ultimately a new generation of large-area X-ray observatories to fully exploit this powerful new tool to get close to black holes.

\subsection{Near-term improvements in methods and data}
There is considerable scope for further advances in the study of X-ray
reverberation using current instruments, particularly {\it XMM-Newton}, which
generally gives the highest count rates and longest continuous exposures, so is optimal for the work.  Longer observations of the
objects which have already shown reverberation signals will improve on
the reverberation spectra and enable changes in the corona to be
tracked \citep[as seen in IRAS 13224-3809 by ][]{kara13b} and 
low-frequency propagation effects to be unravelled from high
frequency reverberation.  An important advance will be to systematically detect and compare the reverberation signatures in a wide range of AGN classes, for example to determine whether there are luminosity-dependent effects or systematic differences in the structure of the corona and inner accretion flow in radio-loud and radio-quiet AGN, or between AGN with distinct spectral energy distributions (which may indicate differences in the inner accretion flow structure).  

The development of techniques \citep{zoghbi13_gaps} to estimate lags from non-contiguous light curves , e.g. with gaps due to the satellite orbit, is an important development which opens up the possibility of lag measurement using a range of X-ray missions such as {\it Suzaku} and the recently launched {\it NuSTAR} mission \citep{harrison13}.  The hard X-ray sensitivity of {\it NuSTAR} also allows the extension of reverberation measurements in AGN to harder X-rays, which is revealing the expected reverberation signature of the reflection continuum, around the so-called `Compton hump' (e.g. \citealt{zoghbi14}, Kara et al., in prep. and see Fig.~\ref{fig:nustarlags}).

Significant advances can also be expected in the study of reverberation in BHXRBs.  As noted in Sect.~\ref{sec:sensitivity}, BHXRB spectral-timing measurements at frequencies where reverberation lags are expected are currently in the regime of few photons per characteristic time-scale, where the signal-to-noise of lag measurements scales {\it linearly} with count rate.  Thus a very significant advance can be made simply by observing brighter sources: a ten times brighter source can yield the same quality data as a hundred times longer exposure time!  Unfortunately however, {\it XMM-Newton} EPIC-pn is limited by instrumental pile-up effects and telemetry limitations to observing sources not much brighter than 0.1-0.2~Crab, so that the crucial higher luminosities where state transitions occur, QPOs appear and jets switch on and off, are probably out of reach for studying in detail using reverberation: these studies will probably have to wait for new missions in the near and longer-term (see Sect.~\ref{sec:future:missions}).  

Another promising line of research is the study of lags in neutron stars, in particular at the frequencies of the so-called `kHz QPOs', typically ranging from 600--1000~Hz.  The combination of a high rms ($\sim 10$~per~cent) concentrated into a narrow frequency range makes these timing signatures especially useful to search for time-lags, with uncertainties of only a few microseconds possible, e.g. in archival data from {\it RXTE} \citep{barret13,deavellar13}.  One advantage for modelling reverberation in neutron star systems is that, at least for the kHz QPOs, it is likely that the continuum emission originates primarily from the boundary layer of accretion close to the neutron star surface \citep{gilfanov03}.  Thus the driving continuum source's geometry and location is probably more tightly constrained than in accreting black holes.  Reverberation mapping of accreting neutron stars could ultimately yield useful constraints on the neutron star mass and radius, and hence provide a useful tool to constrain the neutron star equation of state.

\subsection{Advances in models and model-fitting}
\label{sec:future:models}
 On the theory and modelling side, only the first step has been taken
into fitting the energy dependence of Fe~K lags using the simplest
disc model consisting of an X-ray point source located some height
above the black hole.  In reality the X-ray corona is almost certainly
both radially and vertically extended.  Including such a geometry is
clearly the next step towards building a more realistic model.
Moreover, rather than just considering the Fe~K line, including the
lags expected from the full reflection spectrum will allow for
self-consistent modelling of the lags over a much broader energy
range.  Furthermore, the lag-frequency and lag-energy spectra have
thus far generally been fitted separately.  Moving to an approach
involving fitting the lags as a function of frequency and energy will
provide the best constraints from the current data.  More ambitious
still would involve fitting the cross spectra directly,
self-consistently matching all the properties of the light curves in
each band.  Fitting the cross-spectrum will allow us to use the valuable information contained in the {\it frequency-dependent amplitudes} of variability in each band, which can encode the
smearing effects associated with more extended impulse responses. Such methods will require a merger of current X-ray spectral-fitting tools with timing techniques such as the Fourier transform, so that the required spectral-timing products can be calculated from the energy-dependent model impulse response and also correctly folded through the appropriate detector response matrices (which need to be accounted for to properly fit energy-dependent spectral-timing data).

As is also the case with conventional spectral-fitting, models for spectral-timing data will necessarily include degeneracies, e.g. the degeneracy between black hole mass and continuum source height in the simple lamppost case (Sect.~\ref{sec:model:endeptf}). More realistic models may incur other degeneracies between their larger number of free parameters, however it is important to stress that the degeneracies encountered in fitting spectral-timing data are generally different from (and in some cases orthogonal to) the degeneracies encountered in conventional spectral fitting, e.g. the source height affects the illumination pattern on the disc and hence the detailed shape of the broad iron line, which can be constrained by fitting the line shape.  Therefore there is hope that the full combination of spectral and timing information will break the model degeneracies that limit the power of model fits which use only part of the spectral and timing data.  Ultimately, with good enough data, a completely model-independent measure of the impulse response of the inner disc can be determined, simply by deconvolving the driving continuum (measured from a continuum-dominated band) from the reflection-band light curve.  Deconvolution is a notoriously noisy process however, so this important goal will likely remain out of reach until X-ray observatories with large collecting areas are launched (see Sect.~\ref{sec:future:missions}).

While the Fe~K reverberation lags appear to be relatively consistent between objects, it is already clear from the diversity of soft lag shapes (see Fig.~\ref{lagen_stack}) that the situation in the soft band is complex, with the possibility of multiple components contributing to the soft band, including additional soft continuum components (e.g. disc thermal and/or soft optically thin emission) as well as photoionised reflection. The modelling of continuum lag components is also important for understanding the lower-frequency lags that may be linked to propagation effects in the accretion flow and corona.  In principle these lags could be studied using the same impulse-response approach as the reverberation lags (e.g. following \citealt{kotov01}), although it is not yet clear as to whether a simple linear response of these continuum components can be assumed to generate the impulse response in this case. 

More generally, dealing with non-linear models for spectral variability and reverberation is an important challenge which may need to be addressed if, e.g. the reflection ionisation parameter varies with illuminating flux.  Flux-correlated variations in spectral shape of the primary continuum will also lead to non-linear variations in the shape of the reflection.  Also if X-ray heating of the disc is significant compared to internal (`viscous') heating, the thermal reverberation of the disc will be correlated with changes in blackbody temperature and consequent non-linear variation of the disc blackbody continuum.  Non-linear impulse response functions may be investigated as an approach to modelling this behaviour: they have already been explored in the time-series and signal processing literature in other fields (e.g. see \citealt{potter00}) and these approaches could be adopted for X-ray spectral-timing analysis.

It is also important to consider how to deal with intrinsic incoherence in the light curves, because this will lead to correlated `systematic' type errors between energy-bins in spectral-timing products such as the lag spectrum.  The statistical errors between adjacent energy bins will likely be overestimated in cases where intrinsic incoherence causes a drop in the observed `raw' coherence used to estimate errors in the lags.  However, the broader lag versus energy shape will also be more difficult to interpret in these cases, being a composite of the lag-energy behaviour of the uncorrelated spectral components. This problem could be mitigated by careful selection of a reference energy band dominated by the illuminating continuum, which reverberation signatures are likely to be well-correlated with.

\subsection{Future X-ray observatories}
\label{sec:future:missions}
The power of X-ray reverberation measurements is that they are able to map the emitting regions close to compact objects in terms of the absolute physical scale.  The scales that can be mapped are sub-micro to nanoarcseconds in angular size on the sky - smaller than can ever be accessed with any X-ray imaging technology, for decades into the future.  Large detected count rates are key, especially for XRBs where we have seen that signal-to-noise scales linearly with count rate.  Thus the most important requirement to access these scales is a large effective area instrument, capable of measuring large count rates, combined with good energy resolution (CCD-quality or better) to probe the variable relativistically-smeared reflection.

In the near future, two X-ray instruments will enable important advances in XRB spectral-timing.  The Large Area X-ray Proportional Counter (LAXPC) on the Indian {\it ASTROSAT} mission will pick up where {\it RXTE} left off, bringing a similar effective area at the Fe~K energy, and larger area at higher energies, with good time resolution \citep{paul13}.  The energy resolution of the LAXPC is too coarse for detailed energy-dependent spectral-timing studies, but it could be useful for identifying broader features.  NASA's {\it Neutron star Interior Composition ExploreR}, {\it NICER}, uses X-ray concentrator optics with silicon drift detectors to obtain CCD-quality resolution and a larger soft response than {\it XMM-Newton} \citep{gendreau12}.  Critically, its non-imaging concentrators and detector technology mean that {\it NICER} will not suffer from the restrictions on source flux faced by {\it XMM-Newton}, so that important advances in spectral-timing can be made for bright X-ray binaries, potentially allowing the first X-ray reverberation mapping of black holes during the state transitions.

These new missions will, during this decade, provide useful steps forward in our exploration of reverberation in X-ray binaries, but do not push to significantly larger detector effective areas than previous instruments.  Significant breakthroughs which exploit the full potential of X-ray reverberation will require a step up to square metres of collecting area, which will be attained by the end of the 2020s.  The {\it ATHENA} mission \citep{nandra13} will increase collecting area compared to {\it XMM-Newton} by more than an order of magnitude at soft X-rays, and by a factor of $3-4$ at Fe~K energies, enabling significantly improved X-ray spectral-timing and reverberation measurements, especially in the soft band, for both XRBs and AGN.  The sensitivity of {\it ATHENA} to faint soures, especially in the soft band, will allow the reverberation signal to be discovered in many more fainter objects, so reaching to a much wider luminosity range of sources than is accessible today and opening up the study of the innermost regions of a wide variety of AGN classes. 

Another promising advance, if it can be exploited, is the development of non-imaging, large-area silicon drift detectors, e.g. as suggested for use on the proposed {\it Large Observatory for X-ray Timing}, {\it LOFT} \citep{feroci12}.  This technology could affordably reach collecting areas two orders of magnitude larger than {\it XMM-Newton} at Fe~K and harder energies, while maintaining CCD-like energy resolution.  As with the advances made by {\it ATHENA} in the soft X-ray band, such an instrument would lead to a transformational advance in the study of XRBs and AGN via reverberation and spectral-timing, by enabling Fe~K lags to be measured with extremely high precision. 

\begin{figure}
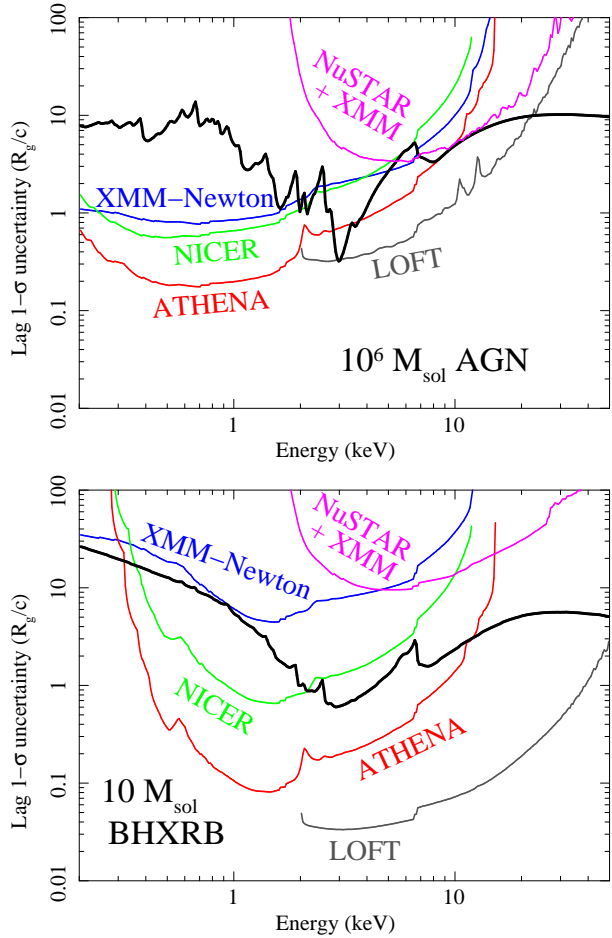

\centering
\includegraphics[angle=270,width=0.7\textwidth]{sensitivity_agn.ps}
\includegraphics[angle=270,width=0.7\textwidth]{sensitivity_xrb.ps}
\caption{{\it Top:} AGN lag-sensitivity curves (and expected lag-energy spectrum in black, with arbitrary absolute lag offset), for the 1--3$\times 10^{-3}$~Hz frequency range for 100~ks exposures with {\it XMM-Newton} EPIC-pn, {\it NuSTAR}, {\it NICER}, {\it ATHENA} and a 10~m$^{2}$ area {\it LOFT}-class mission (the larger background of the {\it LOFT} mission is included in the calculation).  The curves correspond to a 2--10~keV flux of $4\times 10^{-11}$~erg~s$^{-1}$~cm$^{-2}$, with continuum photon index $\Gamma=2$ and Galactic absorbing column $N_{\rm H}=4\times 10^{20}$~cm$^{-2}$.  The assumed black hole mass is $10^{6}$~M$_{\odot}$, corresponding to an NLS1.  However, higher-mass black holes will show the same signal-to-noise over an equivalent mass-scaled frequency range (see Sect.~\ref{sec:sensitivity}). {\it Bottom:} BHXRB lag-sensitivity curves for the 50--150~Hz range and 100~ks exposures, assuming a 10~M$_{\odot}$ black hole with a 2--10~keV flux of 0.2~Crab (1 Crab $=2.4\times10^{-8}$~erg~s$^{-1}$~cm$^{-2}$) in the case of {\it XMM-Newton} and {\it NuSTAR} and 1~Crab for the other missions which can observe significantly brighter sources without significant pileup or deadtime effects.  A typical bright hard state lag-energy spectrum is shown in black.  The assumed photon index is also $\Gamma=2$ and the Galactic absorbing column is $N_{\rm H}=6\times 10^{21}$~cm$^{-2}$.}
\label{fig:sensitivitycurves}
\end{figure}

To illustrate the advances which can be made in the future, we show in Fig.~\ref{fig:sensitivitycurves} the `sensitivity curves' for lag measurements with planned or proposed new missions with CCD-class or better spectral-resolution, also including {\it XMM-Newton} (EPIC-pn) and {\it NuSTAR} for comparison.  The curves assume 100~ks exposure time and typical spectral and variability properties for an AGN (top panel) and a black hole X-ray binary in the hard or intermediate state (lower panel), with the lags measured in the 1--3$\times 10^{-3}$~Hz and 50--150~Hz ranges for the AGN and BHXRB respectively.  The most recent instrument response matrices were used and reference bands were chosen to be 0.5--10~keV for {\it XMM-Newton}, {\it NICER} and {\it ATHENA}\footnote{For {\it ATHENA} we use the Wide Field Imager (WFI) effective area curve for both cases, since this instrument is able to observe sources with flux up to 1~Crab with minimal signal degradation due to pile-up. However, the high-resolution spectrometer (X-ray Integral Field Unit, X-IFU) will be able to observe fainter sources such as AGN, with only slightly reduced sensitivity compared to the WFI.} and 2--20~keV for {\it LOFT} and {\it NuSTAR}.  For {\it NuSTAR}, the lag measurements are also assumed to include reference photons from simultaneous {\it XMM-Newton} data (otherwise the {\it NuSTAR} performance is significantly worse). The lag uncertainties are calculated assuming that the lag-energy spectrum is binned at an energy resolution which is slightly larger than the resolution of CCD-type instruments (the lag bin-size $\Delta E=0.08\times E^{1/2}$, or $\sim 200$~eV at 6.4~keV), and double that for {\it NuSTAR} (to account for the poorer energy resolution).  Further information is given in the figure caption.  For comparison, the expected reverberation lag-energy spectrum is shown in black (assuming a zero-spin black hole with central source height of 4~$r_{g}$, disc thermal reverberation is also included in the BHXRB case).  It should be borne in mind that the relative lags between energies in the lag-energy spectrum are most important, not the absolute scale.  Thus, features will only be clear in the lag-energy spectrum if the lag uncertainty is small {\it compared to the change in lag which defines the feature}.  Hence, detailed Fe~K reverberation studies will remain challenging without the very large areas of a {\it LOFT}-class mission.  

It is clear that the biggest improvements between missions can be seen in the BHXRB case, where the quality of reverberation measurements substantially overtakes that of AGN for the multi~m$^{2}$ class missions.  This is because of the linear-scaling of lag signal-to-noise with count rate that is seen in the XRB regime (Sect.~\ref{sec:sensitivity}).  For example, although {\it NICER} has a similar effective area to the {\it XMM-Newton} EPIC-pn (which is only able to detect disc reverberation at substantially lower frequencies where the origin is more ambiguous), it does not suffer the same flux limits due to pile-up and telemetry dropouts, so is able to observe much brighter XRBs.  Thus NICER should make significant inroads into the study of disc thermal reverberation at high frequencies (e.g. to determine how disc inner radius and radial temperature profile vary with accretion state).  The same bright XRBs can be observed by {\it ATHENA} and the {\it LOFT}-class mission, which have significantly larger area, leading to large improvements in sensitivity.  It is clear also that {\it ATHENA} and {\it LOFT}-class missions are complementary in their energy coverage for AGN and BHXRBs, with {\it LOFT} allowing detailed study of the iron line and reflection continuum, while {\it ATHENA} covers the more complex soft excess (disc thermal emission and photoionised reflection).

\subsection{Concluding remarks}
X-ray reverberation is in some sense a phenomenon discovered ahead of its time.  The original assessments of detection of X-ray reverberation signals were based on time-domain methods and studies of time-dependent spectra from the next generation of instruments (e.g. \citealt{young00}).  Initial approaches using these techniques did not reveal the reverberation signatures hidden in the data \citep{reynolds00,vaughan01}, which were also convolved with other time-dependent spectral variability, e.g. continuum lags on longer time-scales. However, thanks to the combination of Fourier timing and spectral techniques, we now find these signals to be within reach.  Rapid advances can and should be made in modelling the reverberation signatures, but these will require the development of new ways of understanding spectral-timing data.  The requirements for improvements in data are relatively easily achievable with current developments in technology and future missions: large collecting areas coupled with moderate to good spectral resolution.  Given the enormous potential already demonstrated by reverberation measurements over the past few years and the first steps into modelling these signals, it seems likely that over the next two decades spectral-timing, using combined energy-dependent timing products to fit in Fourier frequency and energy space simultaneously, will replace stand-alone spectroscopy or timing as the method of choice for studying the innermost regions around accreting compact objects.  These advances will break through the model degeneracies faced by studying objects in the Fourier or spectral domain alone and allow X-ray studies of the close-environments of compact objects to reach their full potential.

\begin{acknowledgements}
We would like to thank Simon Vaughan for valuable discussions and comments and Zaven Arzoumanian and the {\it NICER} team for providing the latest {\it NICER} instrument response.  ACF acknowledges support from the European Union Seventh Framework Programme (FP7/2007-2013) under grant agreement no. 312789 (STRONGGRAVITY), the UK Science and Technology Facilities Council, and the ERC Advanced Grant FEEDBACK.  EK thanks the Gates Cambridge Scholarship. DRW is supported by a CITA National Fellowship. 

\end{acknowledgements}

\bibliography{agn}   % name your BibTeX data base
\bibliographystyle{aa}      % basic style, author-year citations
%\begin{thebibliography}{}

%\input{analysis_methods_ed_phil.bbl}

% etc
%\end{thebibliography}

\end{document}